\documentclass[twocolumn]{aastex63}
\usepackage{amsmath}

\newcommand\kms{km$\,$s$^{-1}$}
\newcommand\Msol{M$_{\odot}$}

\newcommand{\hi}{H\,{\sc i}}

\accepted{June 7th 2021}
\submitjournal{ApJ}

\shorttitle{UDG formation through tidal interactions}
\shortauthors{Jones et al.}
\graphicspath{{./}{figures/}}

\begin{document}

\title{Evidence for Ultra-Diffuse Galaxy Formation Through Tidal Heating of Normal Dwarfs}

\correspondingauthor{Michael G. Jones}
\email{jonesmg@arizona.edu}

\author[0000-0002-5434-4904]{MICHAEL G. JONES}
\affiliation{Steward Observatory, University of Arizona, 933 North Cherry Avenue, Rm. N204, Tucson, AZ 85721-0065, USA}

\author[0000-0001-8354-7279]{PAUL BENNET}
\affiliation{Space Telescope Science Institute, 3700 San Martin Drive, Baltimore, MD 21218, USA}

\author[0000-0001-9649-4815]{BUR\c{C}\.{I}N MUTLU-PAKD\.{I}L}
\affil{Kavli Institute for Cosmological Physics, University of Chicago, Chicago, IL 60637, USA}
\affil{Department of Astronomy and Astrophysics, University of Chicago, Chicago IL 60637, USA}

\author[0000-0003-4102-380X]{DAVID J. SAND}
\affiliation{Steward Observatory, University of Arizona, 933 North Cherry Avenue, Rm. N204, Tucson, AZ 85721-0065, USA}

\author[0000-0002-0956-7949]{KRISTINE SPEKKENS}
\affiliation{Department of Physics and Space Science, Royal Military College of Canada P.O. Box 17000, Station Forces Kingston, ON K7K 7B4, Canada}
\affiliation{Department of Physics, Engineering Physics and Astronomy, Queen’s University, Kingston, ON K7L 3N6, Canada}

\author[0000-0002-1763-4128]{DENIJA CRNOJEVI\'{C}}
\affil{University of Tampa, 401 West Kennedy Boulevard, Tampa, FL 33606, USA}

\author[0000-0001-8855-3635]{ANANTHAN KARUNAKARAN}
\affiliation{Department of Physics, Engineering Physics and Astronomy, Queen’s University, Kingston, ON K7L 3N6, Canada}

\author[0000-0002-5177-727X]{DENNIS ZARITSKY}
\affiliation{Steward Observatory, University of Arizona, 933 North Cherry Avenue, Rm. N204, Tucson, AZ 85721-0065, USA}



\begin{abstract}

We have followed up two ultra-diffuse galaxies (UDGs), detected adjacent to stellar streams, with Hubble Space Telescope (HST) imaging and \hi \ mapping with the Jansky Very Large Array (VLA) in order to investigate the possibility that they might have a tidal origin. With the HST F814W and F555W images we measure the globular cluster (GC) counts for NGC 2708-Dw1 and NGC 5631-Dw1 as $2^{+1}_{-1}$ and $5^{+1}_{-2}$, respectively. NGC 2708-Dw1 is undetected in \hi \ down to a 3$\sigma$ limit of $\log (M_\mathrm{HI}/\mathrm{M_\odot}) = 7.3$, and there is no apparent \hi \ associated with the nearby stellar stream.
There is a 2$\sigma$ \hi \ feature coincident with NGC 5631-Dw1. However, this emission is blended with a large gaseous tail emanating from NGC 5631 and is not necessarily associated with the UDG. The presence of any GCs and the lack of clear \hi \ connections between the UDGs and their parent galaxies strongly disfavor a tidal dwarf galaxy origin, but cannot entirely rule it out. The GC counts are consistent with those of normal dwarf galaxies, and the most probable formation mechanism is one where these UDGs were born as normal dwarfs and were later tidally stripped and heated. We also identify an over-luminous ($M_\mathrm{V} = -11.1$) GC candidate in NGC 2708-Dw1, which may be a nuclear star cluster transitioning to an ultra-compact dwarf as the surrounding dwarf galaxy gets stripped of stars.

\end{abstract}



\section{Introduction} 
\label{sec:intro}

Examples of very low surface brightness (LSB) galaxies have been known and studied for decades \citep[e.g.][]{Sandage+1984,Impey+1988,Thompson+1993}. 
Improvements in LSB imaging have resulted in larger and more extreme samples of LSB dwarfs \citep[e.g.][]{Jerjen+2000,Conselice+2003,Mieske+2007} over time. Contemporary deep imaging surveys have further extended the accessible LSB dwarf population to more extreme cases of LSB and spatial extent, and since the detection of tens or hundreds of extremely LSB and spatially extended galaxies in the Coma, Virgo, Perseus, and Fornax clusters \citep{vanDokkum+2015,Mihos+2015,Koda+2015,Yagi+2016,Wittmann+2017,Venhola+2017}, dubbed ultra-diffuse galaxies (UDGs), there has been a flurry of interest in the most exceptional LSB dwarf galaxies. 
The typical definition of a UDG (effective radius $>1.5$ kpc and central $g$-band surface brightness $>24$ mag arcsec$^{-2}$) is motivated largely by observational capabilities and as a result they may simply be the extreme of a continuum of dwarf galaxy properties \citep[e.g.][]{Conselice+2018}. However, as they represent the extreme of LSB dwarf galaxies they are a relevant test of galaxy formation models.
Their prevalence across all environments, from clusters \citep{vanderBurg+2016,ManceraPina+2018,Zaritsky+2019,Karunakaran+2020b,Lee+2020,Iodice+2020}, to groups \citep{Merritt+2016,Bennet+2017,Spekkens+2017,vanderBurg+2017,Trujillo+2017,Roman+2017,Karunakaran+2020a}, to the field \citep{Martinez-Delgado+2016,Leisman+2017,Greco+2018a,Greco+2018b,Janowiecki+2019,Prole+2019,Roman+2019,Prole+2020}, demands an explanation for how such extreme galaxies can form in such abundance. As has been proposed by several authors, there are likely multiple formation mechanisms at work \citep{Papastergis+2017,Pandya+2018,Jiang+2019,Liao+2019,Carleton+2019,Wright+2021}, and the UDGs found in different environments may well have formed via different pathways.

A number of potential formation mechanisms have been proposed to date. To try to add clarity we have grouped these hypotheses into 5 categories as follows:
\begin{itemize}
    \item Star formation feedback: In this class of mechanism UDGs are the result of repeated episodes of star formation (SF) feedback driving gas out to large radii, creating a dark matter (DM) halo with a very low central concentration \citep{DiCintio+2017,Chan+2018}, which in turn causes the stellar distribution to expand.
    \item Spin parameter: In this scenario, UDGs may just be the result of galaxies that formed in low-mass halos with particularly large spin parameters. The higher specific angular momentum of the halo would prevent gas from effectively collapsing into a dense structure, resulting in a diffuse galaxy \citep{Amorisco+2016,Rong+2017}. 
    \item Failed L$^\ast$ galaxies\footnote{Here an L$^\ast$ galaxy is intended to mean a galaxy with stellar mass of $\sim$10$^{11}$ \Msol, near the turn over of the galaxy stellar mass function.}: \citet{vanDokkum+2015} originally suggested that UDGs may be failed  L$^\ast$ galaxies that did not form stars at the rate expected for their halo mass and physical size, perhaps due to their SF being stunted upon entering a cluster \citep{Yozin+2015}.
    \item Mergers: Early-time mergers of low-mass galaxies could potentially result in present day UDGs as a result of the additional energy and angular momentum injected by the merger \citep{Wright+2021}.
    \item Tidal origin: This category covers two related but different types of formation scenario; tidal interactions rarefying a satellite galaxy's structure \citep{Conselice+2018,Carleton+2019,Tremmel+2020} and the formation of diffuse tidal dwarf galaxies (TDGs) in a strong interaction event between two other, larger, galaxies \citep{Bennet+2018}.
\end{itemize}
Some authors also propose that the observed population of UDGs is a result of the combination of two or more of these explanations \citep[e.g.][]{Trujillo+2019,Ruiz-Lara+2019}. For example, early SF feedback may make a proto-UDG, which is then more susceptible to tidal perturbations \citep{Martin+2019,Jackson+2021}, or by invoking different mechanisms in the formation of field and group/cluster UDGs \citep{Jiang+2019,Liao+2019,Sales+2020}.

One important tool for understanding the origins of UDGs is their globular cluster (GC) systems. Initial studies indicated that UDGs have unusually large GC populations given their stellar mass \citep{Beasley+2016a,Peng+2016,vanDokkum+2016,Beasley+2016b,vanDokkum+2017}. Both the abundance and the dynamics of the GC systems indicated that these UDGs were heavily DM-dominated galaxies \citep[see also][]{Penny+2009}, but the findings conflicted over whether the GC systems corresponded to $\sim$10$^{11}$ or $\sim$10$^{12}$ \Msol \ halos (i.e. Large Magellanic Cloud-like or Milky Way-like) and therefore whether or not they fit the ``failed L$^\ast$ galaxies" hypothesis. However, there is now mounting evidence from stellar populations and kinematics that most UDGs reside in dwarf-mass halos \citep[e.g.][]{Ruiz-Lara+2018,Ferre-Mateu+2018,Gannon+2020}. More recent studies of larger samples of UDGs also appear to show that UDGs mostly have similar numbers of GCs as other dwarf galaxies \citep{Amorisco+2018,Somalwar+2020}, or at least fewer than initially thought \citep{Lim+2020,Saifollahi+2021}. Though not all such studies agree on this point \citep[c.f.][]{Forbes+2020}. Meanwhile, theoretical work has suggested that cluster UDGs may naturally be expected to host rich GC systems \citep{Carleton+2021}.

The stellar populations of a handful of UDGs have also been analyzed, both through spectral energy distribution modeling \citep{Pandya+2018,Greco+2018b} and deep optical spectra \citep{Ruiz-Lara+2018,Greco+2018b,Ferre-Mateu+2018,Martin-Navarro+2019,Muller+2020}. These studies agree that the stellar populations of cluster UDGs are generally dominated by intermediate to old stellar populations ($\sim$3-7 Gyr), that they are metal-poor, and $\alpha$-enhanced. The results for more isolated UDGs are similar, but these  have ongoing SF \citep{Greco+2018b,Martin-Navarro+2019,Barbosa+2020}. These results could be consistent with most of the proposed formation mechanisms, but an explanation (and more thorough census) of the strong $\alpha$-enhancement seen in some UDGs is still required.

Finally, the \hi \ line width or stellar velocity dispersion measurements for UDGs indicate that they are mostly slowly rotating \citep{Leisman+2017,Greco+2018b,Martin-Navarro+2019,Karunakaran+2020b,Gannon+2020} and the velocity width function of \hi-bearing UDGs is much steeper than for a blind, \hi-selected population \citep{Jones+2018}. This is in tension with the suggestion that UDGs may reside in particularly massive DM halos (for their stellar mass), where rotation velocities would be expected to be higher. Furthermore, resolved kinematics of the \hi \ gas \citep{ManceraPina+2019,ManceraPina+2020} suggest that field UDGs may deviate strongly from the baryonic Tully-Fisher relation (BTFR), in that they rotate very slowly for their baryonic masses.
Although this finding does not rule out the high spin parameter formation scenario\footnote{Even slowly rotating objects can have high spin, if sufficiently extended, and the DM halo spin does not necessarily have to match that of the baryonic matter.}, it raises potentially challenging questions about the dynamics of UDGs.
However, \citet{He+2019} found that edge-on UDGs do fall on the BTFR, suggesting that uncertain inclination corrections may partially explain these apparent deviations.

In this work we focus on the last of the formation scenarios listed above, formation through tidal interactions. Although this cannot be the explanation for UDGs found in relative isolation it may explain at least some of the UDGs found in groups, and therefore potentially those in clusters, as clusters accrete groups over time. Furthermore, a TDG origin would naturally explain the claimed lack of DM in some UDGs \citep{vanDokkum+2018,vanDokkum+2019}, although this may also be explicable by tidal interactions stripping the DM from existing UDGs \citep{Montes+2020}.

In particular, we focus on two UDGs identified by \citet{Bennet+2018} as being in close proximity to tidal streams in loose group environments. \citet{Bennet+2018} performed a semi-automated search \citep{Bennet+2017} for UDGs across a $\sim$150 deg$^2$ portion of the Canada-France-Hawaii Telescope Legacy Survey (CFHTLS), identifying two new UDGs \citep[which fit the criteria of][see Table \ref{tab:udg_props}]{vanDokkum+2015} in close proximity to stellar streams (Figure \ref{fig:CFHT_tails}) connected to their nearby parent galaxies (NGC 2708 and NGC 5631). 
The presence of these two UDGs close to stellar streams has three possible explanations:
\begin{enumerate}
    \item They are TDGs formed from dense clumps of gas stripped from the parent galaxies during some past interactions.
    \item They were normal dwarf galaxies that fell into groups and were subsequently `puffed up' by tidal interactions.
    \item They were UDGs before falling into a group and the proximity to the stellar streams is either coincidental or inconsequential for their diffuse nature.
\end{enumerate}

In order to distinguish between these possibilities we obtained both Jansky Very Large Array (VLA) and Hubble Space Telescope (HST) observations to image the neutral gas environment of the UDGs and to detect any globular clusters (GCs) they may host. If the UDGs are relatively young TDGs then they are likely to still exhibit a clear connection in \hi \ gas with their parent object. Indeed, \hi \ maps of interacting systems frequently contain candidate TDGs \citep[e.g.][]{Duc+2000,Lee-Waddell+2016,Jones+2019,Lee-Waddell+2019,Koribalski2020} and streams of \hi \ in galaxy groups typically remain visible for longer than stellar streams (when both are originally stripped from the same location). On the other hand, GCs provide complementary information as TDGs are not expected to host any GCs, while an otherwise normal dwarf that has undergone tidal heating should host a few GCs \citep[in line with other dwarfs, e.g.][]{Harris+2013}, and some UDGs in dense environments have been found to be extremely rich in GCs given their apparent stellar mass \citep[e.g.][]{Beasley+2016a,vanDokkum+2016}.

This paper is organized as follows. In the next section we present a brief summary of the systems in which the two UDGs reside. Section \ref{sec:obs} provides an overview of the observations, reduction, and GC candidate selection. In Section \ref{sec:results}  we present the results of both the \hi \ imaging and the GC search. In Section \ref{sec:discuss} we discuss the potential formation mechanism for these UDGs, and finally present our conclusions in Section \ref{sec:conclusion}.

\section{Summary of targets}

Here we provide a brief description of the two galaxies assumed to be the hosts of the UDGs presented in \citet{Bennet+2018}.

\subsection{NGC 2708}

NGC 2708 (Figure \ref{fig:NGC2708_mom0}) is a member of the NGC 2698 group, an $N=8$ group identified by \citet{Makarov+2011}. NGC 2708 has two dwarf companions to the north, NGC 2709 and PGC 1075058 (or NGC 2708A). NGC 2709 is an S0 dwarf apparently devoid of gas, while PGC 1075058 is strongly detected in \hi \ \citep{Pisano+2002} and has a faint, blue stellar component. There is a massive \hi \ stream extending from NGC 2708 to the north. \citet{Pisano+2002} argue that it is likely too massive to have originated from PGC 1075058, and suggest that it may have been stripped from NGC 2709. DECaLS \citep[Dark Energy Camera Legacy Survey,][]{Dey+2019} images of NGC 2708 show signs of tidal disturbances in its outer stellar disk. The UDG NGC 2708-Dw1 is to the SE of NGC2708, away from the previously known dwarf companions. It is connected to its host galaxy by a faint, linear stellar stream (Figure \ref{fig:CFHT_tails}, left). For consistency with \citet{Bennet+2018} we adopt a distance of 40.6 Mpc for NGC 2708.

\subsection{NGC 5631}

NGC 5631 (Figure \ref{fig:NGC5631_mom0}) is an elliptical galaxy on the edge of a loose group also containing NGC 5667 and 5678 among various other potential members \citep{Pisano+2004}. The galaxy has previously been studied both in ionized and neutral gas. \citet{Silchenko+2009} found that the stellar and ionized gas components of NGC 5631 are counter-rotating and suggest its present form is the result of an ongoing, gas-rich, minor merger. \citet{Serra+2014} found further evidence of a highly disturbed object in the \hi \ kinematics, which are (globally) not aligned with the ionized gas kinematics, but do twist in the inner region of the galaxy to align where there is overlap. Neutral gas and stellar kinematics are also misaligned \citep{Serra+2014}, varying from counter-rotation to an almost polar orbit, depending on the radius. NGC 5631-Dw1 is to the SW of NGC 5631. It is connected to its host galaxy by a faint, curved stellar stream (Figure \ref{fig:CFHT_tails}, right). For consistency with \citet{Bennet+2018} we adopt a distance of 28.4 Mpc for NGC 5631.

\begin{table}
\centering
\caption{UDG properties from \citet{Bennet+2018} \label{tab:udg_props}}
\begin{tabular}{lcc}
\hline\hline
Name                          & NGC 2708-Dw1    & NGC 5631-Dw1    \\ \hline
R.A.   (J2000)                       & 08:56:12.7      & 14:26:13.6      \\
Dec. (J2000)                         & -03:25:14.8     & +56:31:50.2     \\
$m_g$/mag                     & $19.3 \pm 0.3$  & $20.5 \pm 0.4$  \\
$\mu_{g,0}$/mag arcsec$^{-2}$ & $24.9 \pm 0.6$  & $27.3 \pm 0.7$  \\
$r_\mathrm{h}$/arcsec                  & $13.2 \pm 2.9$  & $15.6 \pm 3.6$  \\
$r_\mathrm{h}$/kpc                     & $2.60 \pm 0.57$ & $2.15 \pm 0.50$ \\
Dist./Mpc                     & 40.6            & 28.4            \\ \hline
\end{tabular}
\end{table}

\section{Observations \& Reduction}
\label{sec:obs}

The two targets were observed with both the VLA and HST in order to constrain the \hi\ content, to search for evidence of nearby neutral gas tidal features, and to provide images with sufficient depth and resolution to identify GCs.

\subsection{VLA observations}

The two targets were observed with the VLA in D-configuration during December 2019 as part of the project 19B-022 (PI: K. Spekkens). Both targets had total on-source integration times of approximately 3.25 hr and the data were recorded (simultaneously) in two correlator configurations. The first had a total of 128 channels over a bandwidth of 8 MHz, giving a channel resolution of 62.5 kHz, or approximately 13 \kms. The second setup consisted of a narrower band, 4 MHz, and finer channel resolution, 7.81 kHz ($\sim$1.7 \kms). Assuming that the velocity width of the targets is $\geqslant13$ \kms \ the coarser frequency resolution suffices for the purposes of a detection experiment, and the wider band extends the redshift range that can be searched to over $\pm$800 \kms \ (centered on the parent galaxy). Therefore, the remainder of this work will focus on the first correlator configuration.

The observations of NGC 2708-Dw1 were impacted by intermittent, broadband radio frequency interference (RFI), and as a result approximately 41\% of the target data were flagged, compared to $\sim$21\% for NGC 5631-Dw1, which had minimal amounts of RFI. The data were Hanning smoothed to lessen the spectral ringing of the RFI at the expense of some spectral resolution. The worst RFI was flagged manually and those flags were supplemented with \texttt{CASA}'s \citep[Common Astronomy Software Applications,][]{CASA} automated flagging algorithms \texttt{tfcrop} and \texttt{rflag} (the latter after initial calibrations). Calibration and imaging used standard \texttt{CASA} tasks. Imaging was performed with Brigg's weighting of robust=2 to maximize the detectability of faint, extended sources. The resulting beam size and rms noise for NGC 2708-Dw1 were 68.3\arcsec $\times$ 54.6\arcsec \ and 0.34 mJy/beam. For NGC 5631-Dw1 these values are 58.7\arcsec $\times$ 53.1\arcsec \ and 0.30 mJy/beam.

\begin{figure*}
    \centering
    \includegraphics[width=0.75\columnwidth]{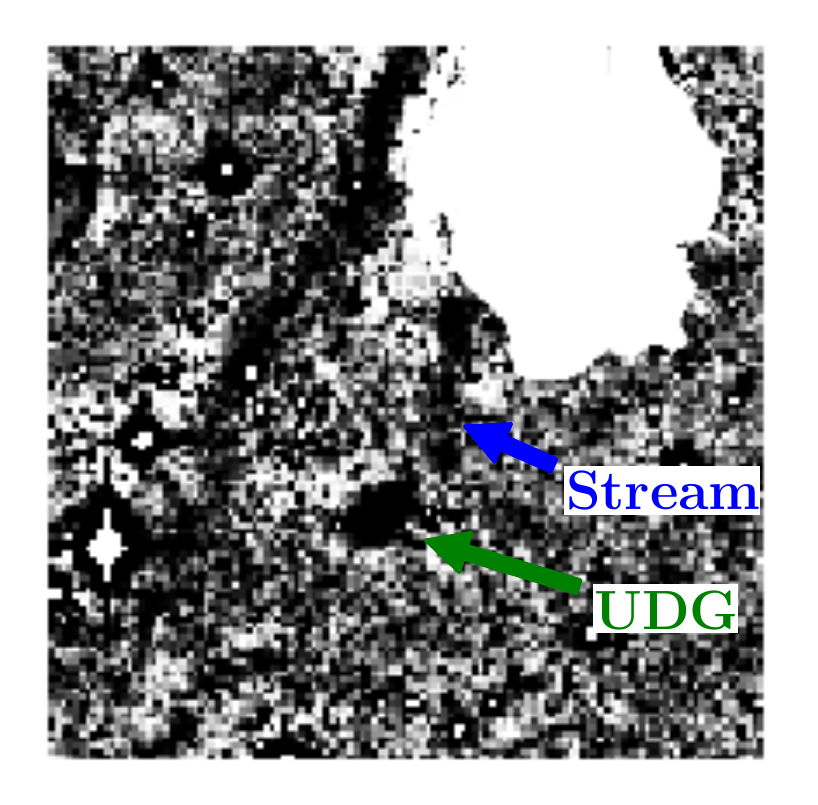}
    \includegraphics[width=0.75\columnwidth]{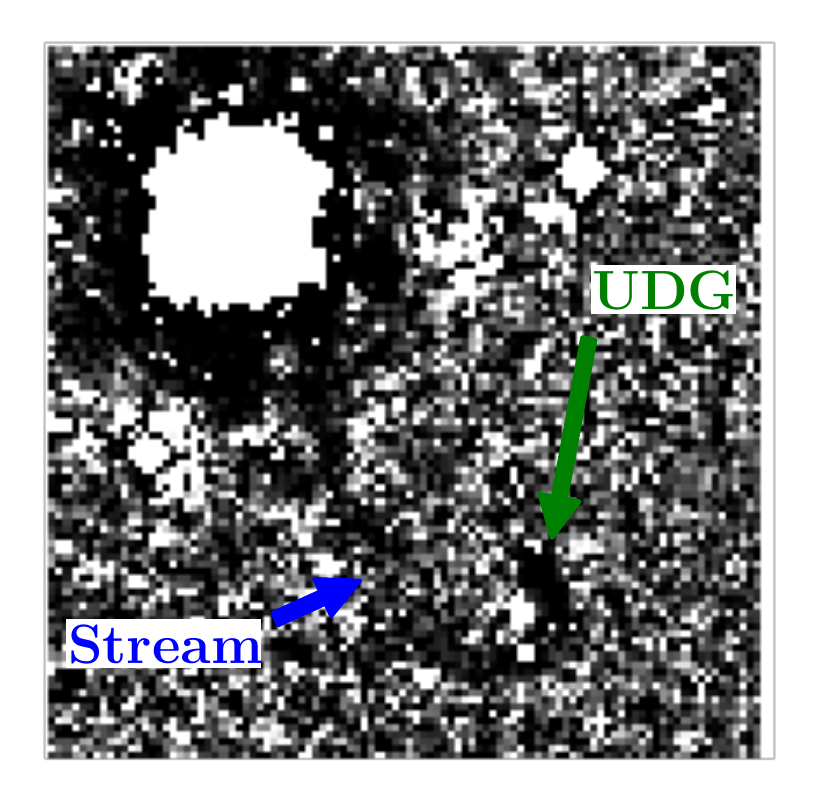}
    \caption{Binned and masked $g$-band CFHT images \citep{Bennet+2018} to enhance LSB features. NGC 2708-Dw1 and its associated stream are shown on the left, NGC 5631-Dw1 and its associated stream on the right. In each panel the green arrow indicates the UDG and the blue arrow the stellar stream. Each panel shows a 5.5\arcmin $\times$ 5.5\arcmin \ FoV.}
    \label{fig:CFHT_tails}
\end{figure*}

\begin{figure*}
    \centering
    \includegraphics[width=\columnwidth]{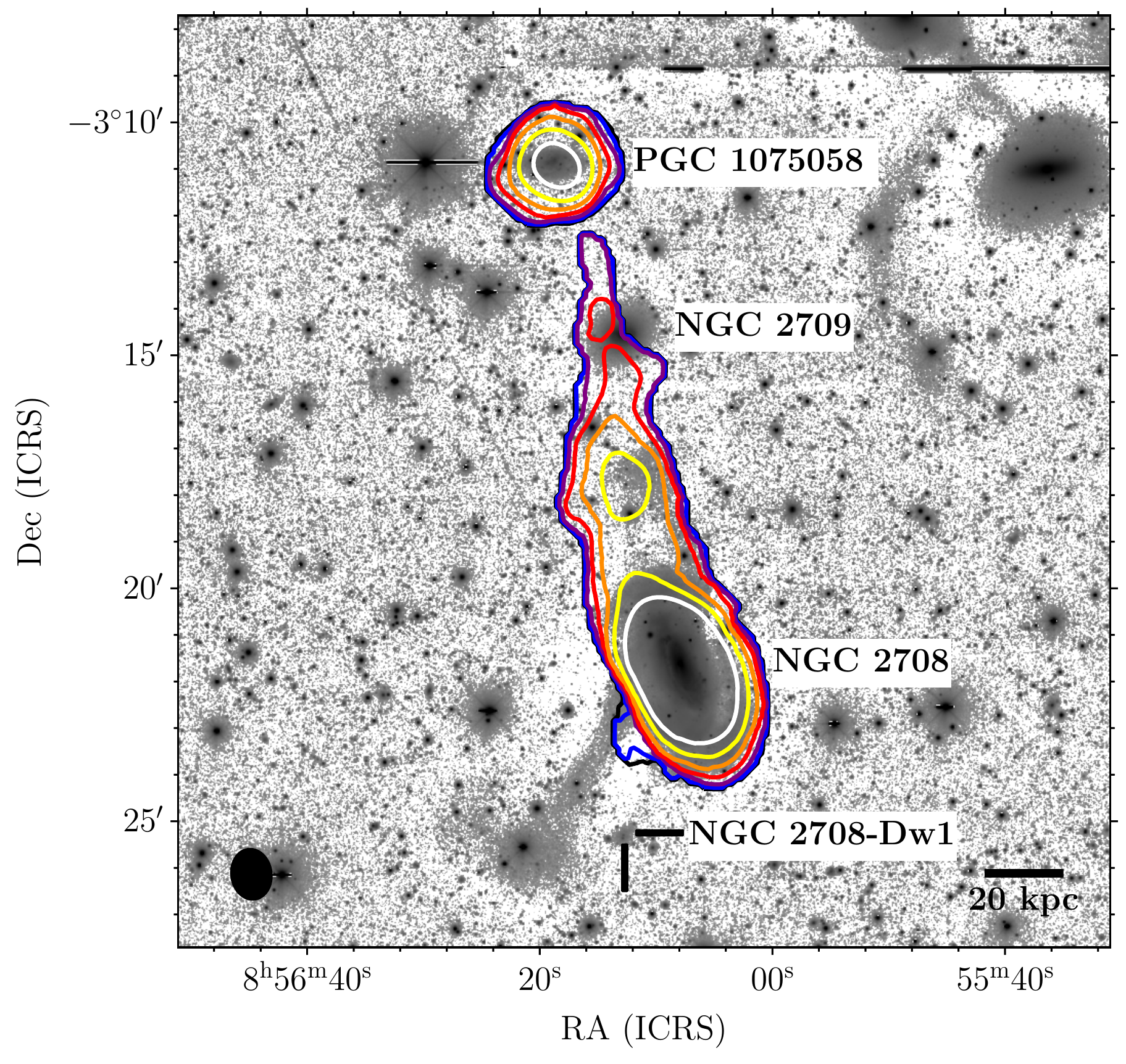}
    \includegraphics[width=\columnwidth]{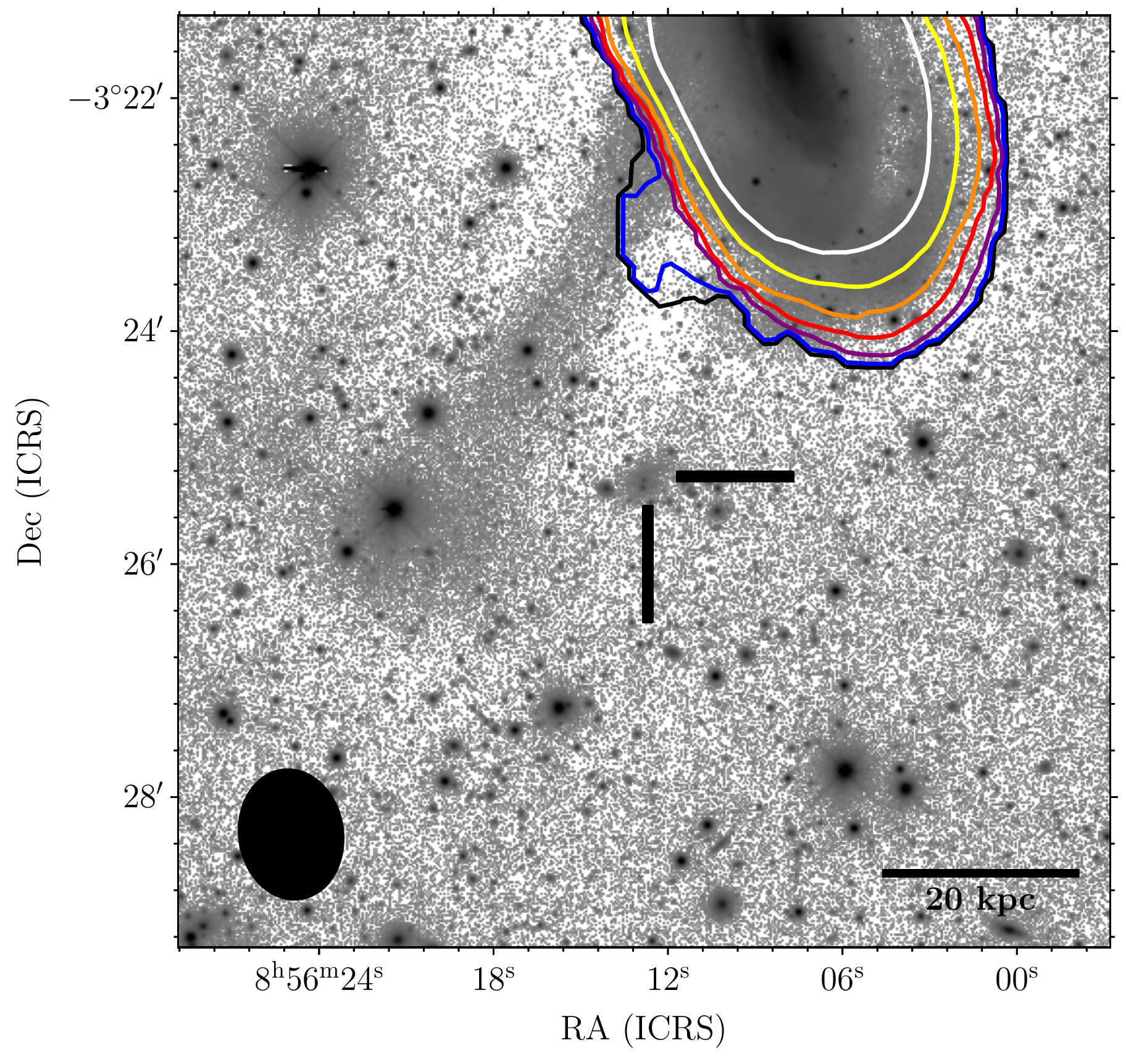}
    \caption{\textit{Left}: Contours of integrated \hi\ emission NGC 2708 (and PGC 1075058) overlaid on an $r$-band image from DECaLS. The contours begin at 3$\sigma$ for one channel (0.136 $\mathrm{Jy\,km\,s^{-1}}$ per beam, 0.0796 $\mathrm{M_{\odot}\,pc^{-2}}$, or $9.96 \times 10^{18} \; \mathrm{cm^{-2}}$) and each subsequent contour is double the value of the previous one. The position of NGC 2708-Dw1 is indicated and the VLA synthesized beam size is shown by the filled black ellipse. \textit{Right}: As before, but showing a zoomed-in FoV centered on NGC 2708-Dw1.}
    \label{fig:NGC2708_mom0}
\end{figure*}

\begin{figure*}
    \centering
    \includegraphics[width=\columnwidth]{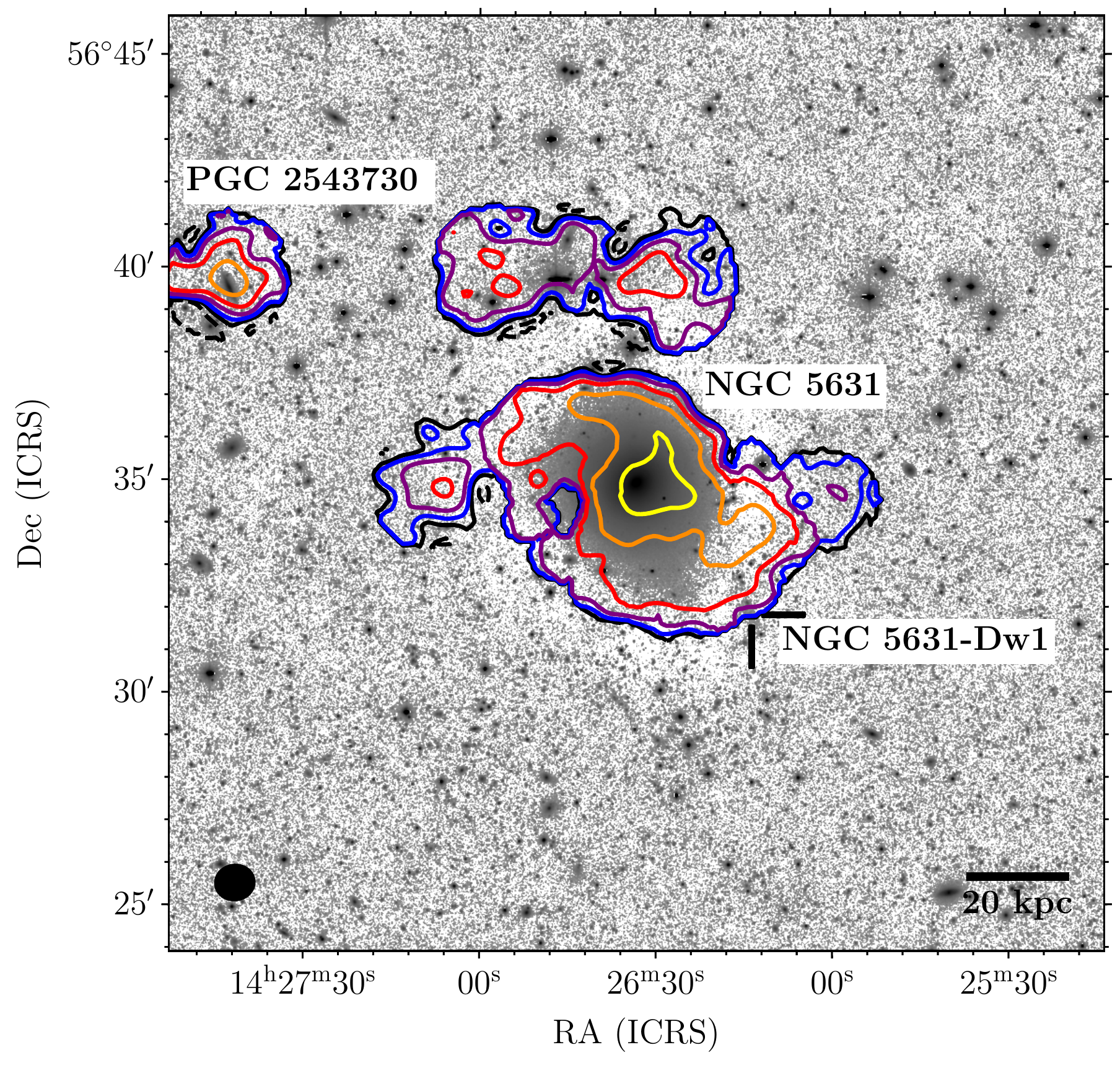}
    \includegraphics[width=\columnwidth]{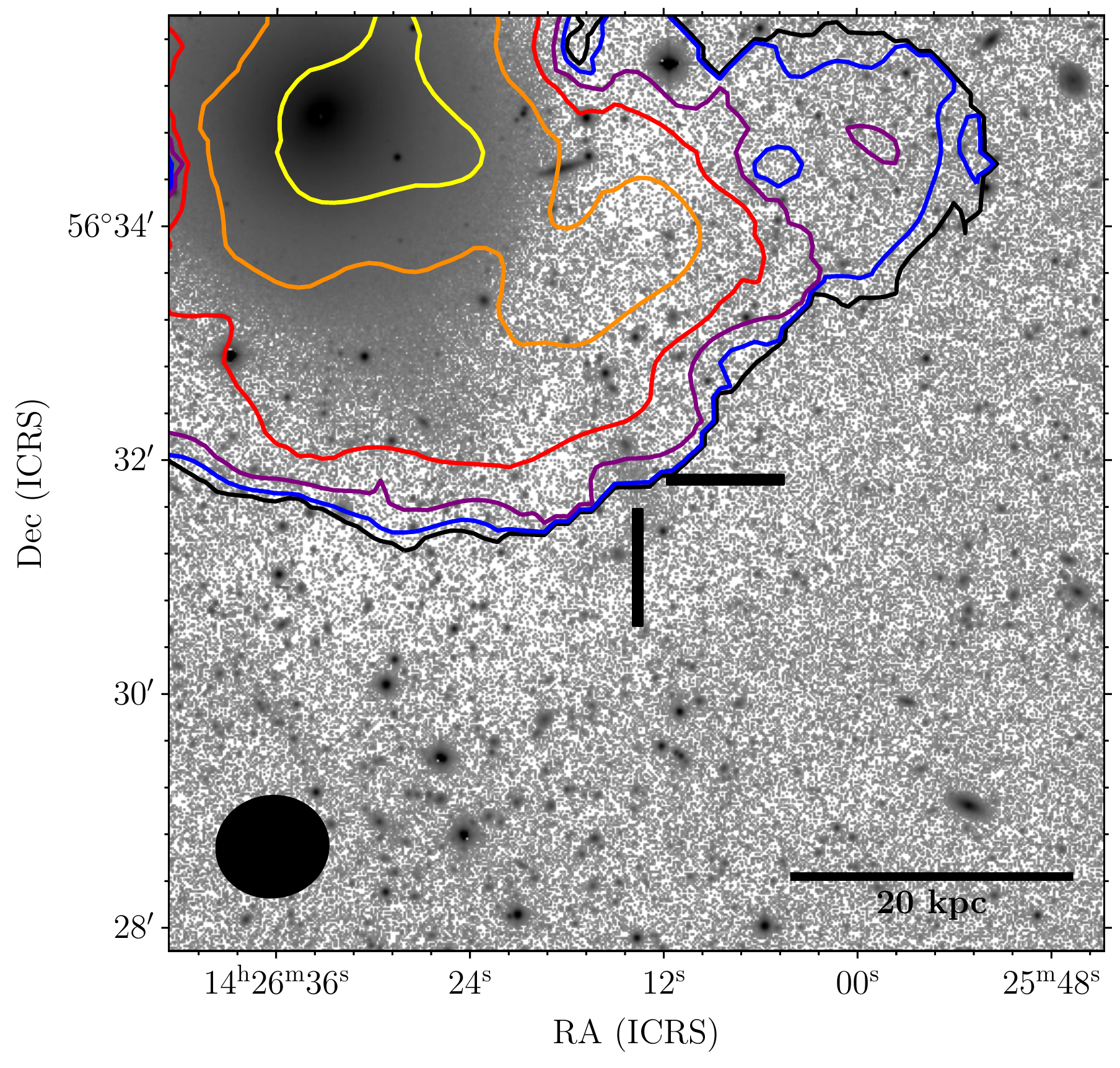}
    \caption{\textit{Left}: Contours of integrated \hi \ emission of NGC 5631 (and PGC 2543730) overlaid on an $r$-band image from DECaLS. The contours begin at 3$\sigma$ for one channel (0.120 $\mathrm{Jy\,km\,s^{-1}}$ per beam, 0.0841 $\mathrm{M_{\odot}\,pc^{-2}}$, or $1.05 \times 10^{19} \; \mathrm{cm^{-2}}$) and each subsequent contour is double the value of the previous one. The position of NGC 5631-DW1 is indicated and the VLA synthesized beam size is shown by the filled black ellipse. \textit{Right}: As before, but showing a zoomed-in FoV centered on NGC 5631-Dw1.}
    \label{fig:NGC5631_mom0}
\end{figure*}

\subsection{HST observations}

NGC 2708-Dw1 and NGC 5631-Dw1 were observed in January and April 2020 as part of HST program 15874 (PI: K.~Spekkens). Both targets were observed with the Advanced Camera for Surveys (ACS) in the filters F814W and F555W, with total integration times of over 2000 s in each filter. In addition, the Wide Field Camera 3 (WFC3) images were obtained simultaneously to provide a nearby reference background field. However, the orientation of the spacecraft was not specified, so this reference field was determined at random.

We use \texttt{DOLPHOT} \citep{Dolphin2000,dolphot} with standard ACS and WFC3 parameters to align the individual exposures and generate a combined source catalog for each of the fields. The V and I band magnitude conversions generated by \texttt{DOLPHOT} are corrected for Galactic extinction \citep{Schlafly+2011} before GC selection. Values of V and I magnitudes quoted throughout this work are extinction corrected, but magnitudes quoted for the F555W and F814W filters are not.

\subsection{Globular cluster candidate selection}
\label{sec:GC_selection}

Assuming a nominal maximum radius for a GC of 20~pc translates to a maximum angular diameter of $\sim$0.2\arcsec \ at the distance of NGC 2708 and $\sim$0.3\arcsec \ at the distance of NGC 5631. From stars in the image we estimate the FWHM of the HST PSF as 0.14\arcsec (for both filters), meaning that for both targets GCs may appear either as point sources or as marginally resolved sources.

Starting with the \texttt{DOLPHOT} photometry we limit the sample to sources classified as stars, with $\mathrm{SNR} > 5$, and no quality flags in their photometry. We require the sharpness parameter to be within the range $-$0.3 to 0.3 (in both filters) to remove any very extended sources (or overly compact ones), and the roundness parameter to be less than 0.3 in both filters (where 0 corresponds to circular). Crowding is limited to a maximum of 0.5 mag in either filter (the additional flux removed due to overlapping sources), and the magnitude uncertainty is restricted to be less than 0.3 mag (in both bands). Finally, we use a color cut of $0.5 < V-I < 1.5$ which encompasses the colors of almost all GCs \citep[e.g.][]{Brodie+2006}, but will eliminate some blue star forming clumps that may have been identified within galaxies in the field.

In addition to the above selection we calculate the concentration index of each GC candidate by comparing the magnitudes calculated in circular apertures (on the background subtracted F814W image) of diameters 4 and 8 pixels \citep[similar to the approach of][]{Peng+2011,Beasley+2016b}. As the GCs may not be exactly point sources we allow a very broad range of concentrations (0.3--0.85 mag), which encompasses the locus of point sources, but also extends well below this and is only strict enough to eliminate remaining background galaxies with clear structure as well as any remaining diffraction spike artifacts.

For NGC 2708 a magnitude floor of $\mathrm{F814W} < 25.95$ mag is imposed, which corresponds to 90\% completeness based on artificial star tests. For NGC 5631 the limit is $\mathrm{F814W} < 25.59$. These limits were estimated by placing $\sim$10000 artificial stars, of similar colors to the GC candidates, throughout the ACS FoV and measuring the recovered fraction as a function of apparent magnitude.
Following \citet{Peng+2011} we assume that the globular cluster luminosity function (GCLF) is Gaussian in shape and consider two possible forms, one for massive galaxies with a mean of $\mu_\mathrm{I} = -8.12$ mag ($\mu_\mathrm{I,Vega} = -8.56$ mag) and a width of $\sigma_\mathrm{I} = 1.37$ mag \citep{Peng+2011}, and another for dwarf galaxies with a mean of $\mu_\mathrm{I} = -7.67$ mag ($\mu_\mathrm{I,Vega} = -8.1$ mag) and a width of $\sigma_\mathrm{I} = 1.1$ mag \citep{Miller+2007}. Assuming the former for the GCLF, 85\% of GCs would be brighter than our completeness limit for NGC 2708 Dw1, and 91\% for NGC 5631 Dw1. For the latter case 81\% and 90\% of GCs would lie above the completeness limit. Due to the minimal corrections implied we make no correction to the raw GC counts.

\section{Results}
\label{sec:results}

\subsection{\hi \ content and connection to host}
\label{sec:hi_props}

Both of the \hi \ cubes were visually inspected to identify any potential detection of either UDG, or an \hi \ feature that is likely connected or associated with them. In addition to the visual inspection we used \texttt{SoFiA} \citep[Source Finding Application,][]{SoFiA,Serra+2015} to create a source mask with a 4$\sigma$ threshold (and 90\% reliability threshold) after smoothing over kernels of approximately 1 and 2 times the beam size, as well as over 0 and 2 spectral channels. The moment 0 maps generated with the \texttt{SoFiA} masks are shown overlaid on DECaLS images in Figures \ref{fig:NGC2708_mom0} and \ref{fig:NGC5631_mom0}.

\subsubsection{NGC 2708-Dw1}

\begin{figure}
    \centering
    \includegraphics[width=\columnwidth]{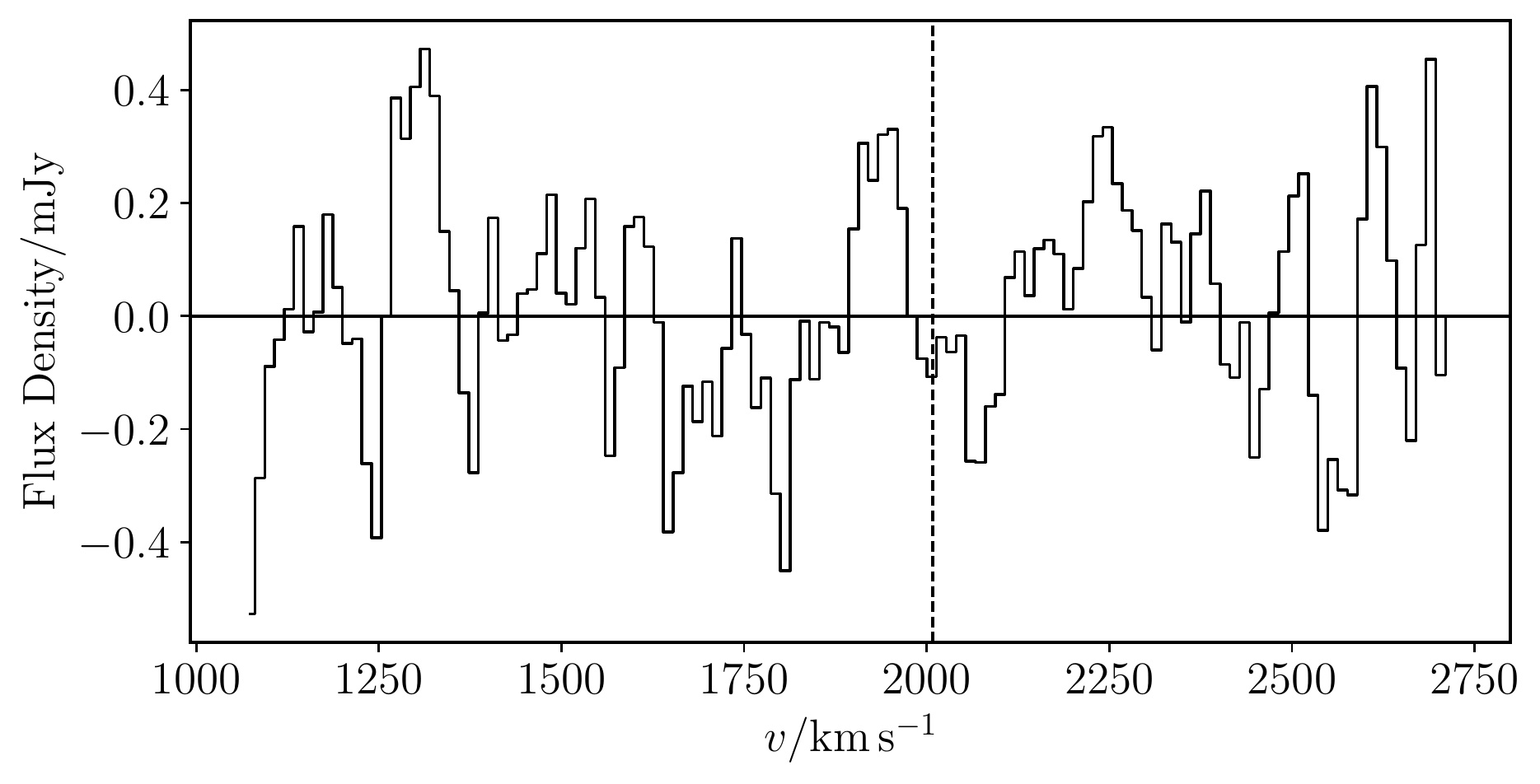}
    \caption{\hi \ spectrum at the position of NGC 2708-Dw1 extracted over a single synthesized beam from the \hi \ cube. The vertical dashed line indicates the central velocity of NGC 2708.}
    \label{fig:NGC2708DW1_spec}
\end{figure}

Figure \ref{fig:NGC2708_mom0} shows the \hi \ emission contours in the vicinity of NGC 2708-Dw1. The UDG is undetected within the source mask. There is an extension on the SE edge of the disk of NGC 2708 which might be associated with the stellar stream adjacent to NGC 2708-Dw1, but inspection of the \hi \ cube (see channel maps in Appendix \ref{sec:chan_maps}) suggest, albeit at low S/N, that this \hi \ feature is more likely oriented almost perpendicular to the stellar stream. This would indicate that the stellar stream probably originated from a gas-poor object rather than from the disk of NGC 2708. A spectrum at the position of NGC 2708-Dw1 covering the full velocity range of the data (extracted over 1 synthesized beam area) is shown in Figure \ref{fig:NGC2708DW1_spec}, but this shows no statistically significant feature.

To place a limit on the \hi \ mass of NGC 2708-Dw1 we assume that any \hi \ emission would be point-like at the stated resolution and take a fiducial velocity width of 30 \kms. At a distance of 40.6 Mpc, with an rms noise of 0.34 mJy/beam, this gives a 3$\sigma$ detection limit of $\log (M_\mathrm{HI}/\mathrm{M_\odot}) = 7.3$.\footnote{Note that a factor of $\sqrt{8/3}$ is included to account for the fact that the data are Hanning smoothed.}

\subsubsection{NGC 5631-Dw1}
\label{sec:hi_props_ngc5631dw1}

Figure \ref{fig:NGC5631_mom0} (right) shows the \hi \ emission in the vicinity of NGC 5631-Dw1. In this case the moment map provides relatively little information as the position of the UDG overlaps with the \hi \ emission associated with the parent galaxy. 
However, no clear dense knot or kinematically distinct feature is present. 
The \hi \ tail overlapping the UDG's position at first glance appears to emanate from the southern side of the galaxy, but on close inspection of the cube the tail to the NE and that to the SW appear to form one continuous structure that wraps most of the way around the galaxy. It is even possible that the dense clump (for which we couldn't identify any optical counterpart) to the north of the galaxy (Figure \ref{fig:NGC5631_mom0}) is also part of this same structure, however, deeper \hi \ observations would be required to know for certain.

\begin{figure}
    \centering
    \includegraphics[width=\columnwidth]{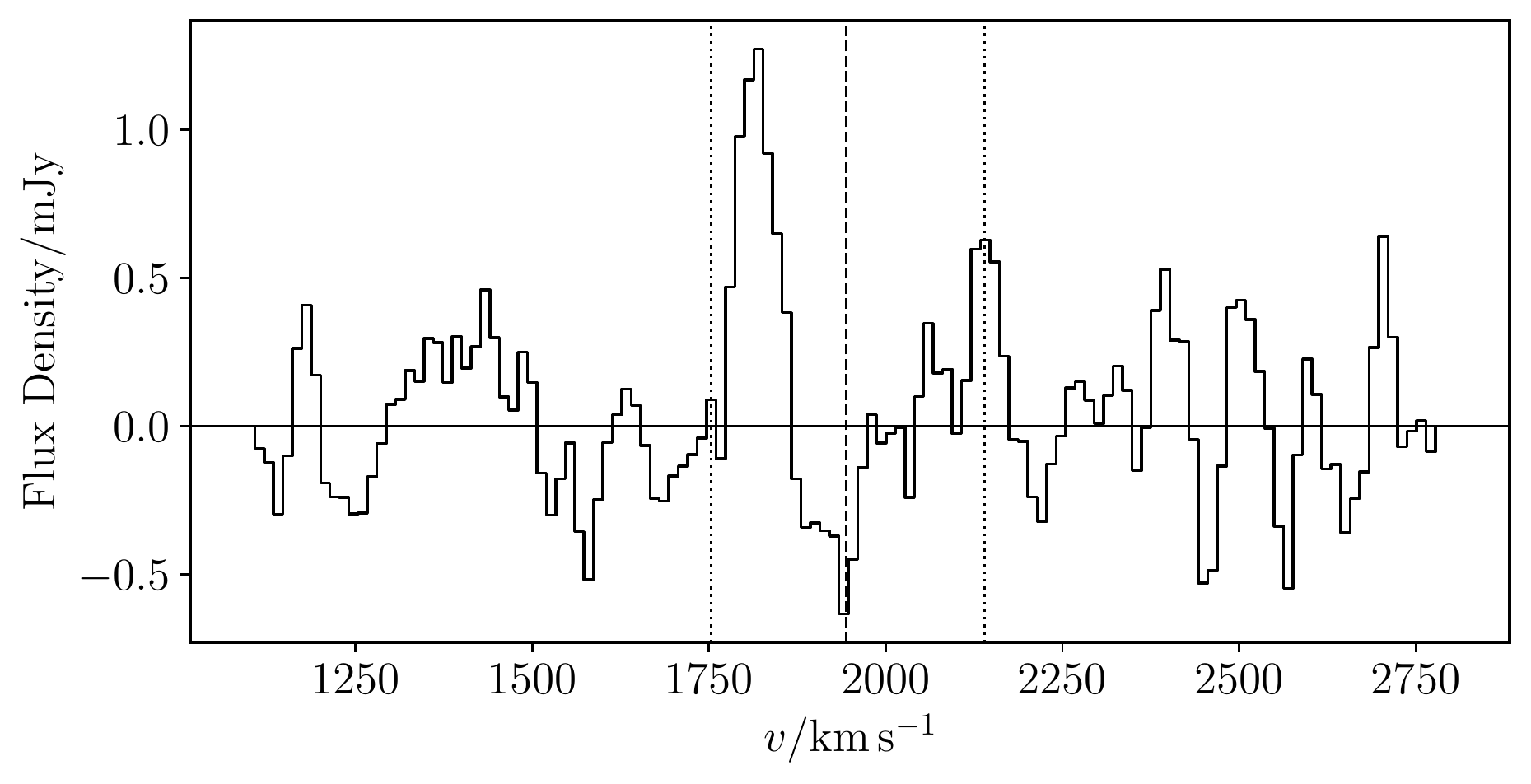}
    \caption{\hi \ spectrum at the position of NGC 5631-Dw1 extracted over a single synthesized beam from the \hi \ cube. The vertical dashed line indicates the central velocity of NGC 5631 and the dotted lines indicate the approximate range over which any \hi \ emission from NGC 5631-Dw1 might be confused with that of NGC 5631.}
    \label{fig:NGC5631DW1_spec}
\end{figure}

Figure \ref{fig:NGC5631DW1_spec} shows the \hi \ spectrum within a single synthesized beam at the location of NGC 5631-Dw1. The prominent feature at $\sim$1800\kms \ corresponds to approximately the velocity of the SW tail and so is not necessarily associated with the UDG. However, an inspection of the channel maps (Figure \ref{fig:NGC5631DW1_chan_maps}, in particular at 1794\kms) indicates that there is a low S/N feature that is coincident with the UDG. The nature of this feature will be discussed further in Section \ref{sec:TDGs?}.

Integrating the flux in this feature gives an \hi \ mass of $\log (M_\mathrm{HI}/\mathrm{M_\odot}) = 7.2$. This should be treated as an upper limit on the \hi \ mass of NGC 5631-Dw1 as this feature is certainly blended with the SW tail at the resolution of the D-array VLA data. 
Taking the $g$-band absolute magnitude, $M_{g} = -11.8$, and color, $g-r \approx 0.4$, from \citet{Bennet+2018} gives a mass-to-light ratio of 0.62 \citep{Zibetti+2009}, resulting in a stellar mass estimate of $\log (M_\ast/\mathrm{M_\odot}) = 6.6$.\footnote{Solar magnitudes from \citet{Willmer+2018}.} Therefore, if the co-incident \hi \ does belong to NGC 5631-Dw1 then it has a factor of several times more \hi \ than stellar mass. Although such gas fractions are not uncommon for dwarf galaxies, they are generally indicative of star forming galaxies \citep[e.g.][]{Huang+2012}. Whereas the non-detection of NGC 5631-Dw1 (and NGC 2708-Dw1) in GALEX \citep[Galaxy Evolution Explorer,][]{Martin+2005} NUV or FUV implies a star formation rate of $\lesssim 3 \times 10^{-3}$ \Msol$\,\mathrm{yr^{-1}}$ \citep{Bennet+2018}.

\subsection{Optical Properties}

We measured the observational properties of the UDGs in the HST data using \texttt{GALFIT} \citep[][]{Peng+2002} with a procedure identical to that presented in our previous work \citep[][]{Bennet+2017,Bennet+2018}. Briefly, we use a fitting region of a 40\arcsec \ (800 pixel) square, chosen to be slightly more than twice the expected half light radius of the UDGs. All observational parameters were allowed to vary without restriction for NGC 2708-Dw1. However, the S\'ersic index was fixed to $n$=1 for NGC5631-Dw1 and spatial binning was required to obtain high quality fits, due to its low surface brightness. This was also the case with fits derived from the CFHT data \citep[][]{Bennet+2018}. The uncertainties were determined by implanting 100 simulated dwarfs with the best-fit properties into our images and re-measuring each with GALFIT; the scatter in these measurements is our quoted uncertainty \citep[][]{Bennet+2017,Bennet+2018}. We present these results alongside the properties derived from the CFHT data in Table \ref{tab:opt_prop}. We find generally good agreement between the two datasets and have therefore decided to use the structural parameters from the CFHT images \citep[][]{Bennet+2018}, which have better surface brightness sensitivity, to construct the apertures used for counting GC candidates in the following section.

\floattable
\begin{deluxetable}{ccccccc}
\tablecaption{Comparison of UDG properties in CFHT and HST ACS images \label{tab:opt_prop}}
\tablehead{
\colhead{Name} & \colhead{V-band} & \colhead{V-band} & \colhead{F555W} & \colhead{F814W} & \colhead{Half light} & \colhead{Half light}\\
\colhead{} & \colhead{Magnitude} & \colhead{Magnitude} &\colhead{Magnitude} & \colhead{Magnitude} & \colhead{radius (CFHT)} & \colhead{radius (HST)}\\
\colhead{} & \colhead{(CFHT)} & \colhead{(HST)} & \colhead{(HST)} &\colhead{(HST)} & \colhead{(arcsec)} & \colhead{(arcsec)\tablenotemark{a}}}
\colnumbers
\startdata
NGC 2708-Dw1 & 19.0$\pm$0.4 & 18.8$\pm$0.2 & 18.9$\pm$0.2 & 17.9$\pm$0.2 & 13.2$\pm$2.9 & 16.1$\pm$3.6\\
NGC 5631-Dw1 & 20.3$\pm$0.5 & 20.3$\pm$0.3 & 20.4$\pm$0.3 & 19.8$\pm$0.2 & 15.6$\pm$3.6 & 14.8$\pm$3.3\\
\enddata
\tablenotetext{a}{Derived from F555W images}
\tablenotetext{}{ NOTE -- col(1): Candidate name. col(2): V-band magnitude, based on CFHTLS imaging \citep{Bennet+2018}, converted from $g$ and $r$ following \citet{Jester+2005}. col(3): V-band magnitude, based on the F555W HST imaging, converted via the relation from \cite{Sahu+2014}. col(4) \& col(5): F555W and F814W magnitude, based on HST imaging. col(6): The half-light radius of the candidates, based on CFHTLS imaging \citep{Bennet+2018}. col(7): The half-light radius of the candidates, based on the F555W HST imaging. }
\end{deluxetable}

\subsection{Globular cluster counts}
\label{sec:GC_counts}

\begin{figure*}
    \centering
    \includegraphics[width=\columnwidth]{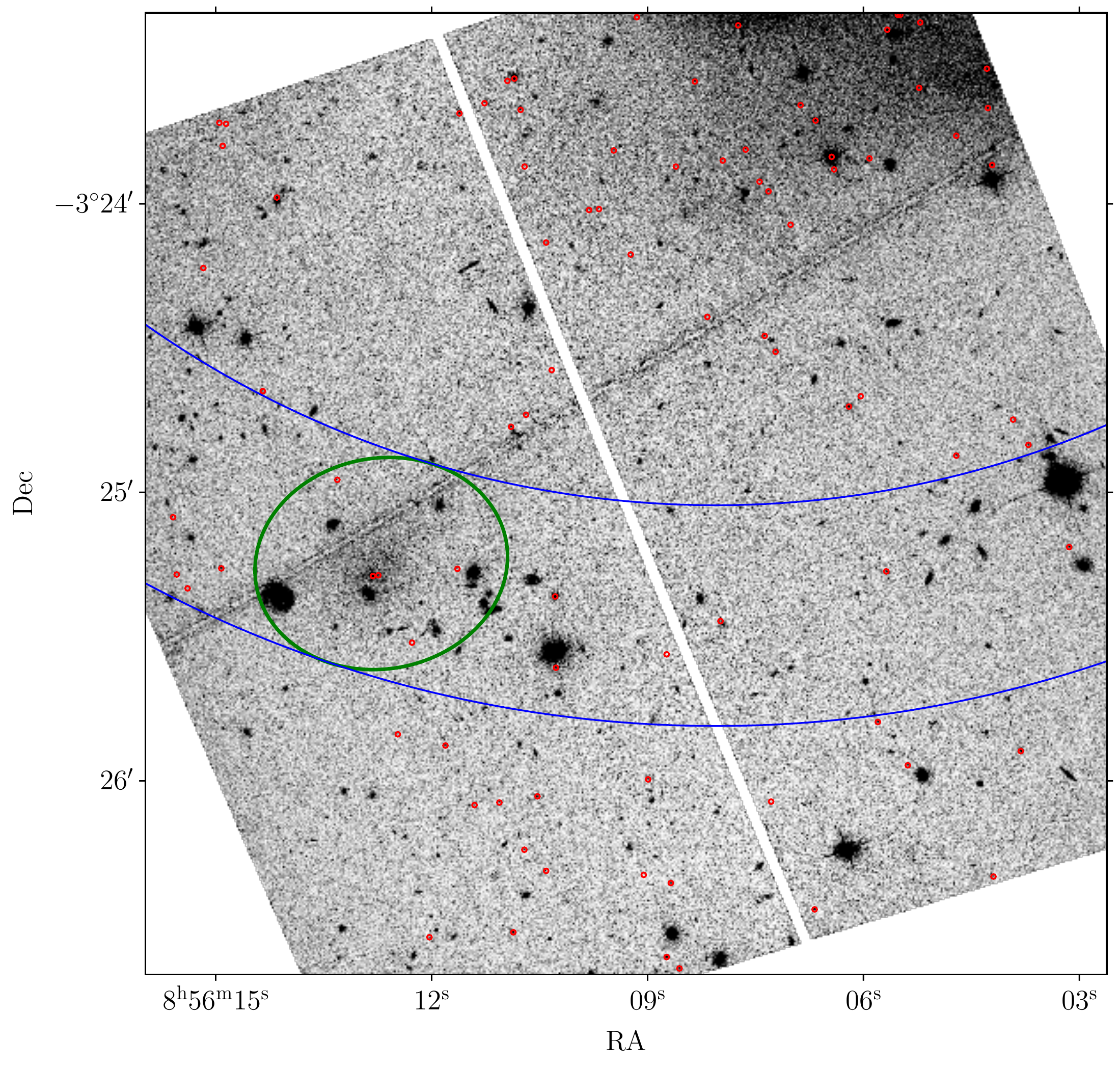}
    \includegraphics[width=1.035\columnwidth]{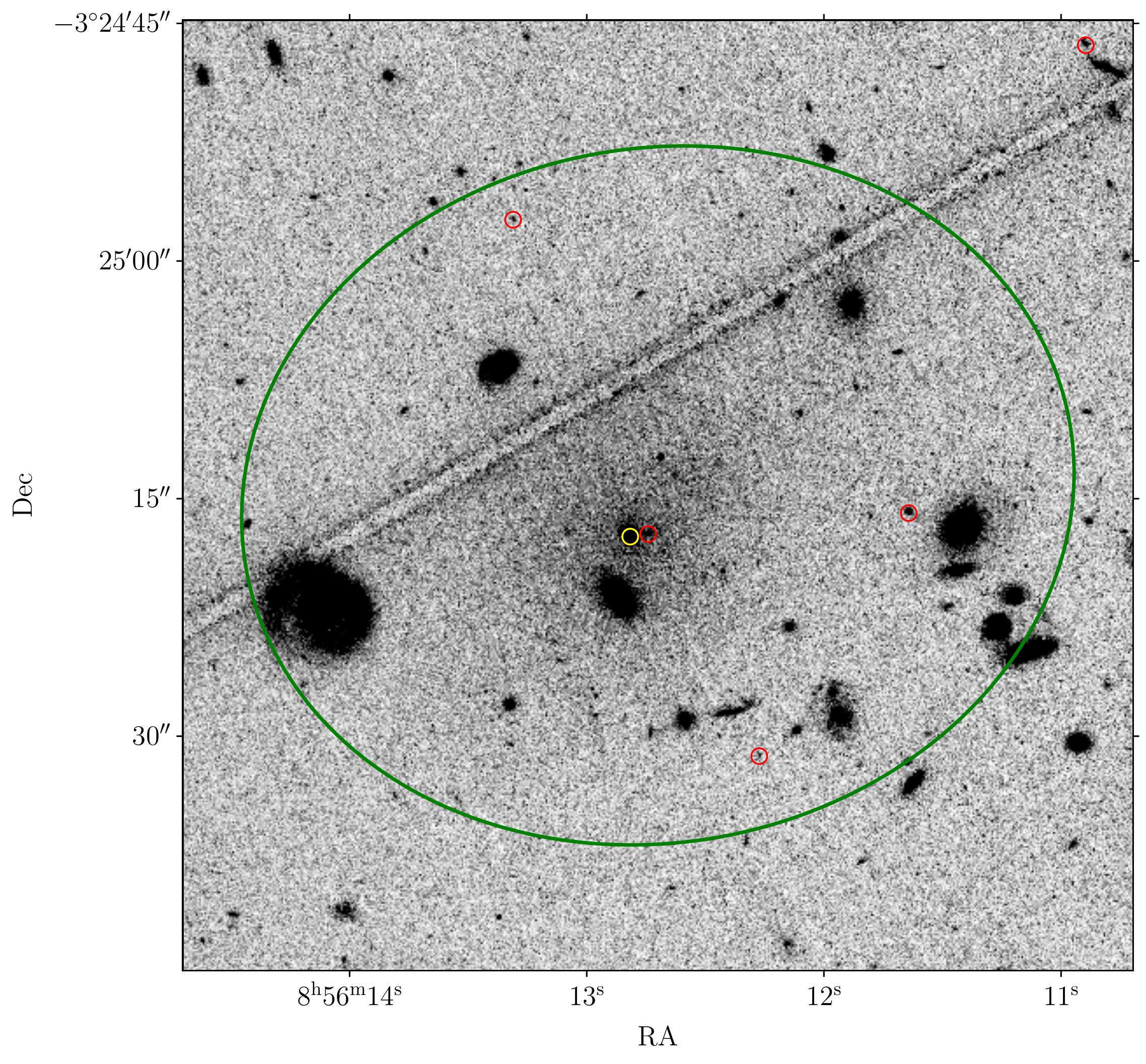}
    \caption{\textit{Left}: HST ACS F814W image of NGC 2708-Dw1 and the edge of the disk of NGC 2708 to the NW (N is up, E is left). All GC candidates are highlighted with small red circles. The green ellipse shows the aperture used to count the GC candidates associated with the UDG, while the partial annulus outlined in blue shows the sky area used to estimate background counts. Due to the poorer surface brightness sensitivity of this image compared to the CFHT images used in \citet{Bennet+2018} the stellar stream adjacent to the UDG is not apparent here. 
    The straight streaks stretching across the entire image are residuals from a satellite trail in one of the exposures.
    \textit{Right}: A smaller FoV showing the GC candidates in the vicinity of NGC 2708-Dw1 in more detail. The one yellow circle shows the ultra-luminous GC candidate that has an absolute magnitude of $M_\mathrm{V} = -11.1$ and is likely a NSC rather than a GC.}
    \label{fig:NGC2708DW1_GCC}
\end{figure*}

\begin{figure*}
    \centering
    \includegraphics[width=\columnwidth]{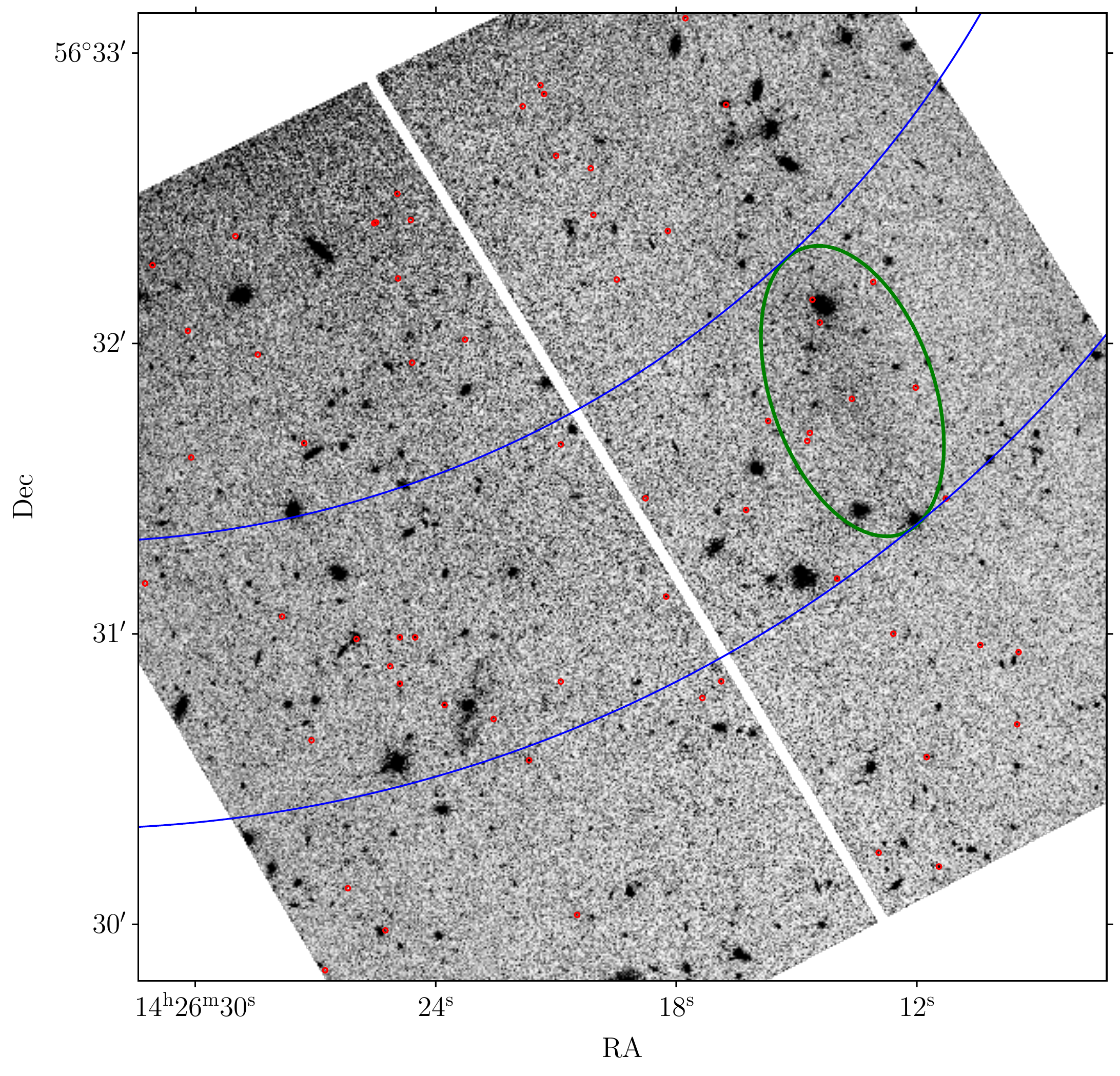}
    \includegraphics[width=1.04\columnwidth]{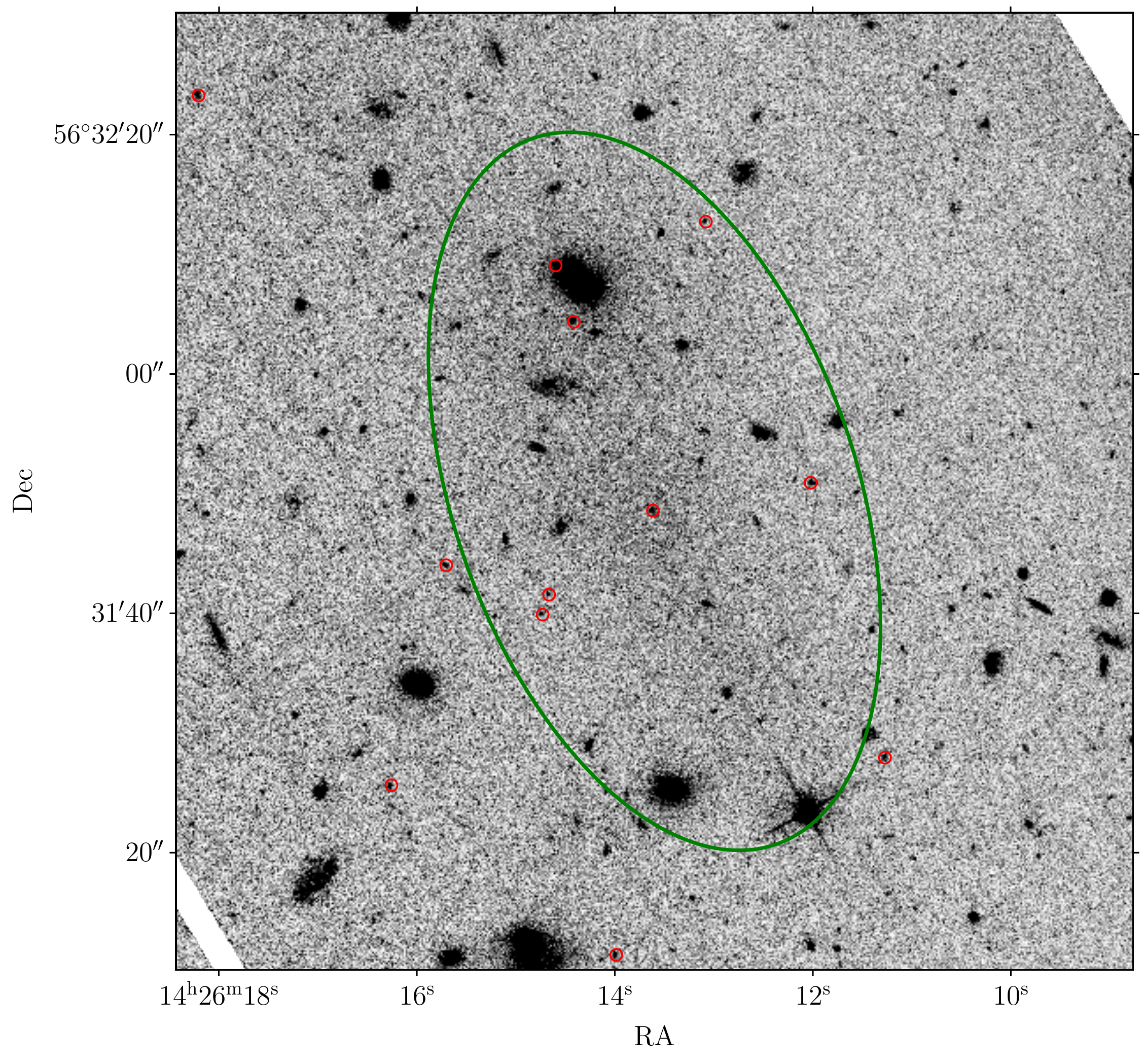}
    \caption{\textit{Left}: HST ACS F814W image of NGC 5631-Dw1, where NGC 5631 is located outside the FoV to the NE (N is up, E is left). All GC candidates are highlighted with small red circles. The green ellipse shows the aperture used to count the GC candidates associated with the UDG, while the partial annulus outlined in blue shows the sky area used to estimate background counts. \textit{Right}: A smaller FoV showing the GC candidates in the vicinity of NGC 5631-Dw1 in more detail. Due to the poorer surface brightness sensitivity of this image compared to the CFHT images used in \citet{Bennet+2018} the UDG itself is scarcely visible here.}
    \label{fig:NGC5631DW1_GCC}
\end{figure*}

\begin{figure}
    \centering
    \includegraphics[width=\columnwidth]{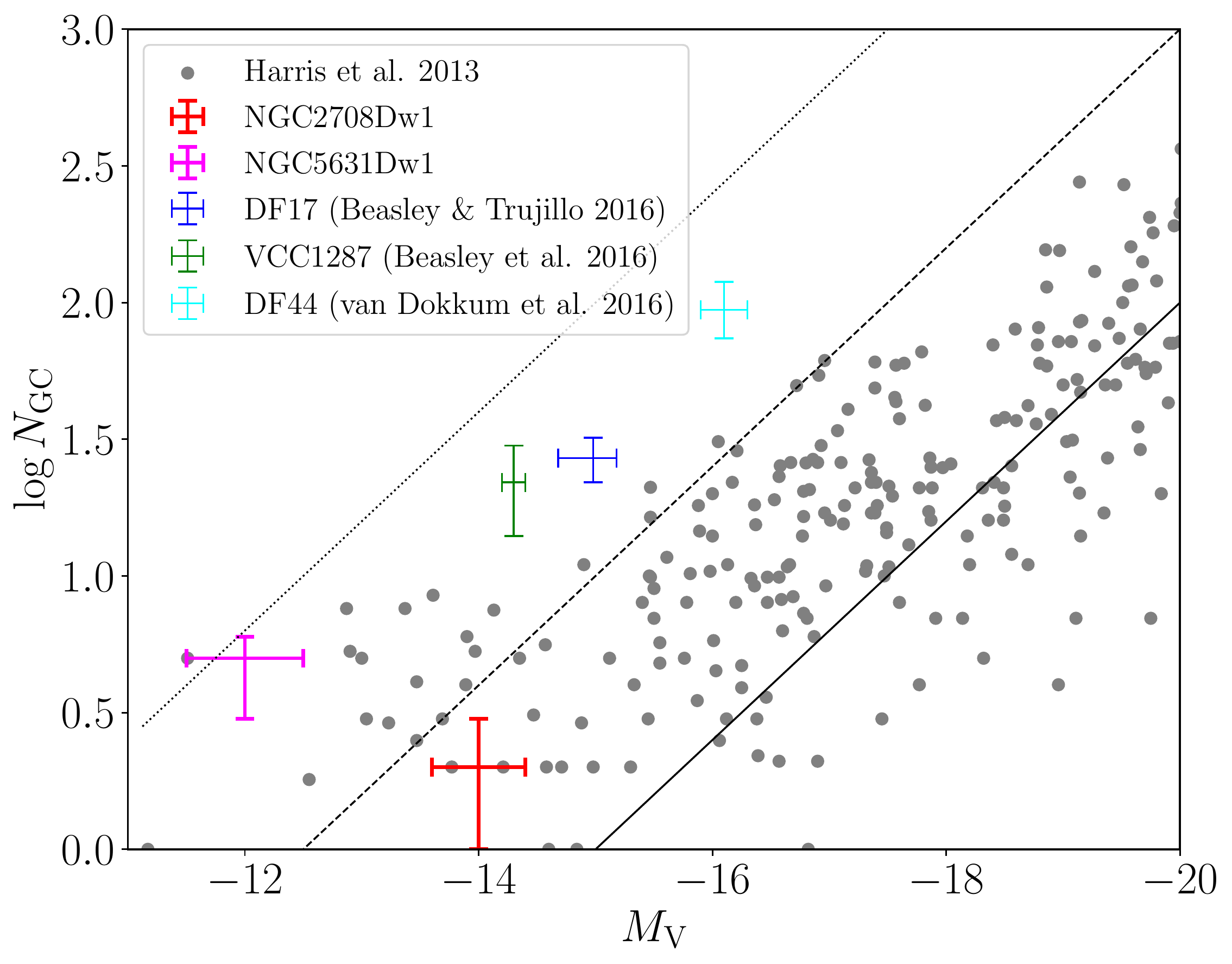}
    \caption{Globular cluster counts and absolute V-band magnitudes of NGC 2708-Dw1 (red) and NGC 5631-Dw1 (magenta) compared to other UDGs \citep{Beasley+2016a,Beasley+2016b,vanDokkum+2016}, reported to have exceptionally rich GC systems, and a general catalog of nearby galaxies \citep{Harris+2013}. The solid, dashed, and dotted lines indicate specific frequency values of 1, 10, and 100 respectively. As \citet{Beasley+2016a} did not estimate uncertainties for the magnitude of VCC 1287, we adopt the magnitude measurements from \citet{Pandya+2018} and convert $g$ and $u$ to V following Lupton 2005 (\url{http://classic.sdss.org/dr4/algorithms/sdssUBVRITransform.html}). For DF17 and DF44 we use the $g$-band uncertainties from \citet{vanDokkum+2015} and assume these are equivalent to those in V-band as \citet{Beasley+2016b} and \citet{vanDokkum+2016} do not report uncertainties in their magnitudes.}
    \label{fig:NGC_MV}
\end{figure}

After restricting the catalog of point-like objects within the field to those that could plausibly be GCs, based on their colors and concentration indices (Section \ref{sec:GC_selection}), we estimate the number of GCs associated with NGC 2708 Dw1 using an elliptical aperture that has the same axial ratio as given in Table 1 of \citet{Bennet+2018} and a position angle of $-$80$^\circ$. We set the semi-major axis length to be double the half-light radius, that is, 26.4\arcsec. We count a total of 5 GC candidates within this aperture. To estimate the background count rate of false GC candidates we initially used the remainder of the ACS field, however, assuming a specific frequency of GCs of unity \citep[e.g.][]{Harris+1981,Harris+2013}, NGC 2708 is expected to host on the order of 270 GCs, given its absolute V-band magnitude \citep[$M_\mathrm{V} = -21.07$,][]{RC3}. These genuine GCs could significantly contaminate the background count estimates over the image, as well as falsely contribute to the counts for NGC 2708-Dw1.

Following \citet{Zibetti+2009} we estimate the stellar mass of NGC 2708 as $\log (M_\ast/M_\odot) = 11.0$, based on its B-V color (0.92) and V-band magnitude. Then using the relation \citep{Hudson+2018} between a galaxy's stellar mass and the extent of its GC system we estimate the 50\% radius of the GC system to be 14.8 kpc, or 1.25\arcmin. As NGC 2708-Dw1 is less than 4\arcmin \ away from NGC 2708, a significant gradient in the background counts of GC candidates is expected across the ACS image. Furthermore, although the parallel WFC3 image (taken in tandem with the ACS exposures) was intended to be used as an additional blank field to improve the estimate of background GC candidate counts, the center of this image is separated from NGC 2708 by over 9\arcmin, meaning that the GC candidate counts will be significantly lower due to the lack of contamination from the GC system of NGC 2708. Therefore, to estimate the background counts for NGC 2708-Dw1 we measure the counts per unit area in an annulus centered on NGC 2708 and an extent matching that covered by the elliptical aperture centered on NGC 2708-Dw1 (Figure \ref{fig:NGC2708DW1_GCC}).

As the counts both within the UDG aperture and the background annulus are so low (5 and 15 respectively) the statistics governing the uncertainties are decidedly Poissonian rather than Gaussian. Under such conditions the number of GC candidates within the UDG aperture that are above the background level can be estimated analytically in a Bayesian manner. The number of spurious GC candidates within the UDG aperture follows a Poisson distribution with an unknown mean of $b$, the background false positive count rate from GCs associated with the host galaxy plus other spurious sources.

The observed GC candidate counts within the annulus can be used to estimate $b$. For this problem the analytic Bayesian solution (with a flat prior) is 
\begin{equation}
    \label{eqn:background_rate}
    p(b | N_\mathrm{back}, A_\mathrm{back}) \propto b^{N_\mathrm{back}} \exp(-b A_\mathrm{back})
\end{equation}
where $N_\mathrm{back}$ is the count within the annulus (excluding those assigned to the UDG) and $A_\mathrm{back}$ is the area of the annulus (minus the area of the UDG aperture). Here we have chosen the units of $A_\mathrm{back}$ to be multiples of $A_\mathrm{UDG}$, the area of the UDG aperture, which normalizes $b$ such that it is the rate of background or spurious objects within an aperture of area $A_\mathrm{UDG}$. 

With this expression for $b$ we can write down the probability of there being $N_{\mathrm{f}}$ false GC candidates in the UDG aperture as:
\begin{equation}
    p(N_{\mathrm{f}} | b) = \frac{(bA_\mathrm{back})^{N_{\mathrm{f}}} \exp(-b))}{N_{\mathrm{f}}!}
\end{equation}
that is, a normal Poisson process. Changing variables from $N_{\mathrm{f}}$ to $N_{\mathrm{UDG}}$, the total number of GC candidates within the UDG aperture, and $N_{\mathrm{GC}}$, the number of those that are genuine GCs, gives:
\begin{equation}
    p(N_{\mathrm{GC}} | b) = \frac{(bA_\mathrm{back})^{N_{\mathrm{UDG}}-N_{\mathrm{GC}}} \exp(-b))}{(N_{\mathrm{UDG}}-N_{\mathrm{GC}})!}
\end{equation}
Finally, this must be multiplied by Equation \ref{eqn:background_rate} and $b$ marginalized. Thus the final expression for the probability mass function of $N_{\mathrm{GC}}$ is:
\begin{align} 
    &p(N_{\mathrm{GC}} | O) \nonumber \\
    &\propto \int_{0}^{\infty} \frac{b^{N_\mathrm{back} + N_{\mathrm{UDG}} - N_{\mathrm{GC}}} \exp(-b(A_\mathrm{back}+1))}{(N_{\mathrm{UDG}}-N_{\mathrm{GC}})!} \, \mathrm{d}b \\
    &= \frac{\Gamma(N_\mathrm{back} + N_{\mathrm{UDG}} - N_{\mathrm{GC}} + 1)}{(A_\mathrm{back}+1)^{N_\mathrm{back} + N_{\mathrm{UDG}} - N_{\mathrm{GC}} + 1} (N_{\mathrm{UDG}}-N_{\mathrm{GC}})!}
\end{align}
where here we have denoted all the observations ($N_\mathrm{UDG}$, $A_\mathrm{UDG}$, $N_\mathrm{back}$, $A_\mathrm{back}$) as $O$ for brevity.

We identify the highest posterior probability region in the above expression equaling at least 68\% in order to estimate the uncertainties in the number of GCs around the most likely value. For NGC 2708-Dw1 this gives $N_{\mathrm{GC}} = 3^{+1}_{-2}$, with a probability of $N_{\mathrm{GC}} = 0$ of 12\%.

One of the GC candidates associated with NGC 2708-Dw1 is visibly much brighter than the others (yellow circle in Figure \ref{fig:NGC2708DW1_GCC}, right). It has an absolute V-band magnitude of $-$11.1, which would make it an extraordinarily luminous GC. We therefore consider the possibility that it is not a GC (discussed further in Section \ref{sec:discuss_bright_gcc}) and set a minimum absolute magnitude limit of $M_\mathrm{V}=-11$ for all GC candidates (both those assigned to the UDG and background candidates) and recalculate $N_\mathrm{GC}$ as above, arriving at the value $N_{\mathrm{GC}} = 2^{+1}_{-1}$, with a probability of $N_{\mathrm{GC}} = 0$ of 16\%.

For NGC 5631-Dw1 (Figure \ref{fig:NGC5631DW1_GCC}) we follow the same steps as above (also imposing the same $M_\mathrm{V}=-11$ limit, although this removes no GC candidates near the UDG) using an elliptical aperture with a semi-major axis of 31.2\arcsec \ and a position angle of 19$^{\circ}$. A total of 7 candidates fall within this aperture and applying the same analysis as above gives the value $N_{\mathrm{GC}} = 5^{+1}_{-2}$, with a probability of $N_{\mathrm{GC}} = 0$ of 2\%.

The GC counts for both UDGs are shown in Figure \ref{fig:NGC_MV} along with other UDGs found to harbor rich GC systems and a sample of nearby galaxies for reference. In contrast to the UDGs DF44, DF17 and VCC 1287 \citep{Beasley+2016a,Beasley+2016b,vanDokkum+2016}, we find that NGC 2708-Dw1 and NGC 5631-Dw1 are entirely consistent with the GC counts of normal galaxies \citep[from the][sample]{Harris+2013}.

\subsection{Specific frequency of globular clusters}

Using the magnitudes from CFHT in Table \ref{tab:opt_prop} and the assumed distances of 40.6 and 28.4 Mpc, the absolute V-band magnitudes of NGC 2708-Dw1 and NGC 5631-Dw1 are $-14.0 \pm 0.4$ and $-12.0 \pm 0.5$, respectively \citep[converted from $g$ and $r$ following][]{Jester+2005}.
This gives specific frequency values of $S_\mathrm{N} = 3.3^{+5.6}_{-1.0}$ and $S_\mathrm{N} = 54^{+49}_{-33}$ (with respective probabilities of $S_\mathrm{N} = 0$ of 16\% and 2\%).\footnote{As this calculation involves the multiplication of two non-negative variables, one discrete and one continuous, there is a discrete probability of the result being exactly zero. While all positive values follow a continuous (but non-Gaussian) probability density function, for which we quote the 68\% highest density region.} We corrected for the portion of the GCLF below the completeness limit of the HST F814W images (though the correction factor is close to unity in both cases).
In the case of NGC 2708-Dw1 we consider the GC count with the over-luminous GC candidate excluded.

\section{Discussion}
\label{sec:discuss}

As described in Section \ref{sec:intro} we aim to examine three potential formation scenarios of NGC 2708-Dw1 and NGC 5631-Dw1: 1) tidal dwarf galaxies, 2) normal dwarfs `puffed up' by interaction, and 3) UDGs prior to joining the group. There is also a 4th possibility, that the UDGs are not associated with their assumed hosts, NGC 2708 and NGC 5631. However, this is unlikely given the configuration of the stellar streams (Figure \ref{fig:CFHT_tails}). Nevertheless we checked for possible background structure \citep[as in][]{Roman+2019} that the UDGs could be associated with, but there are no plausible candidates where the half-light radii of the UDGs would be $<5$ kpc \ \citep[as noted by][very extended UDGs candidates are likely due to overestimated distances]{Zaritsky+2019}. Therefore, we do not further consider this possibility.

In this section we discuss each of the three above scenarios in turn as well as the nature of the ultra-luminous GC candidate found near the center of NGC 2708-Dw1.

\subsection{How did these UDGs form?}

Even with the multiple data sets collected it is challenging to say with certainty how these two UDGs formed. However, as we will argue below, there are strong indicators which favor or disfavor certain formation scenarios.

\subsubsection{Are they tidal dwarf galaxies?}
\label{sec:TDGs?}

As tidal dwarf galaxies form from gas expelled from larger, interacting galaxies \citep[e.g.][]{Mirabel+1992} they are not expected to have either GC systems or DM halos. A GC count of zero would therefore be the only result consistent with this formation scenario. In both cases there is still a small probability (16\% for NGC 2708-Dw1 and 2\% for NGC 5631-Dw1) that none of the GC candidates identified are genuinely associated with the UDG in question. Thus the GC count disfavors this formation scenario, but it cannot rule it out at high confidence.

The VLA \hi \ observations of NGC 2708-Dw1 find no trace of \hi \ in the UDG itself and no clear indication of a tidal connection with NGC 2708 in neutral gas. In their simulations of TDG formation, \citet{Bournaud+2006} state that the tidal streams responsible for forming TDGs typically dissolve after 300--500 Myr. \hi \ tidal tails (outside of galaxy clusters) are frequently long-lived \citep[e.g.][]{Hibbard+1999} and it is improbable that the stellar component of a tail with sufficient gas to form a TDG would remain visible for longer than the \hi \ component. The stellar stream connected to NGC 2708-Dw1 \citep{Bennet+2018} appears to be devoid of gas, indicating that it is unlikely that it originated from the gas-rich disk of NGC 2708 (which would be required for a TDG origin), as a tidal interaction strong enough to strip stars from NGC 2708 would presumably also strip \hi. Alternatively the origin of the tail could be from the disruption of a gas-poor, dwarf satellite -- probably NGC 2708-Dw1 itself.

In the case of NGC 5631-Dw1 there is a tentative detection of \hi \ emission that coincides with the UDG (Section \ref{sec:hi_props} \& Figure \ref{fig:NGC5631DW1_chan_maps}) and is connected  to the (highly disturbed) \hi \ content of NGC 5631. This \hi \ feature in the SW tail of NGC 5631 may be an indication of a tidal connection between UDG and host. Long-lived TDGs are generally expected to form at the tips of \hi \ tails \citep[e.g][]{Bournaud+2006}, but NGC 5631-Dw1 is midway along the (\hi) tail in question (Figure \ref{fig:NGC5631DW1_chan_maps}), making this an unlikely explanation for the gas or the UDG. NGC 5631-Dw1 is, however, at the tip of the optical stellar stream (Figure \ref{fig:CFHT_tails}), suggesting that the \hi \ tail and the stellar tail may simply be superposed, rather than physically connected. Given that the \hi \ mass of this feature is greater than the stellar mass of NGC 5631-Dw1 it seems improbable that it could have stripped this gas from NGC 5631, although it may have been able to attract loosely bound gas as it passed by an existing tail. There also remains the possibility of a chance superposition between a random clump in the SW tail and the UDG.

Finally, we note that if these UDGs are TDGs then they must be old. TDG candidates in the act of formation are universally blue, but neither of the UDGs is particularly blue ($g-r \approx 0.5$) and neither is detected in the UV in GALEX data. Therefore, if these UDGs did originate as TDGs then they must now be sufficiently old for their stellar populations to have reddened \citep[e.g.][]{Maraston2005,Schawinski+2009} and for the \hi \ tail that formed them to have dissipated, probably on the order of 500 Myr. However, the stellar streams which these UDGs are connected to are still coherent structures, whereas if they were the streams that the UDGs formed from (i.e. as TDGs) $\sim$500 Myr ago, then they probably would have dispersed already \citep{Bournaud+2006}.

Although there is no single piece of evidence that directly disproves the TDG formation hypothesis, it is disfavored by the GC counts and by the absence of any conclusive support for the hypothesis from the \hi \ observations.

\subsubsection{Were they UDGs prior to falling into their current groups?}

Without strong evidence for an alternative formation mechanism, the possibility that these UDGs were ultra-diffuse prior to joining their current groups will always remain, especially as it is still debated how UDGs form in the field \citep[e.g.][]{Papastergis+2017,Jones+2018,Janowiecki+2019,Jiang+2019,Liao+2019}. However, what is apparent from Figure \ref{fig:NGC_MV} is that the number of GCs associated with either of these UDGs is broadly consistent with the value for normal dwarfs. Therefore, what can be concluded is that these UDGs likely did not form through the same pathway as objects like DF 17, DF 44, or VCC 1287, which exhibit very rich GC systems given their luminosities \citep{Beasley+2016a,Beasley+2016b,vanDokkum+2016}.

This assertion may be questioned by the fact that in terms of specific frequency, NGC 5631-Dw1's GC system is similarly rich to those of DF 17, DF 44, and VCC 1287. However, the relation between $M_\mathrm{V}$ and $N_\mathrm{GC}$ has a break for low luminosity dwarfs, likely because the GC count is more directly related to halo than stellar mass \citep[e.g.][]{Harris+2013,Zaritsky+2016}. Thus the specific frequencies of very low luminosity dwarfs that host any GCs are higher than would be expected by comparison with more luminous galaxies. Ideally we would make this plot with halo mass on the horizontal axis, but given the currently available data this is not possible without highly subjective extrapolations regarding the halo masses of these UDGs. 

Another caveat to this interpretation is that the aforementioned UDG formation scenarios are not mutually exclusive. It is possible that these galaxies were UDGs (with rich GC systems) prior to joining their current groups, but have since undergone tidal stripping that involved the removal of some of their GCs \citep[c.f. Sagittarius dwarf, e.g.][]{Bellazzini+2020}. Hence, giving the appearance of the GC systems of normal dwarfs. However, we note that the claimed exceptional richness of the GC systems of some UDGs has been debated \citep[e.g.][]{Saifollahi+2021}, and that even these UDGs may turn out to be more in line with the normal dwarf population than previously thought.

Ultimately, further studies of the GC systems of field UDGs are needed to facilitate a more robust discussion of this point. It may be that field UDGs host GC systems typical of dwarf galaxies, which would strengthen the case for these two galaxies being UDGs prior to joining their current groups, rather than normal dwarfs that are `puffed up' by interactions, the scenario that we consider next.

\subsubsection{Are they normal dwarfs that have been `puffed up' by interactions?}

This is the interpretation that the GC counts favor most, as they are neither rich nor do they likely have zero GCs, with both UDGs falling in line with normal dwarfs in Figure \ref{fig:NGC_MV}. Furthermore, without a clear \hi \ counterpart to the stellar tails connected to the UDGs (particularly NGC 2708), the most likely origin for these tails is from the stripping or disruption of a gas-poor satellite, probably the UDGs themselves. Otherwise, the positions of the UDGs at the ends of the stellar tails would be coincidence, which seems extremely unlikely given the apparent configuration (Figure \ref{fig:CFHT_tails}). Therefore, we take these streams as evidence that the hosts galaxies are exerting strong tidal influence on these UDGs.

The importance of tidal forces and harassment has long been recognized as a driver of galaxy evolution and morphological change for galaxies in dense environments \citep[e.g.][]{Moore+1996,Moore+1998,Mastropietro+2005}. \citet{Errani+2015} showed that the tidal stripping of satellite dwarf spheroidals leads to their half-light radii increasing and them becoming more diffuse.
\citet{Carleton+2019} and \citet{Tremmel+2020} build on these finding to describe a potential UDG formation mechanism where tidal stripping and heating of (cored) dwarf-mass DM halos, causes the half-light radii of the stellar populations to expand, driving them towards the UDG parameter space. An analogous process in the group, rather than cluster, environment could be responsible for NGC 2708-Dw1 and NGC 5631-Dw1 \citep[which are also less massive that many of the UDGs considered in][]{Carleton+2019,Tremmel+2020}. However, \citet{Carleton+2021} expanded on their previous findings by predicting that UDGs formed through this mechanism would host rich GC systems, which these two UDGs do not. This apparent disagreement may be explained by the fact that \citet{Carleton+2021} focused on cluster UDGs, whereas these UDGs are in a group environment. \citet{Carleton+2021} argues that UDGs which formed their stars at higher redshift (such as those in highest density regions of the Universe) will host more GCs. However, this would not necessarily be true of UDGs found in less extreme environments (such as groups). 

\citet{Jiang+2019} and \citet{Liao+2019} focus on less dense environments, and although they disagree on the formation mechanism for field UDGs (favoring stellar feedback and halo spin, respectively), both works estimate that approximately 50\% of the UDGs in groups are the result of normal dwarf satellites on highly eccentric orbits that undergo major tidal stripping and heating. Furthermore, \citet{Jiang+2019} predict that the pericenter passage that causes them to become UDGs will be accompanied by the complete removal of cold gas (as well as a considerable amount of DM). This is consistent with NGC 2708-Dw1 being undetected in \hi. NGC 5631-Dw1 may also be consistent with this prediction if the \hi \ emission (Section \ref{sec:hi_props_ngc5631dw1}) is superposed on the UDG, or if we are seeing it at a special point in its evolution, right as it is losing its gas. Although the latter is unlikely given that it has shown no signs of recent SF.

\subsection{Globular cluster, nuclear star cluster, or ultra-compact dwarf?}
\label{sec:discuss_bright_gcc}

The brightest GC candidate in the vicinity of NGC 2708-Dw1 (Figure \ref{fig:NGC2708DW1_GCC}, yellow circle) has an apparent V-band magnitude of 21.9. This object has a FWHM of $\sim$0.2\arcsec \ and is thus marginally resolved. At the assumed distance of NGC 2708 (40.6 Mpc) this equates to an absolute V-band magnitude of $M_V = -$11.1. For either of the GCLFs discussed in Section \ref{sec:GC_selection}, less than 4\% of GCs should be this bright or brighter. This is also almost 1 mag brighter than the brightest GC identified around DF2 \citep{vanDokkum+2018b} and similarly brighter than any Milky Way GC. Therefore, it is  unlikely that this object is a normal GC.

Given the limited information available we propose and discuss four possibilities of what this object may be: 1) a superposed ultra-compact dwarf (UCD) or compact background galaxy (at much higher redshift), 2) a long-lived nuclear star cluster (NSC) in a stable configuration, 3) a bright GC (or merged GCs) in the process of in-spiraling, and 4) a NSC that is in the process of losing its diffuse stellar component and forming a UCD.

The first of these is the most mundane, but is difficult to fully exclude without a spectrum of the object. There are 6 other GC candidates throughout the full ACS image that are as bright or brighter than this object, so chance superposition is certainly possible, although not particularly likely. 
The full FoV of ACS is 40804 arcsec$^{2}$, thus with a total of 7 such objects throughout the image, the probability of one randomly lying within 5\arcsec \ of the center of the UDG is 1.3\%.
Even with a spectrum it may still be difficult to say for certain whether the UDG and this cluster are physically connected, as there may be UCDs that are satellites of NGC 2708, and would thus have similar redshifts whether or not they were associated with NGC 2708-Dw1. However, a redshift measurement would be able to rule out a high redshift background object.

About 20\% of normal dwarfs (with $M_\ast \approx 10^7 \mathrm{M_\odot}$) host NSCs \citep{Neumayer+2020}, and they have also been seen in UDGs before \citep{Lim+2020}, so upon first inspection this seems like a plausible scenario. However, the cluster in question is $\sim$3\arcsec \ ($\sim$0.5 kpc) away from the optical center of the UDG, as fit by \citet{Bennet+2018}. Therefore, this cannot be a normal NSC, and even a superficial consideration of the dynamics makes it apparent that such a configuration could not be stable.  That said, a similar, potential NSC was found around the Virgo Cluster UDG VLSB-D, offset spatially and in velocity from the galaxy as a whole \citep{Toloba+2018}. VLSB-D also showed a velocity gradient among its GC population, possibly indicating tidal disruption of the dwarf, another point in common with NGC~2708-Dw1.

UCDs generally have a (stellar) mass-to-light ratio of $\sim$2 \citep[e.g.][]{Taylor+2010}. If we assume this value for the over-luminous GC candidate, which has a luminosity of $\log L_\mathrm{V}/L_{\mathrm{V,}\odot} = 6.36$, then its stellar mass is approximately $\log M_\ast/M_\odot \approx 6.7$. For the UDG itself, using a mass-to-light ratio of unity \citep{Zibetti+2009} and the $g$-band magnitude from \citet{Bennet+2018}, we estimate the stellar mass as $\log M_\ast/M_\odot \approx 7.5$. Although these estimates are approximate this indicates that the diffuse stellar component of the UDG is less than an order of magnitude more massive than the star cluster. Such a massive cluster embedded off-center in a diffuse stellar envelope would surely not be a stable configuration.

This leads us to the third and fourth possibilities, which posit that we are capturing this object at a special time. \citet{Leaman+2020} argue that a bright GC in the Pegasus dwarf galaxy may be in the process of in-spiraling to form a NSC, indicating that this mechanism is a valid formation pathway for NSCs in faint dwarfs as well as more massive galaxies \citep[e.g.][]{Tremaine+1975,Antonini+2012,Gnedin+2014}. If we are catching NGC 2708-Dw1 at a special time in its history it is possible we are seeing the same here. Alternatively, we may be witnessing the transformation of an NSC into a UCD. In this scenario, the surrounding stellar envelope, identified as a UDG, would merely be a transient feature. A similar scenario was hypothesized previously (for cluster UDGs) by \citet{Mihos+2017} and \citet{Janssens+2017}, and for which there is circumstantial evidence in the distribution of UCDs and UDGs within clusters \citep{Janssens+2019}. The ACS Virgo cluster survey \citep{Cote+2006} found that the typical nuclear star cluster of a nucleated dwarf is approximately 3.5 magnitudes brighter than a typical GC, which is remarkably close to the 3.8 magnitude gap between the over-luminous GC candidate and any of the other GC candidates associated with NGC 2708-Dw1. Ultra-compact dwarfs (UCDs) were first identified in the Fornax cluster \citep{Drinkwater+2000} and one proposed formation for these objects is by the `threshing' of nucleated dwarf galaxies until only the nucleus remains \citep{Bekki+2001}. This mechanism is qualitatively similar to the `puffing up' mechanism that has been proposed for the formation of UDGs, and in this particular scenario we suggest that they may be one and the same.

\section{Conclusions}
\label{sec:conclusion}

We have followed-up the detection of two UDGs associated with stellar streams \citep{Bennet+2018} with both HST imaging and \hi \ mapping with the VLA, with the goal of studying their GC populations and identifying any tidal connection between the UDGs and their hosts in neutral gas. The GC counts of both UDGs are found to be consistent with those of normal dwarf galaxies of similar luminosities. There is no \hi \ connection found between NGC 2708-Dw1 and its host (NGC 2708) and the UDG itself is also undetected in \hi \ down to a 3$\sigma$ limit of $\log (M_\mathrm{HI}/\mathrm{M_\odot}) = 7.3$. There is low significance \hi \ emission that is coincident with NGC 5631-Dw1, but this feature is blended with the edge of an \hi \ tail extending from NGC 5631 and may or may not be associated with the UDG. 

We consider three formation scenarios for these UDGs: 1) that they are TDGs, 2) that they were UDGs before becoming satellites of their current hosts, and 3) that they were normal dwarfs that have been `puffed up' by interactions with their hosts. 

Due to the small number of GCs detected and the high background counts, the possibility of there being zero true GCs (as would be expected for TDGs) cannot be entirely ruled out, although it is disfavored by the data at 84\% and 98\% confidence for the two UDGs, respectively. The lack of an \hi \ connection between NGC 2708-Dw1 and NGC 2708 also disfavors this formation hypothesis. In the case of NGC 5631-Dw1, although there is a tentative \hi \ connection to NGC 5631, this UDG is located midway along the \hi \ tail, whereas TDGs are typically expected at tail tips. Furthermore, this \hi \ tail also appears to be much larger scale than the stellar stream connected to the UDG, implying that the two are probably not physically connected. Taken together these results make a TDG formation scenario extremely unlikely. 

If NGC 2708-Dw1 and NGC 5631-Dw1 were UDGs prior to joining their current groups then one possibility is that they might display the rich GC systems that have been found in some other UDGs \citep{Beasley+2016a,Beasley+2016b,vanDokkum+2016}. The fact that their GC systems appear normal for dwarfs galaxies somewhat disfavors this hypothesis. However, given that there is little information on the richness of the GC systems of field UDGs, this hypothesis is difficult to reliably evaluate at present. Therefore, with this caveat in mind, we conclude that there is currently no evidence that directly supports this hypothesis.

The fact that the GC systems of these UDGs imply they were once normal dwarfs and their proximity to stellar streams, means that the most favored formation scenario is that these were normal dwarfs that have been made diffuse through interactions with their host. \citet{Jiang+2019} and \citet{Liao+2019} have studied this pathway as a means to form UDGs in groups through tidal stripping and heating, and this mechanism appears qualitatively consistent with our findings for NGC 2708-Dw1 and NGC 5631-Dw1. However, these simulations lack predictions for GC richness and a more complete analysis based on a larger sample of objects (both from simulations and observations) is needed before robust conclusions can be drawn.

\acknowledgements{
We thank the anonymous referee for constructive comments that helped to improve this paper.
Based on observations made with the NASA/ESA Hubble Space Telescope, obtained at the Space Telescope Science Institute, which is operated by the Association of Universities for Research in Astronomy, Inc., under NASA contract NAS5-26555. These observations are associated with program \# HST-GO-15874.  Support for program \# HST-GO-15874 was provided by NASA through a grant from the Space Telescope Science Institute, which is operated by the Association of Universities for Research in Astronomy, Inc., under NASA contract NAS5-26555.
The work used data observed with the Karl G. Jansky Very Large Array. The National Radio Astronomy Observatory is a facility of the National Science Foundation operated under cooperative agreement by Associated Universities, Inc. 

KS acknowledges support from the Natural Sciences and Engineering Research Council of Canada (NSERC).
B.M.P. is supported by an NSF Astronomy and Astrophysics Postdoctoral Fellowship under award AST-2001663.
Research by DJS is supported by NSF grants  AST-1821967 and AST-1813708. DC is
supported by NSF grant AST-1814208.

The work used images from the Dark Energy Camera Legacy Survey (DECaLS; Proposal ID 2014B-0404; PIs: David Schlegel and Arjun Dey). Full acknowledgment at \url{https://www.legacysurvey.org/acknowledgment/}. We also acknowledge the use of the HyperLeda database \citep{HyperLeda}.

\facilities{VLA, HST (ACS), Blanco, CFHT, GBO:300ft}
\software{\texttt{DOLPHOT} \citep{Dolphin2000}, \texttt{CASA} \citep{CASA}, \texttt{astropy} \citep{astropy2013,astropy2018}, \texttt{APLpy} \citep{aplpy2012,aplpy2019}, \texttt{Photutils} \citep{photutils}, \texttt{astroquery} \citep{astroquery}, \texttt{reproject} \citep{reproject}, \texttt{DS9} \citep{DS9}, \texttt{Aladin} \citep{Aladin2000,Aladin2014}}
}

\bibliography{refs}{}
\bibliographystyle{aasjournal}



\appendix

\section{Channel maps}
\label{sec:chan_maps}

\begin{figure}
    \centering
    \includegraphics[width=0.32\textwidth]{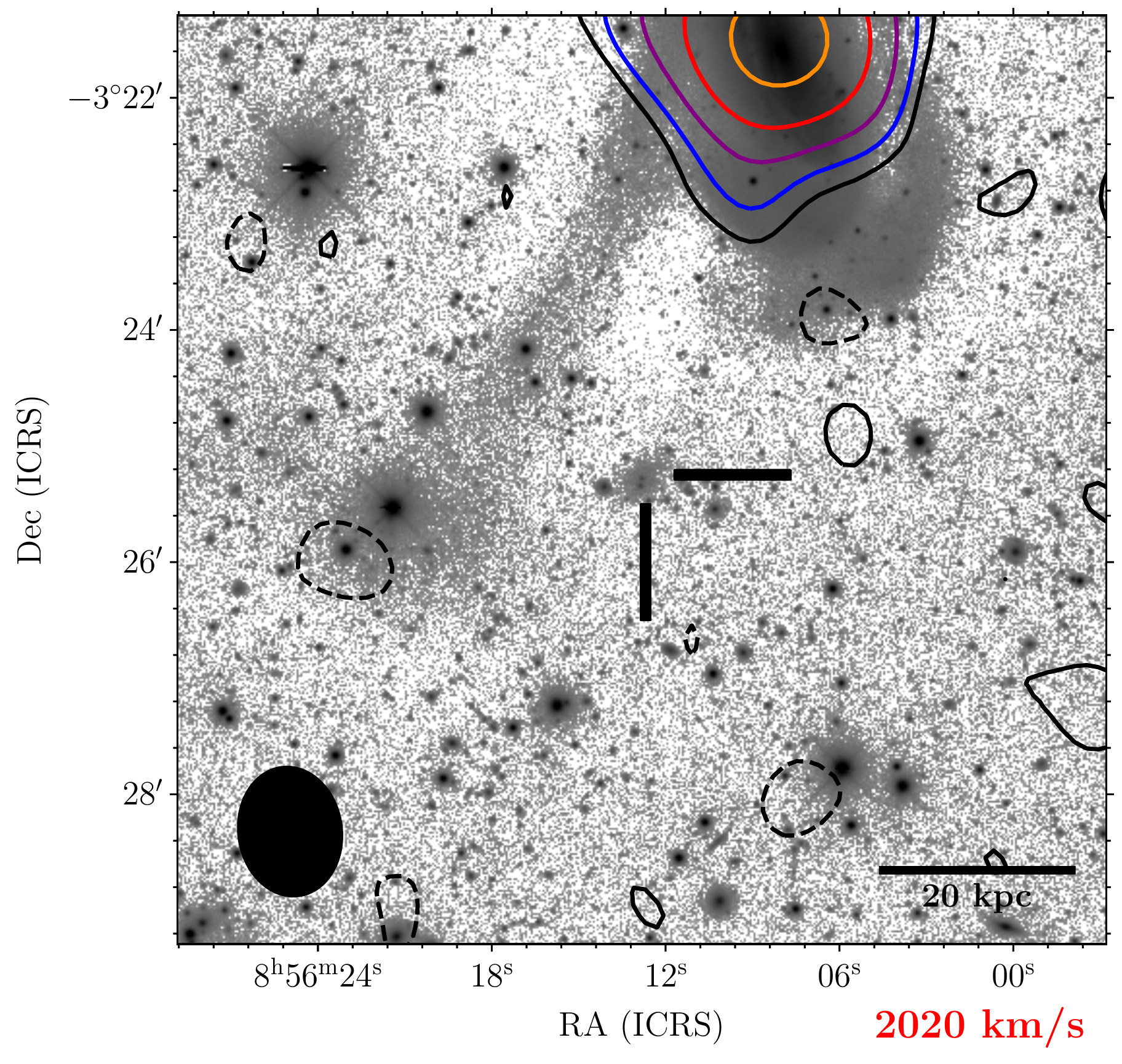}
    \includegraphics[width=0.32\textwidth]{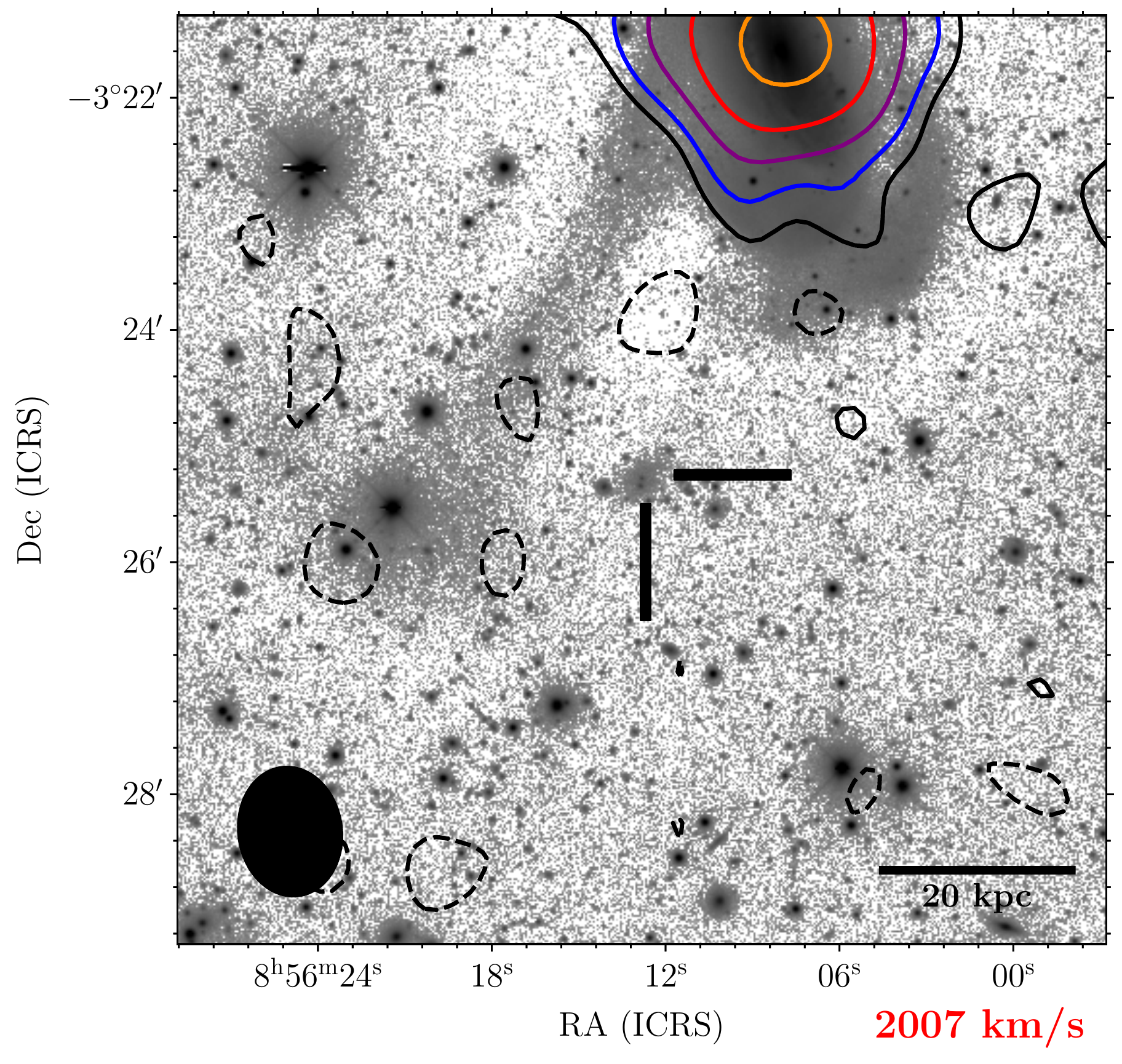}
    \includegraphics[width=0.32\textwidth]{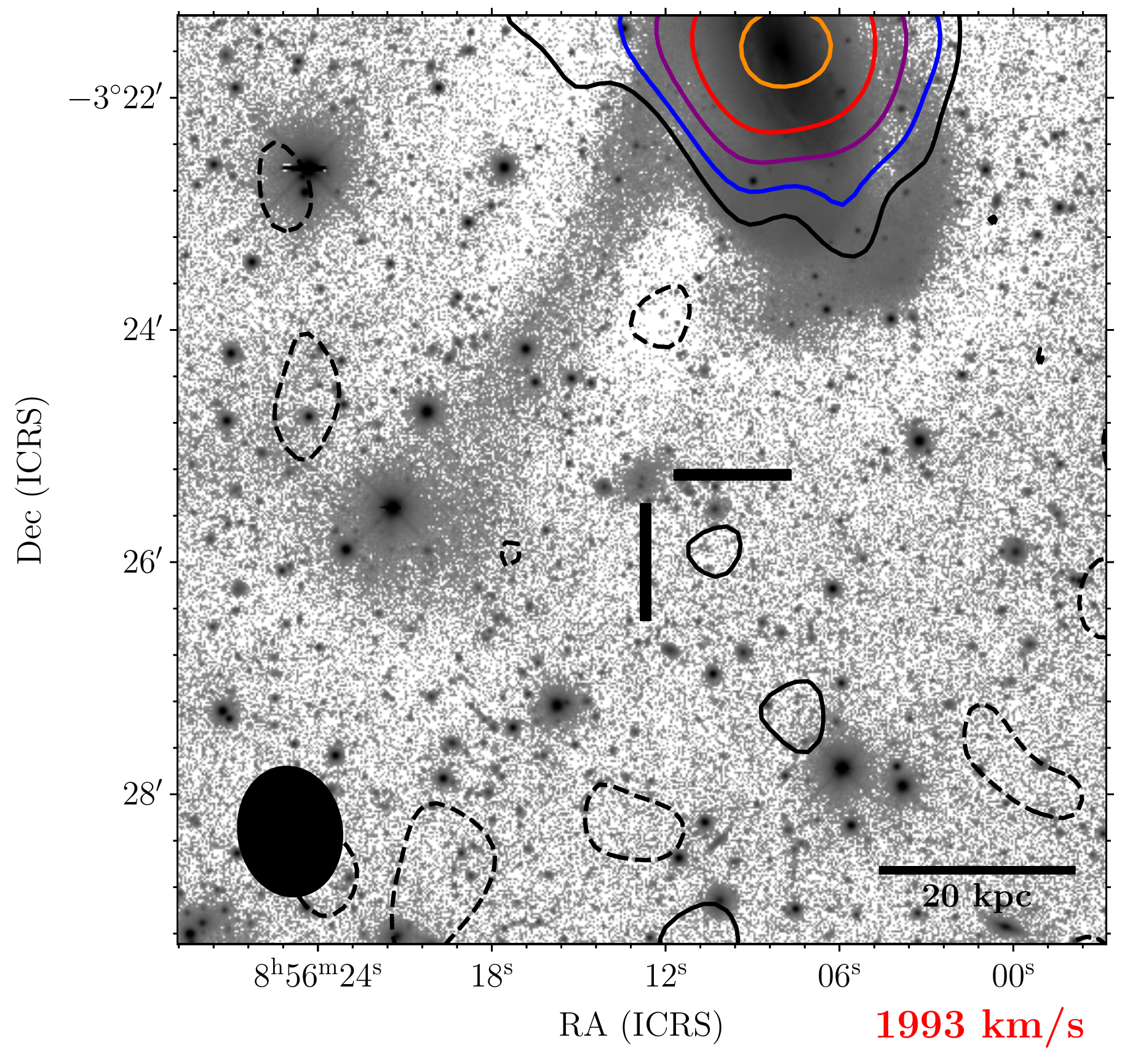}
    
    \includegraphics[width=0.32\textwidth]{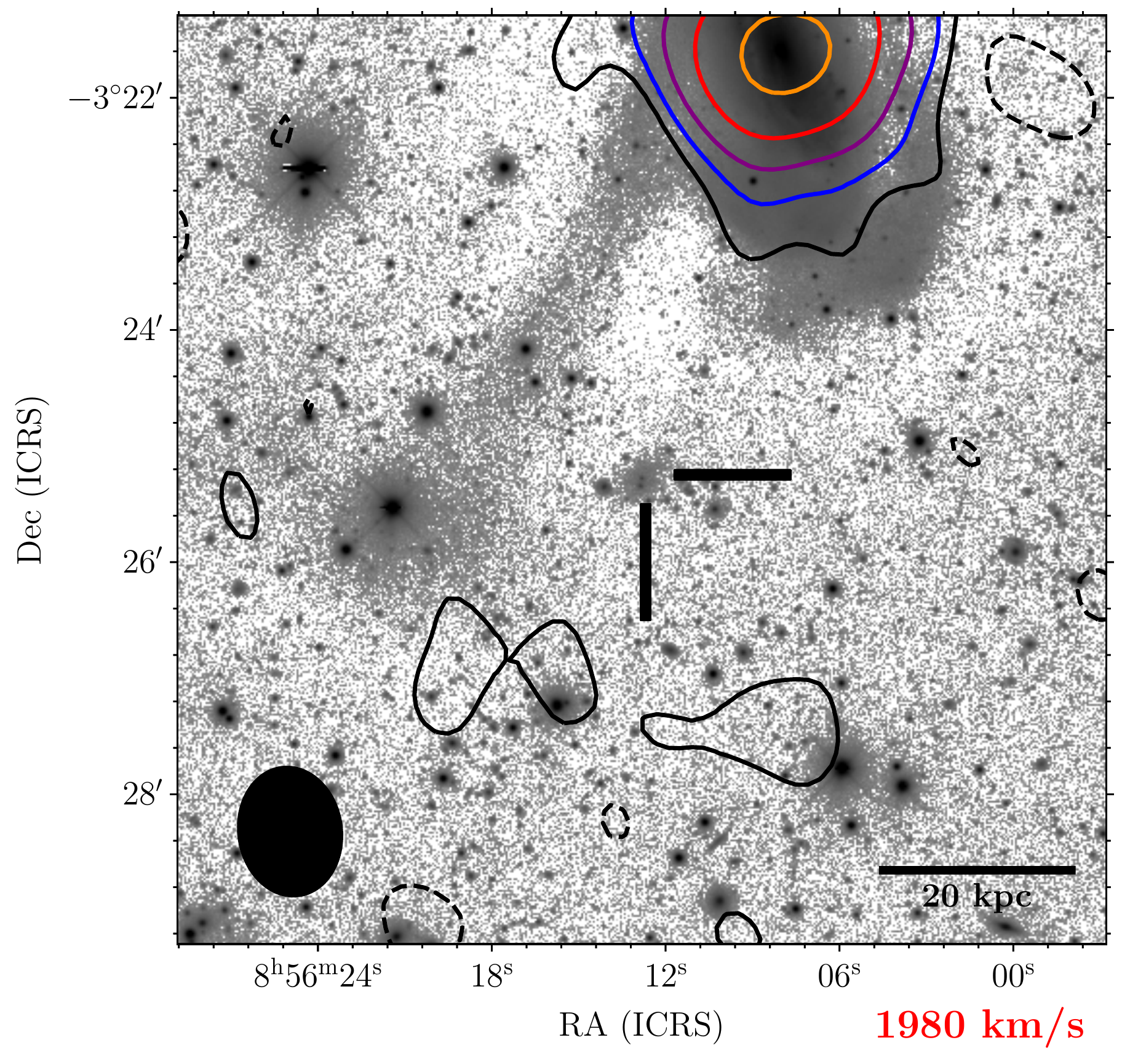}
    \includegraphics[width=0.32\textwidth]{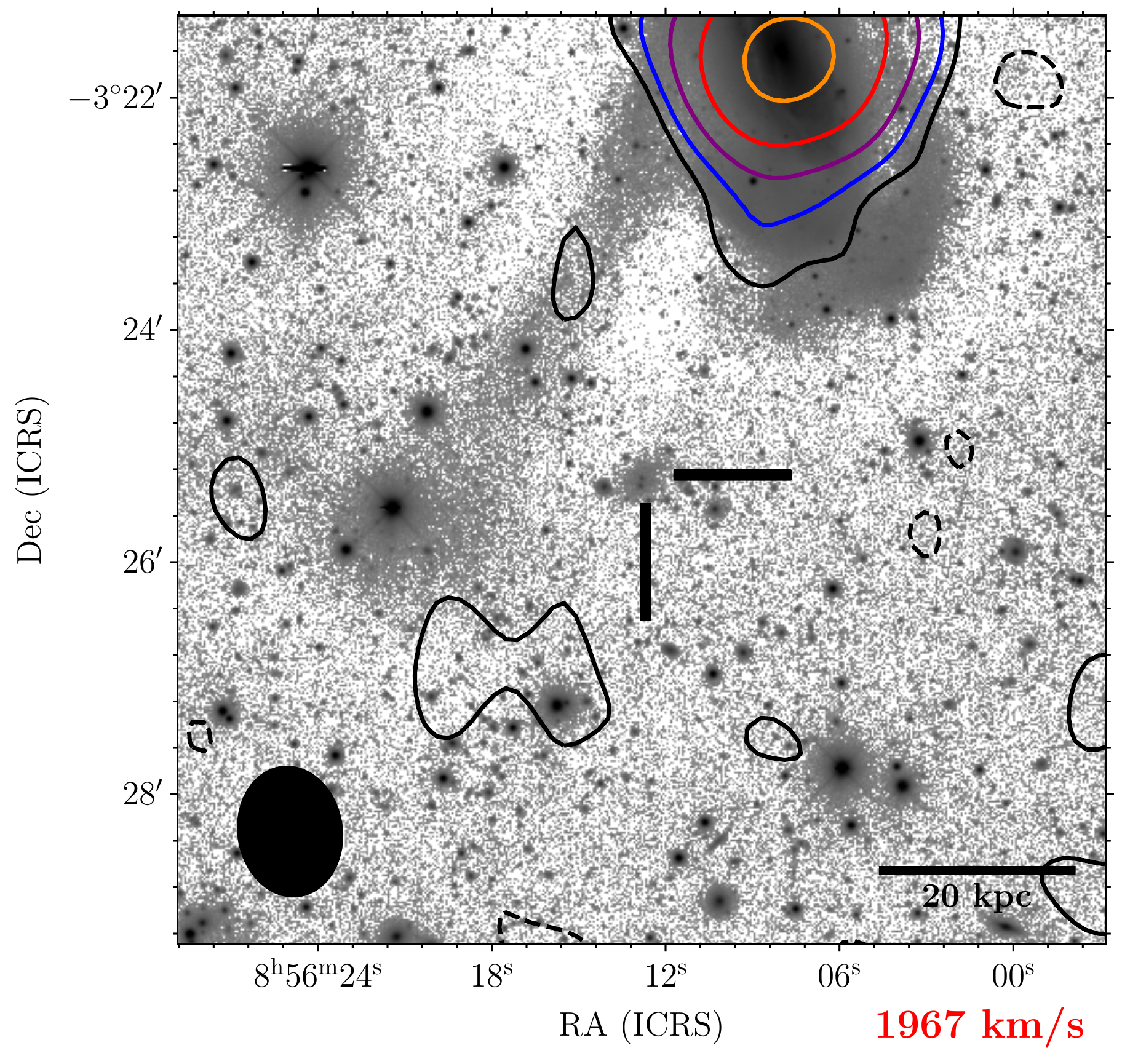}
    \includegraphics[width=0.32\textwidth]{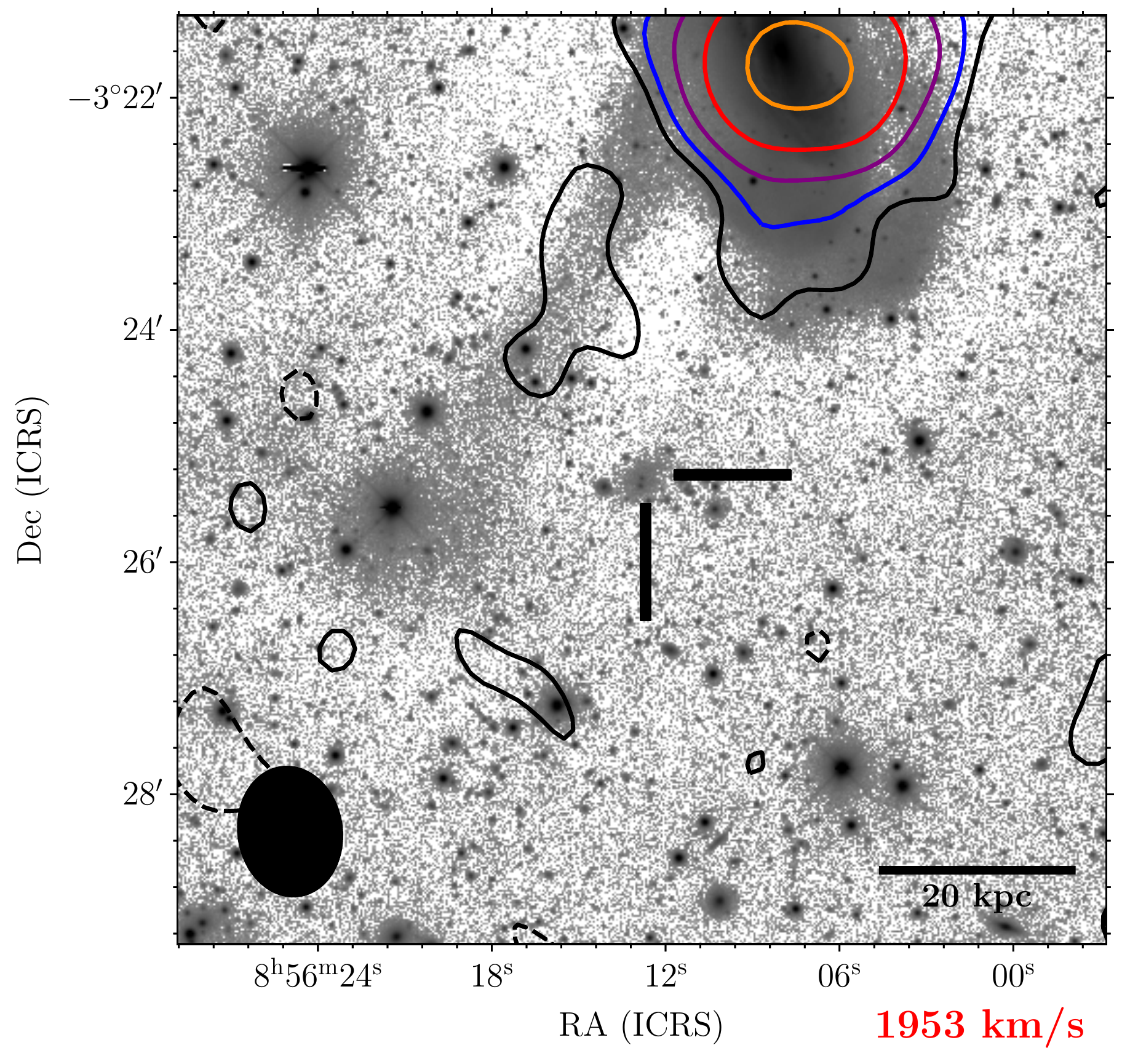}
    
    \includegraphics[width=0.32\textwidth]{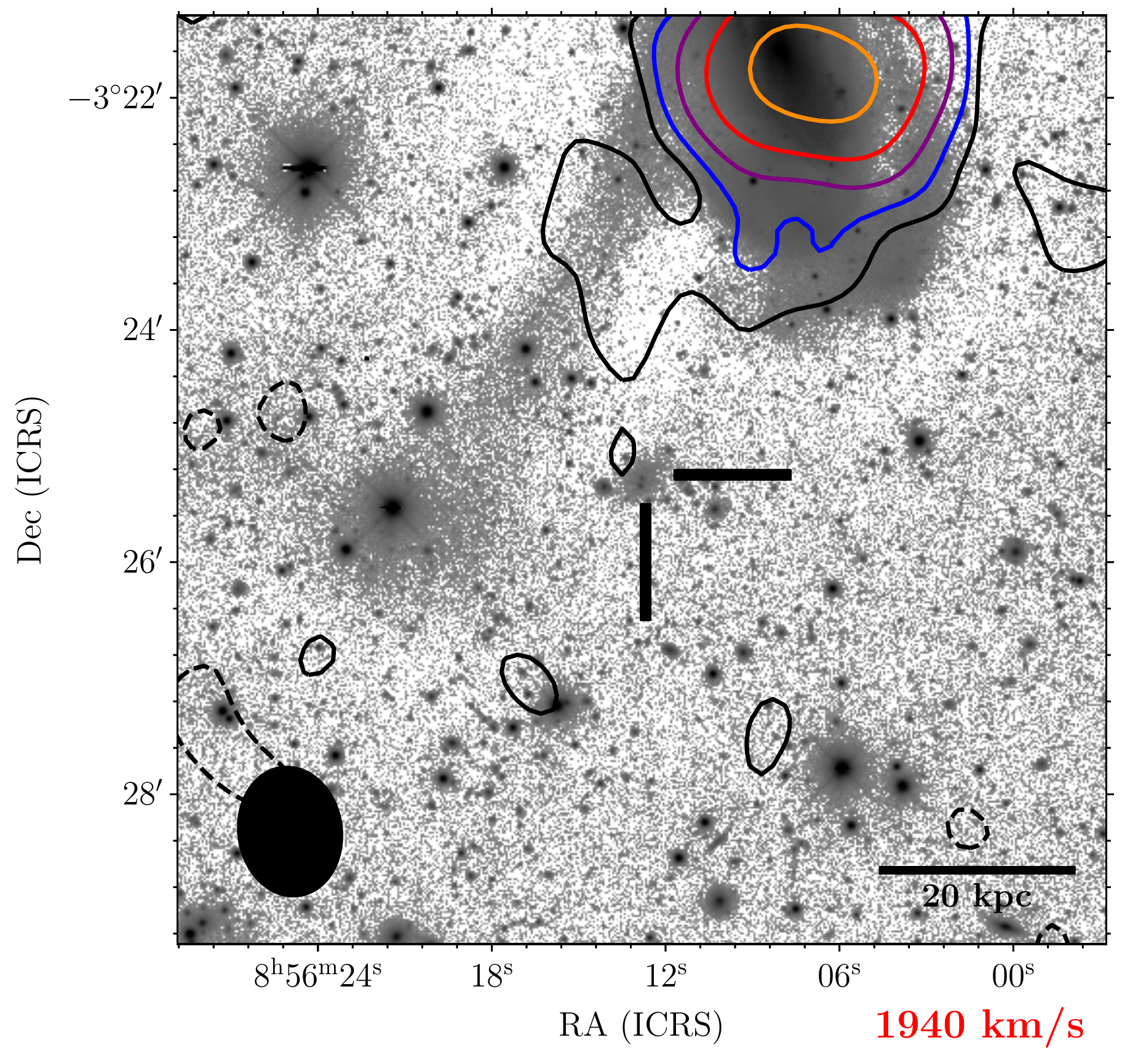}
    \includegraphics[width=0.32\textwidth]{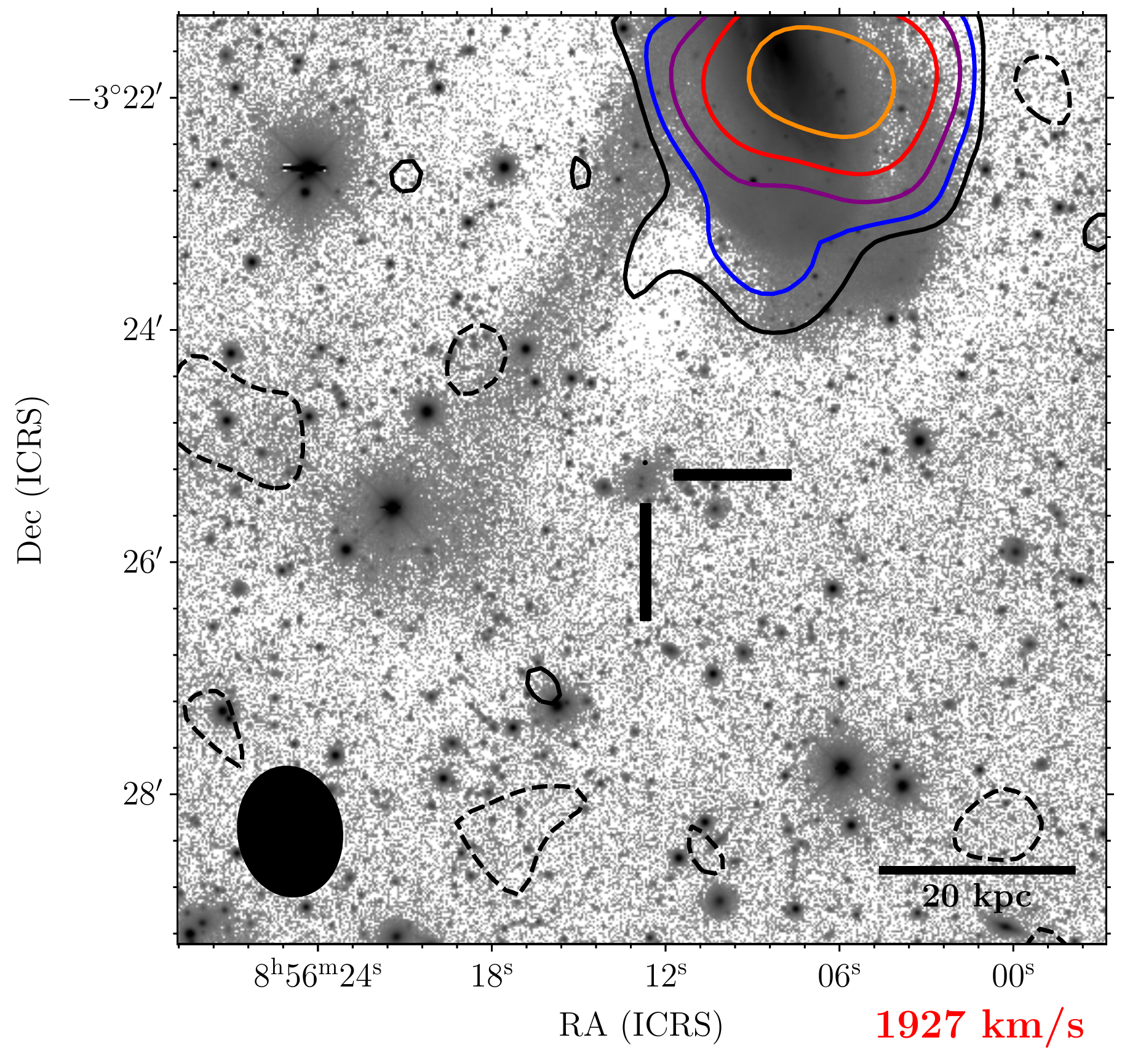}
    \includegraphics[width=0.32\textwidth]{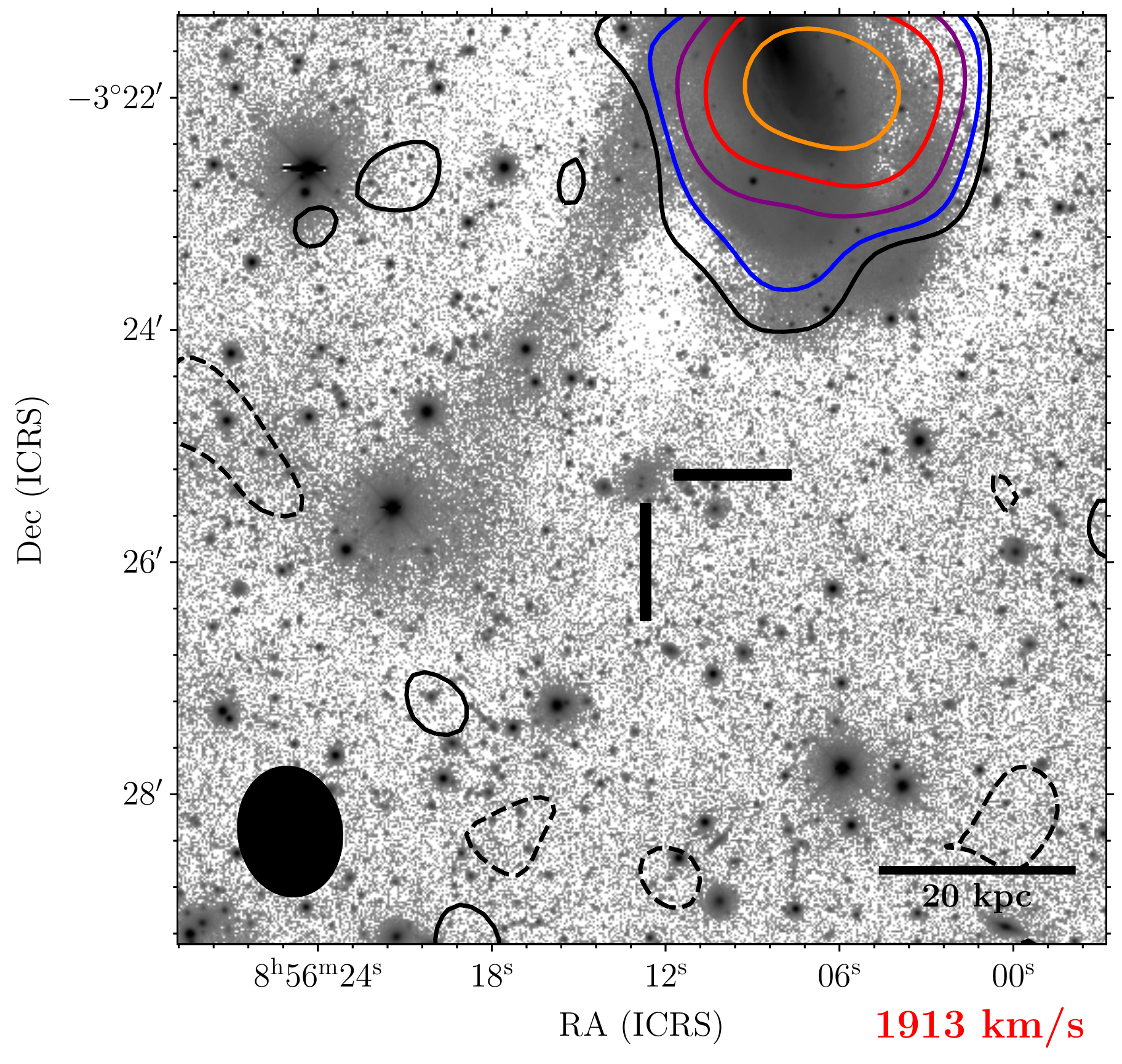}
    
    \caption{Channel maps in the vicinity of NGC 2708 DW1 covering the velocity range of the disturbance on the SE side of the disk of NGC 2708. Contours levels: -1, 1, 2, 4 $\times 0.091 \; \mathrm{Jy\,km\,s^{-1}}$ per beam, 0.053 $\mathrm{M_{\odot}\,pc^{-2}}$, or $6.6 \times 10^{18} \; \mathrm{cm^{-2}}$ (over 1 channel, $\sim$13 \kms). The lowest (positive) contour corresponds to approximately 2$\sigma$, which is slightly lower than those in Figure \ref{fig:NGC2708_mom0}.}
    \label{fig:NGC2708DW1_chan_maps}
\end{figure}

\begin{figure}
    \centering
    \includegraphics[width=0.32\textwidth]{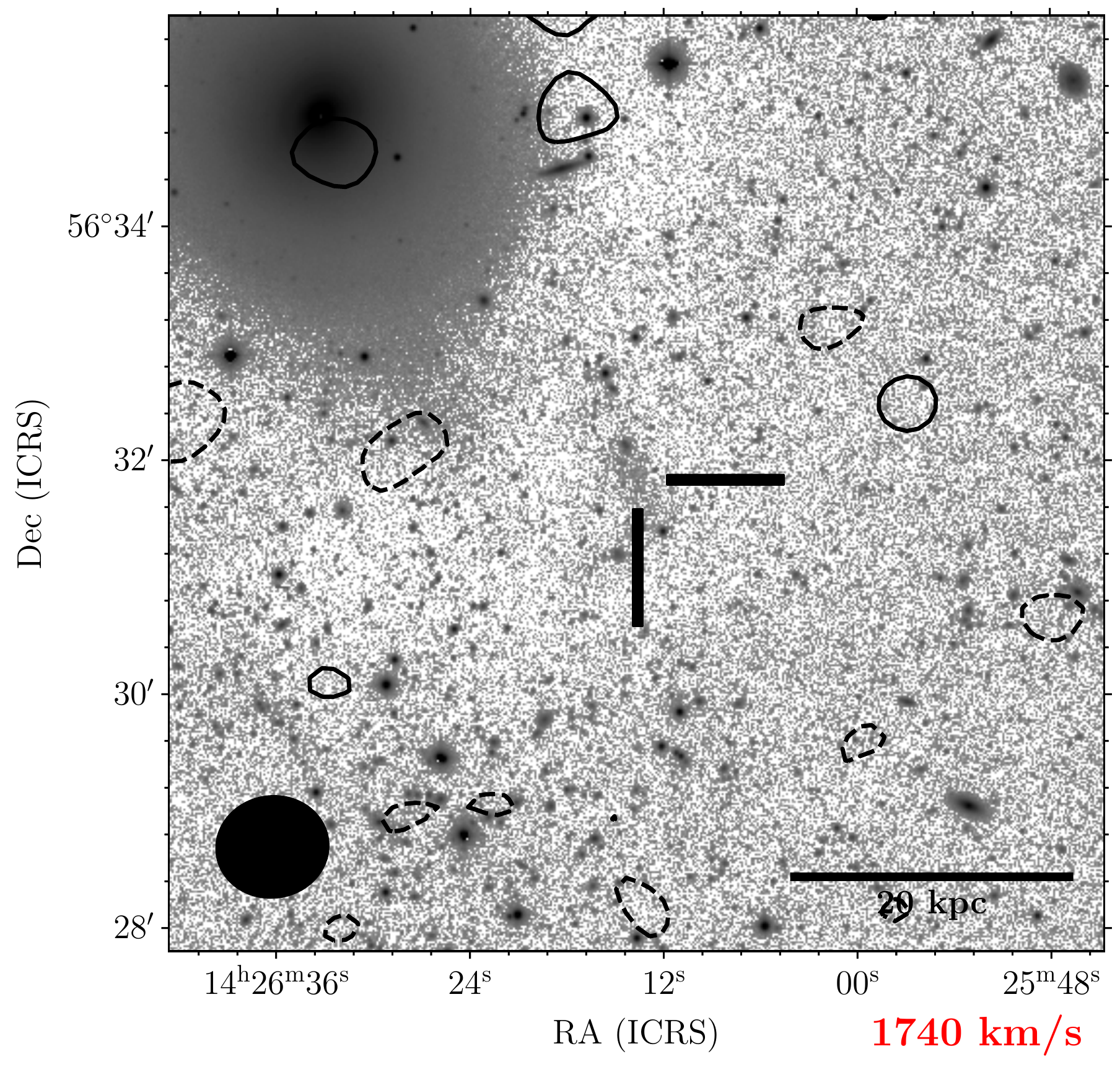}
    \includegraphics[width=0.32\textwidth]{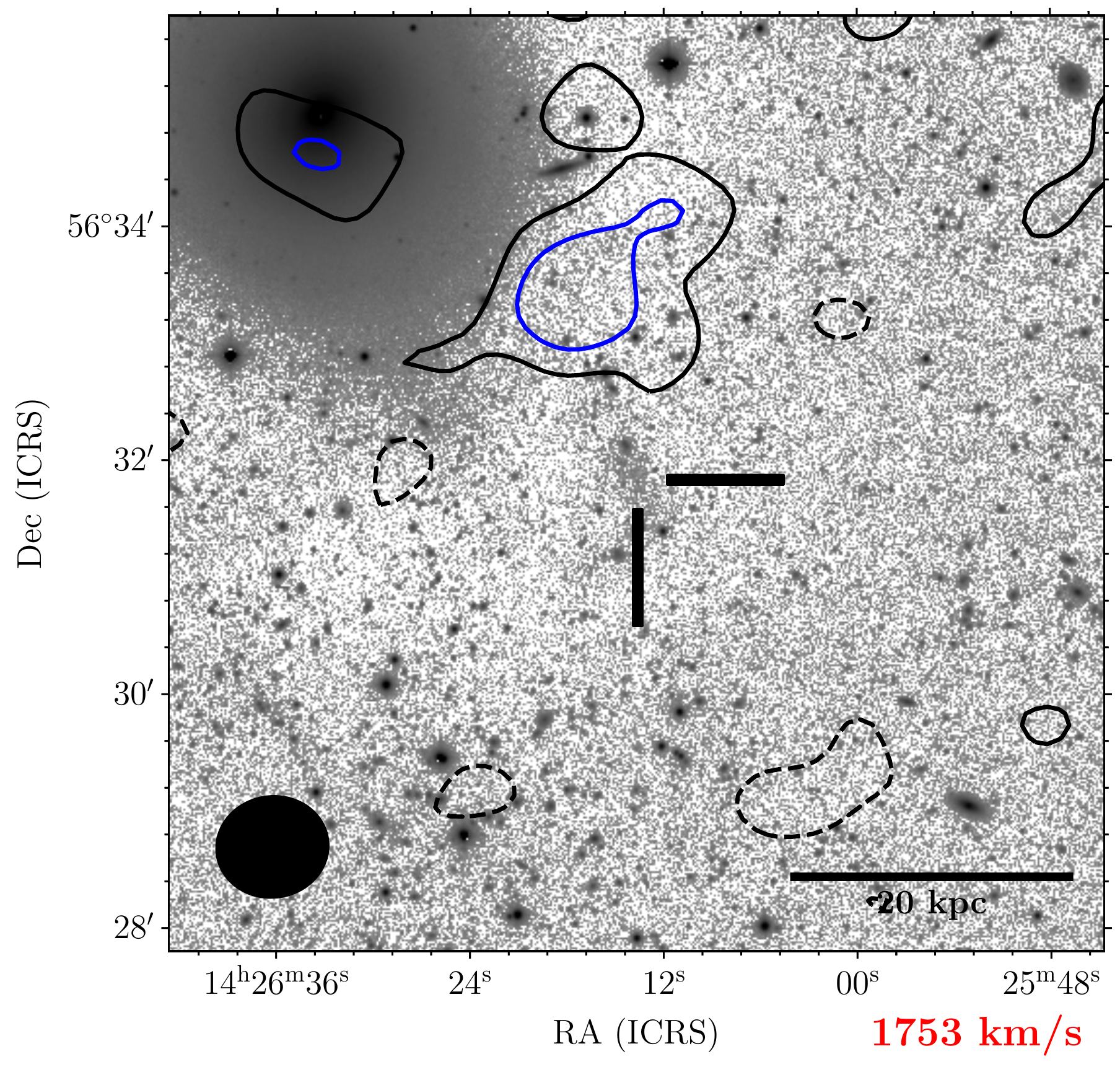}
    \includegraphics[width=0.32\textwidth]{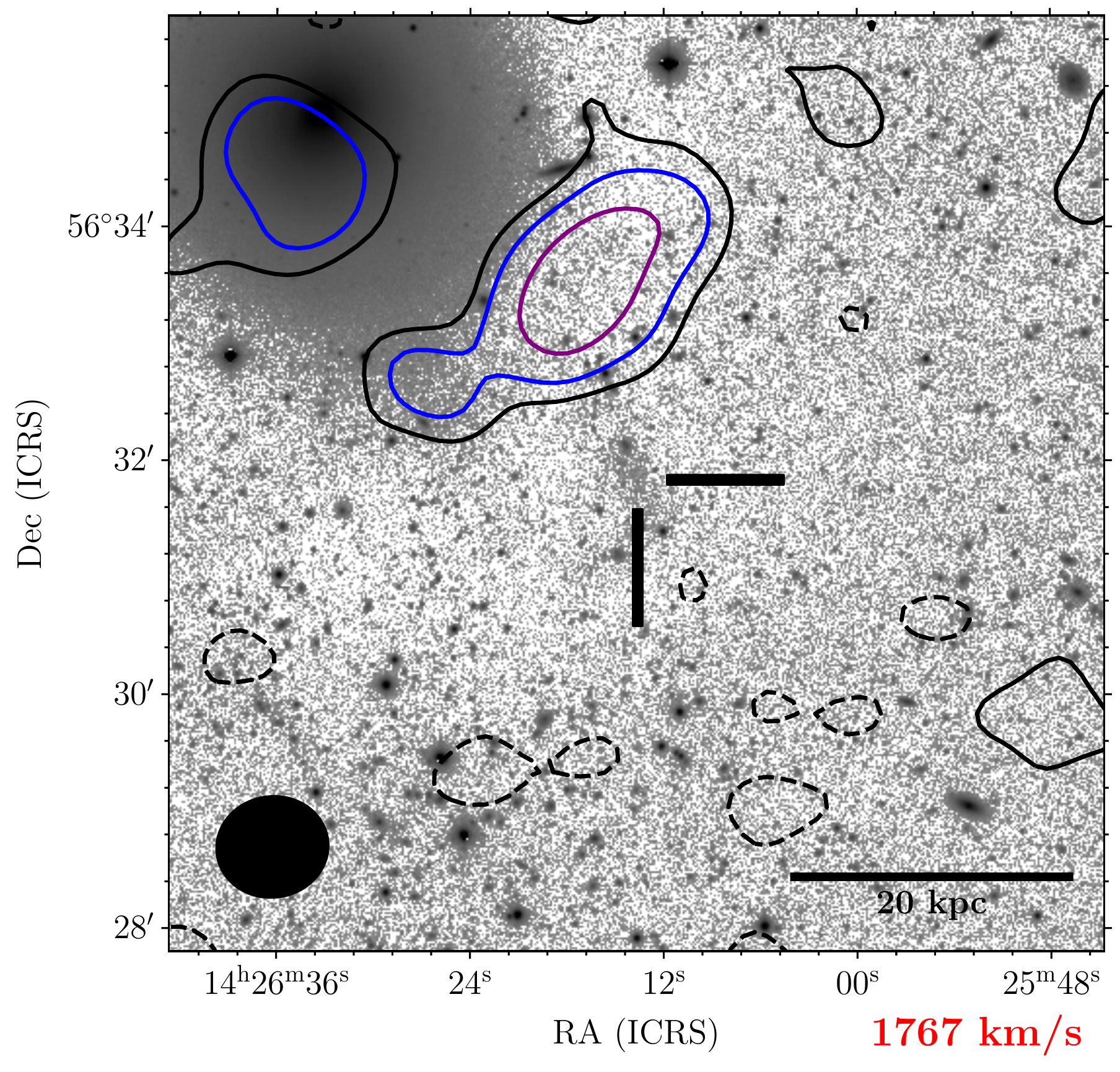}
    
    \includegraphics[width=0.32\textwidth]{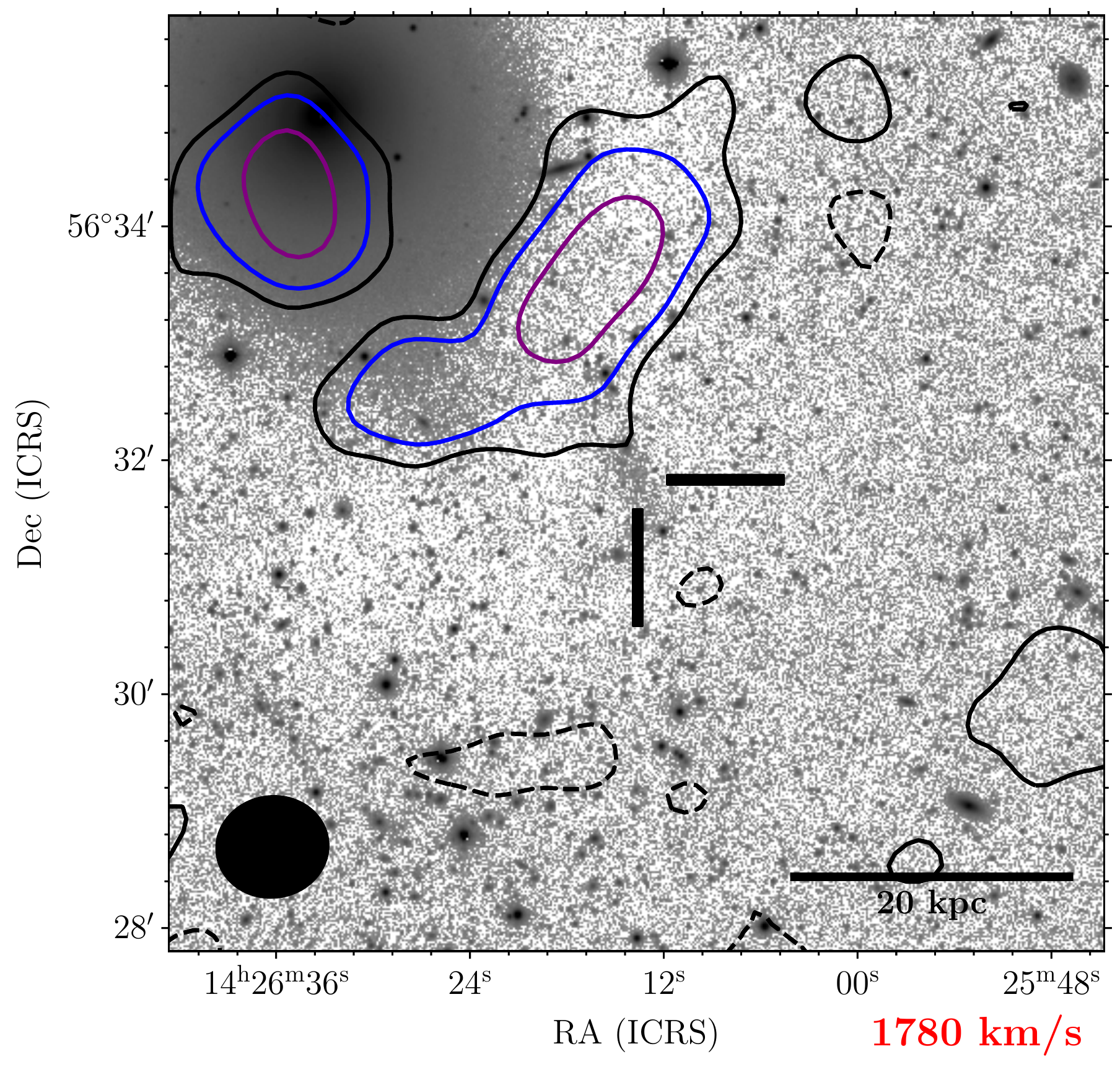}
    \includegraphics[width=0.32\textwidth]{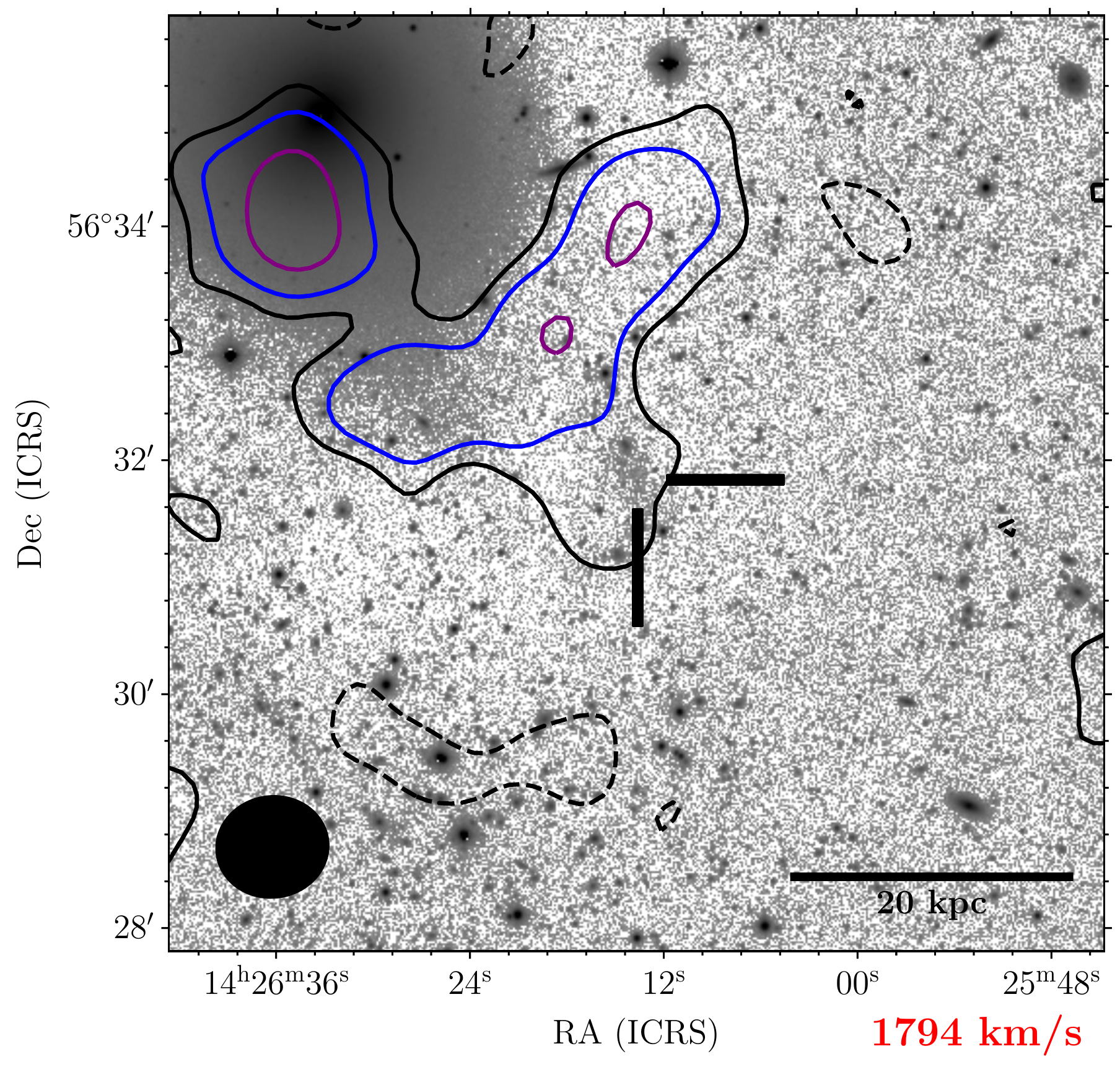}
    \includegraphics[width=0.32\textwidth]{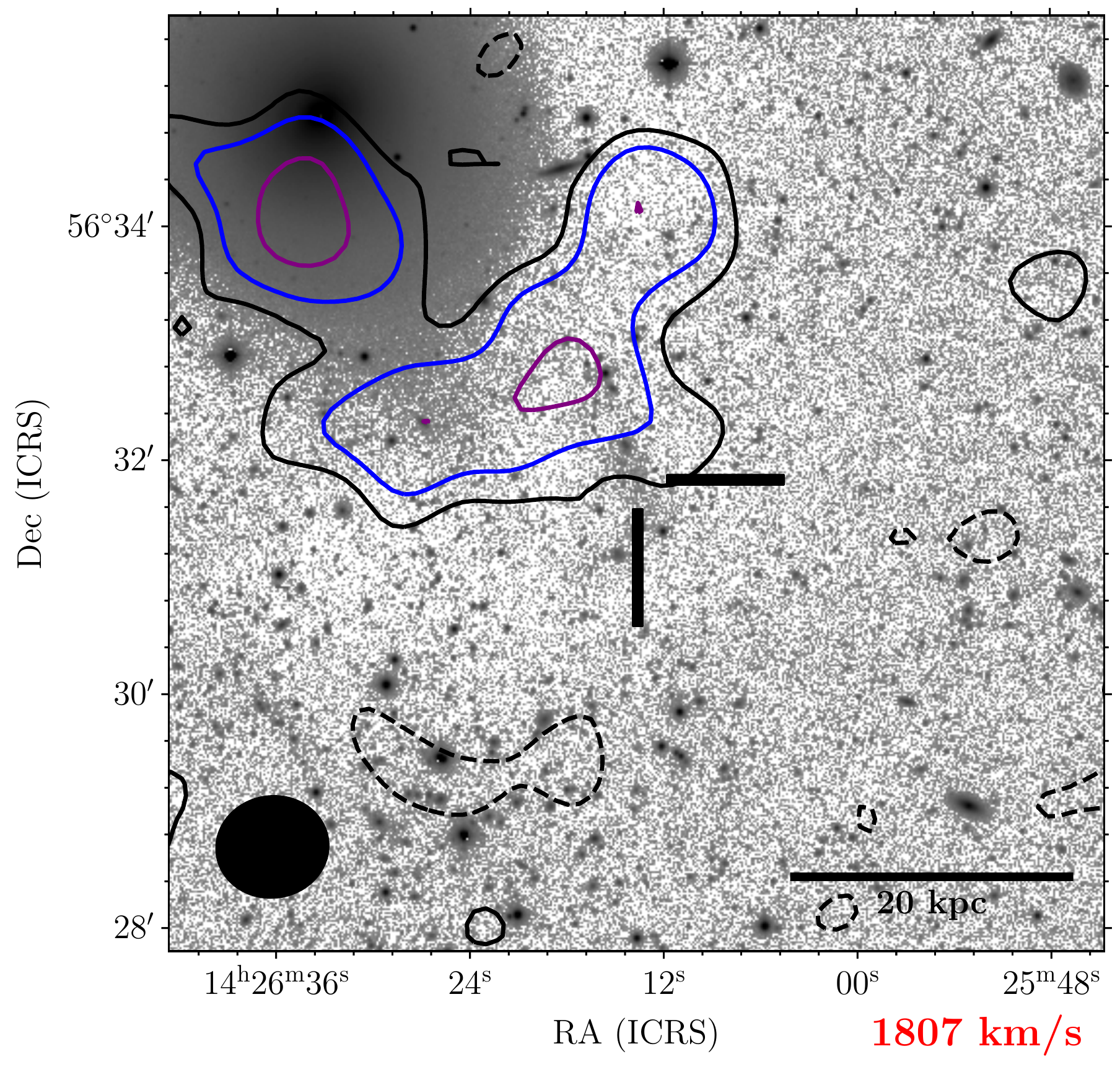}
    
    \includegraphics[width=0.32\textwidth]{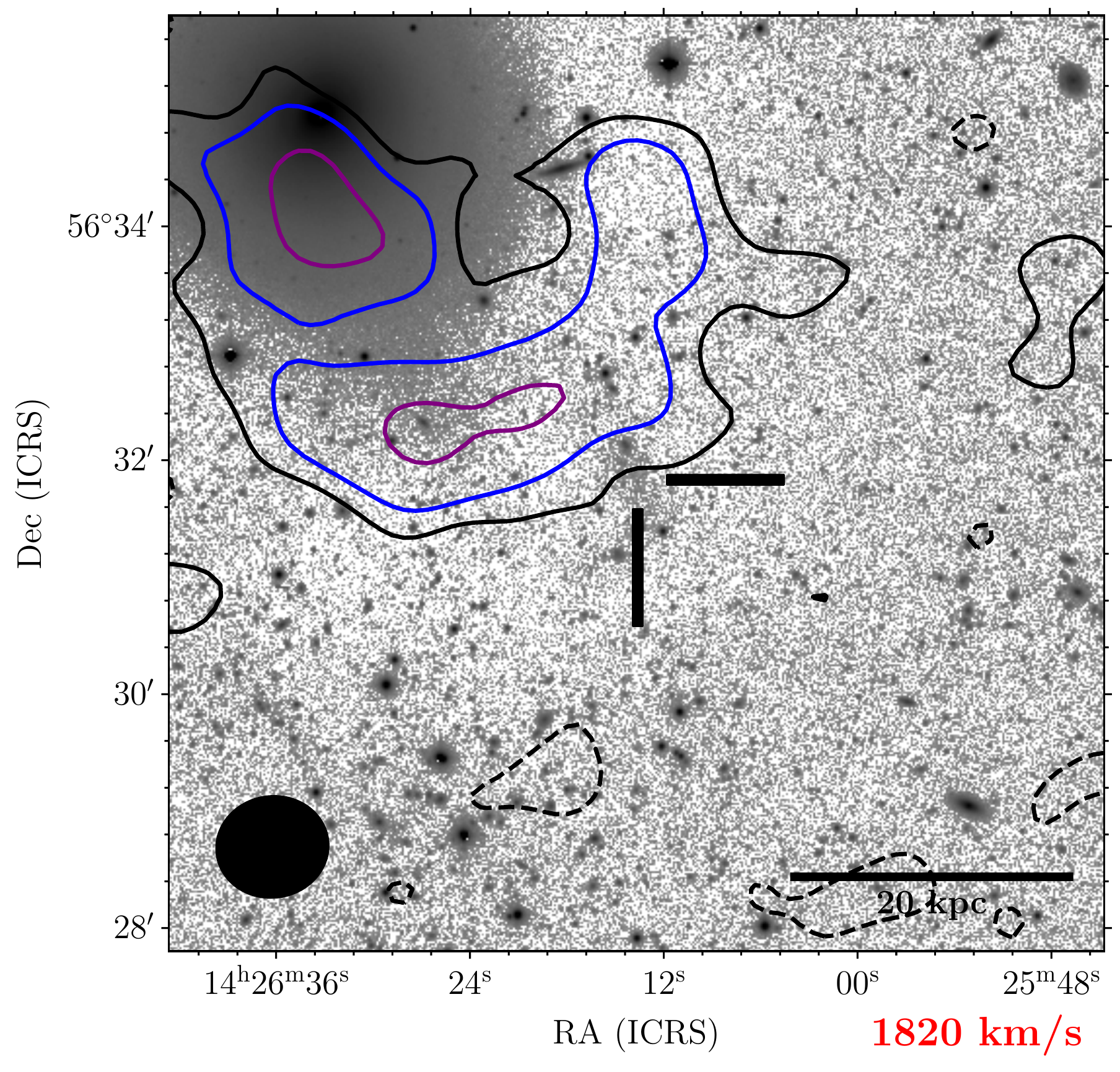}
    \includegraphics[width=0.32\textwidth]{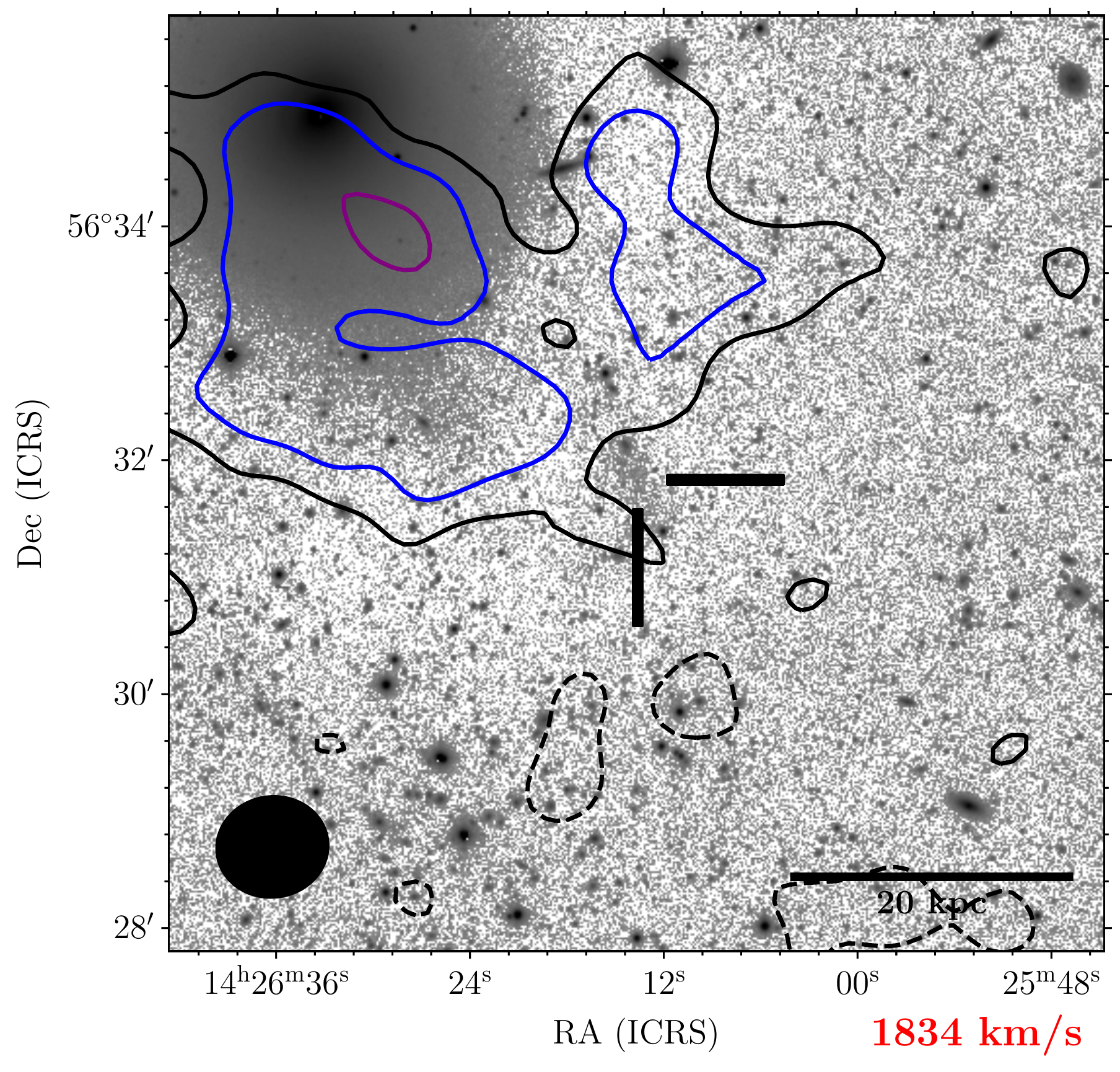}
    \includegraphics[width=0.32\textwidth]{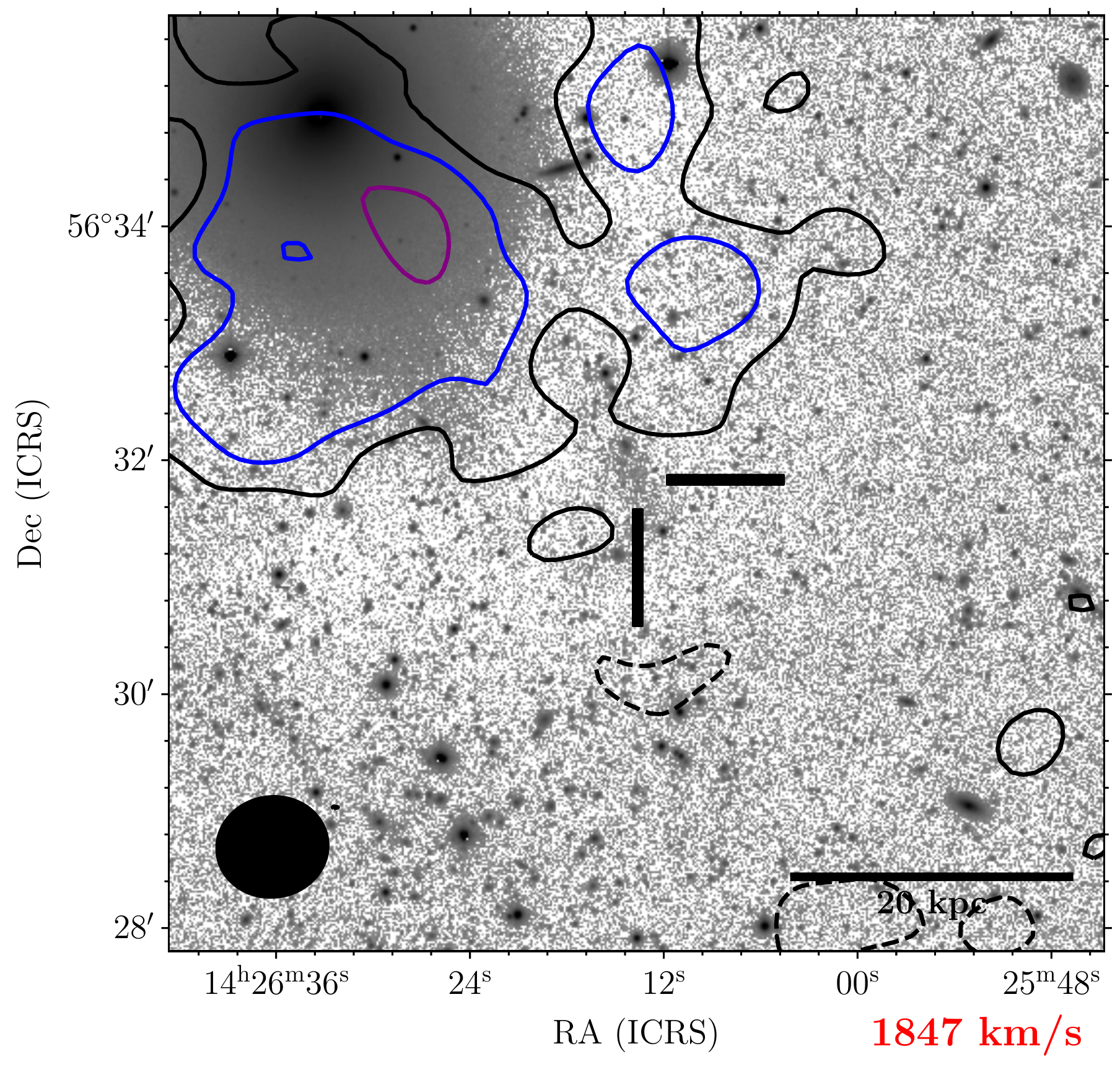}
    
    \caption{Channel maps in the vicinity of NGC 5631 DW1 covering the velocity range of the SW tail of NGC 5631. Contours levels: -1, 1, 2, 4 $\times 0.080 \; \mathrm{Jy\,km\,s^{-1}}$ per beam, 0.056 $\mathrm{M_{\odot}\,pc^{-2}}$, or $7.0 \times 10^{18} \; \mathrm{cm^{-2}}$ (over 1 channel, $\sim$13 \kms). The lowest (positive) contour corresponds to approximately 2$\sigma$, which is slightly lower than those in Figure \ref{fig:NGC5631_mom0}.}
    \label{fig:NGC5631DW1_chan_maps}
\end{figure}

Figures \ref{fig:NGC2708DW1_chan_maps} and \ref{fig:NGC5631DW1_chan_maps} show the channel maps for the VLA \hi \ observations of NGC 2708-Dw1 and NGC 5631-Dw1 respectively.

\begin{figure}
    \centering
    \includegraphics[width=0.32\textwidth]{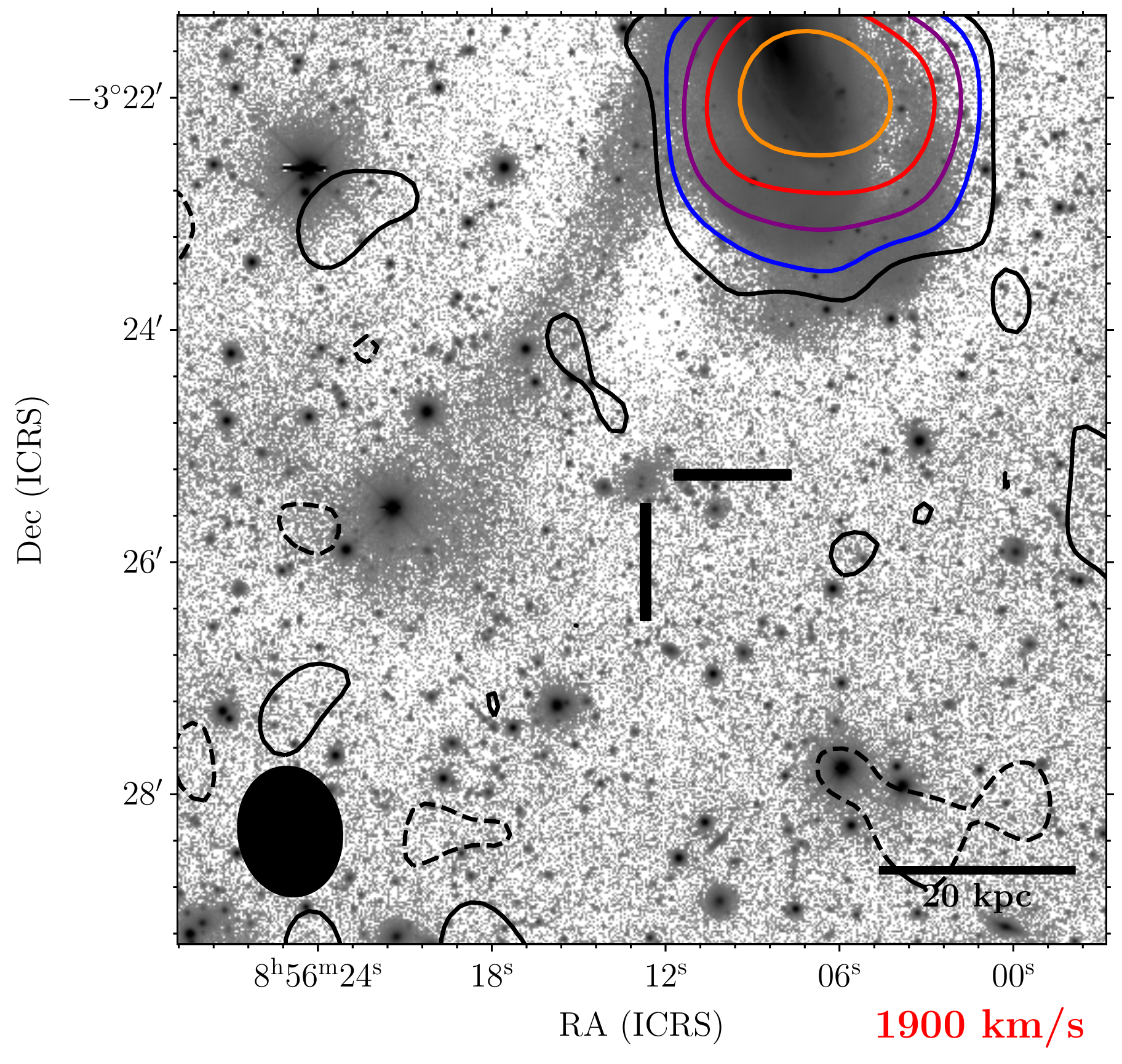}
    \includegraphics[width=0.32\textwidth]{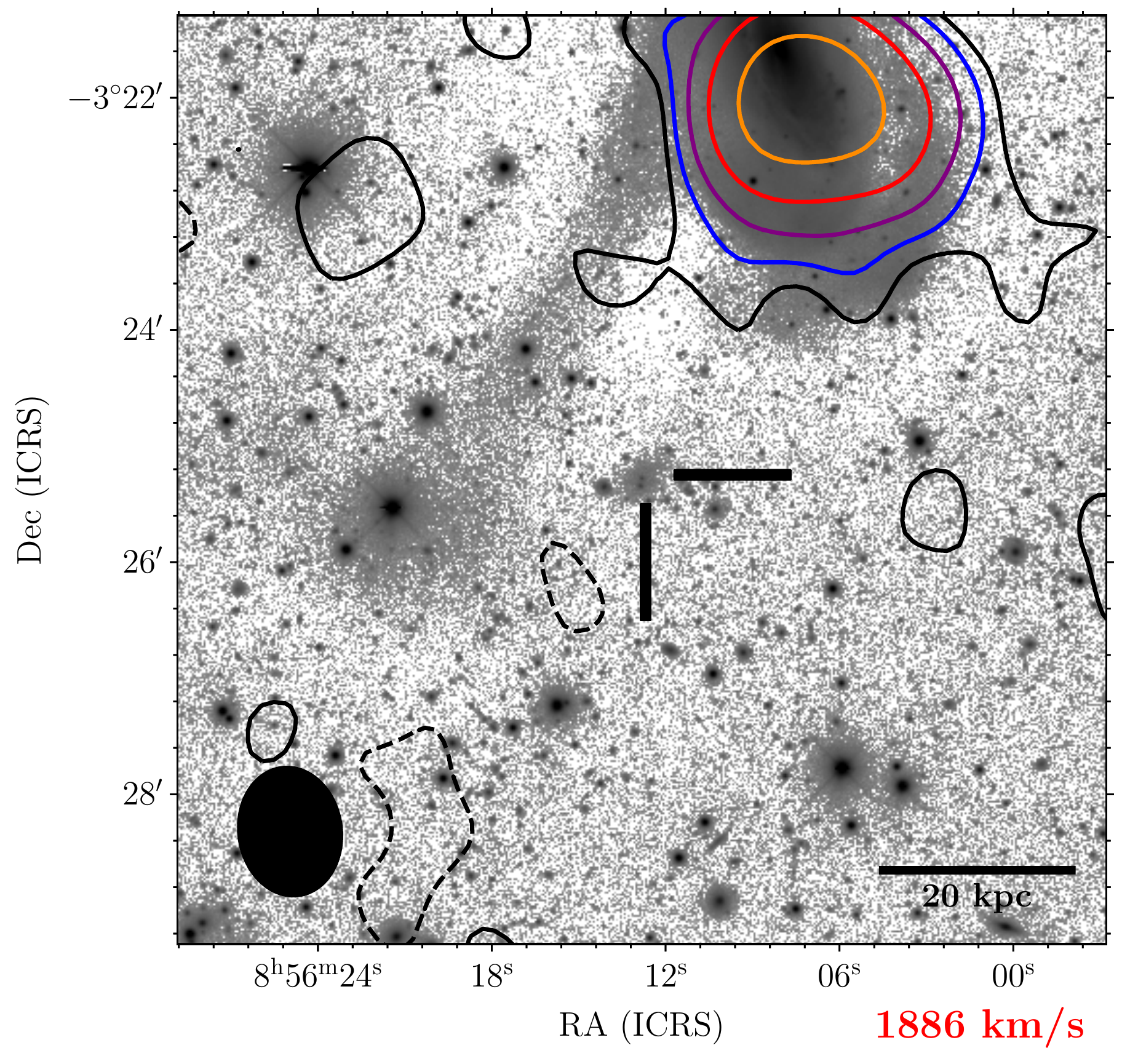}
    \includegraphics[width=0.32\textwidth]{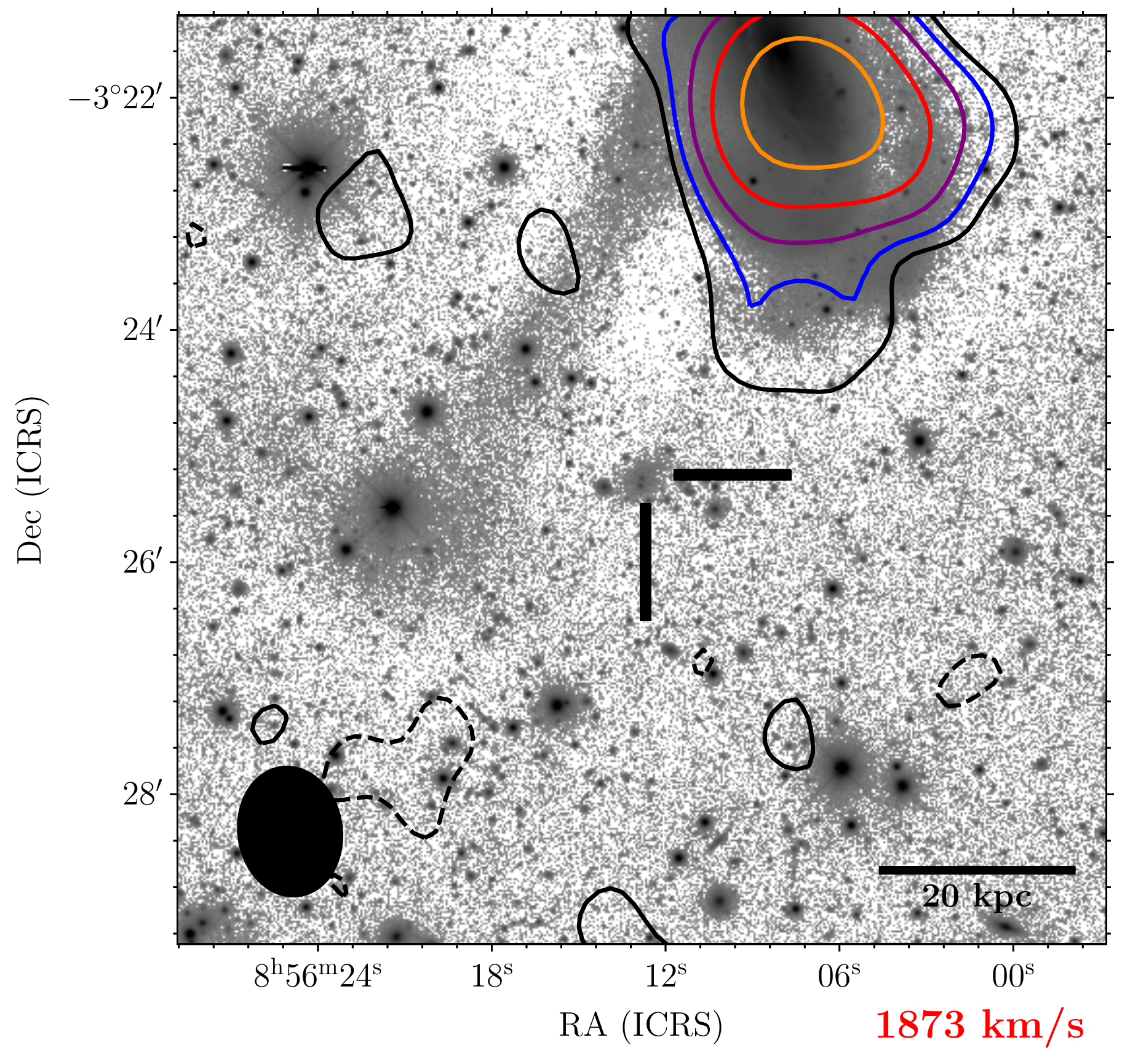}
    
    \includegraphics[width=0.32\textwidth]{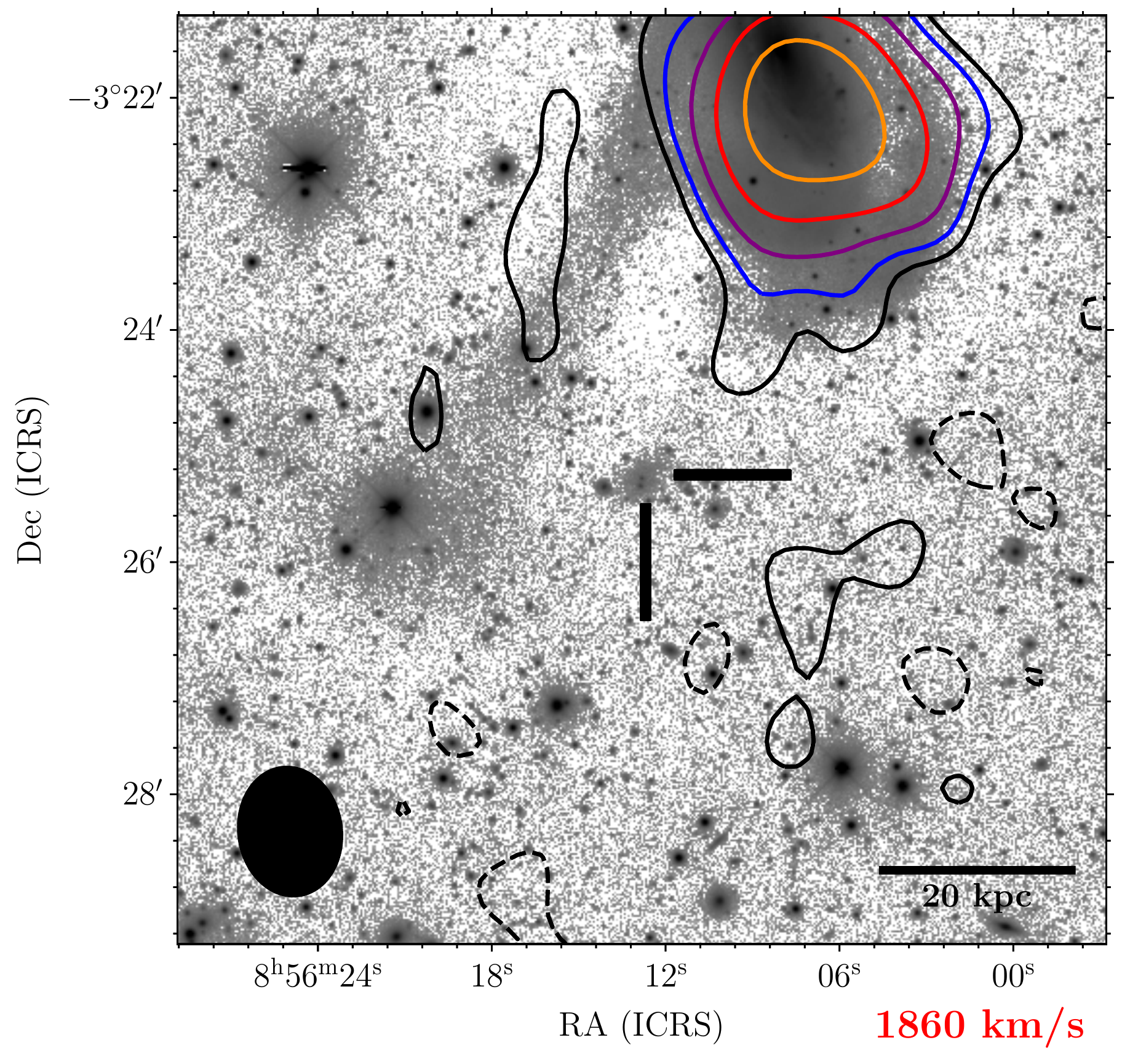}
    \includegraphics[width=0.32\textwidth]{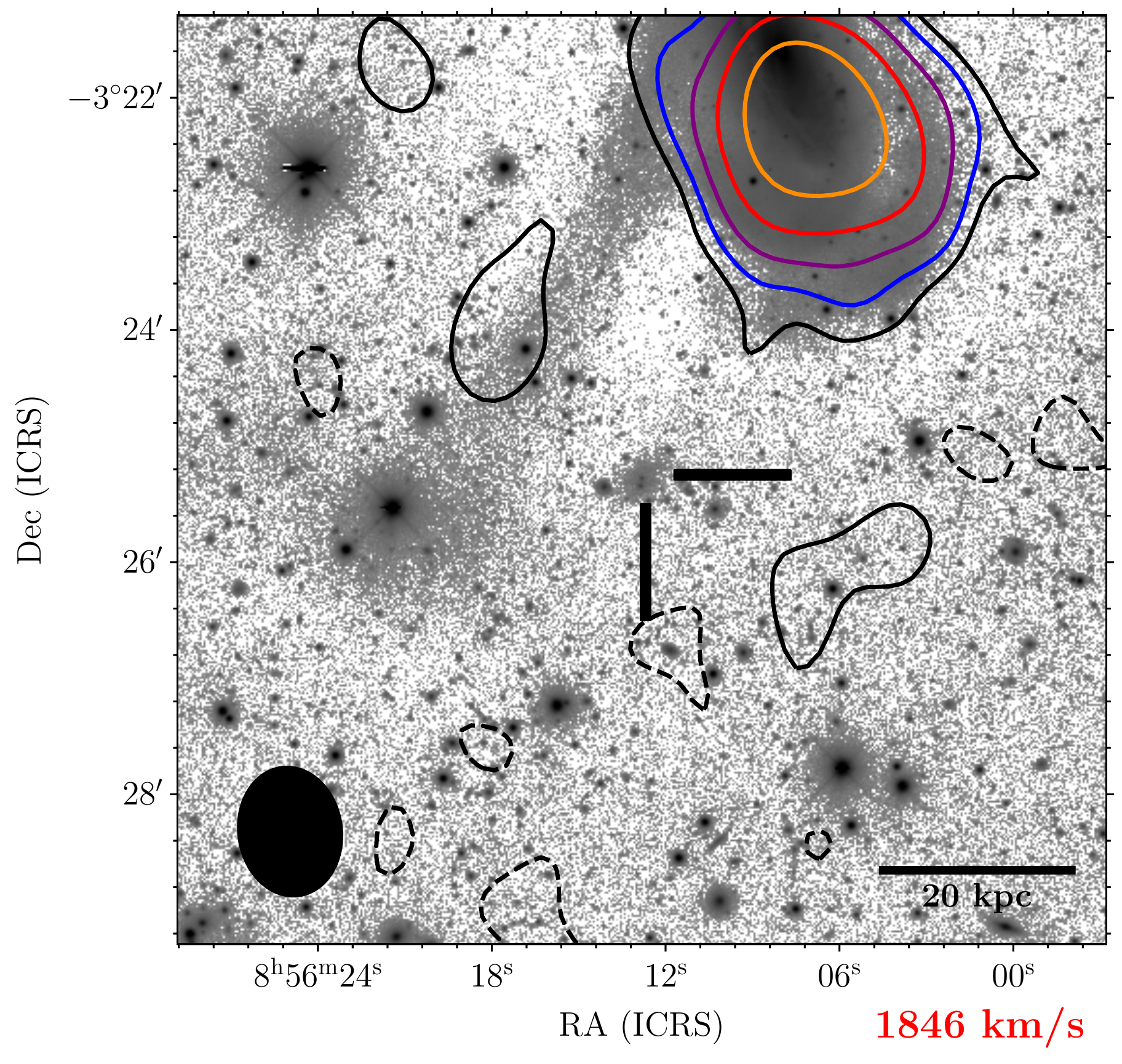}
    \includegraphics[width=0.32\textwidth]{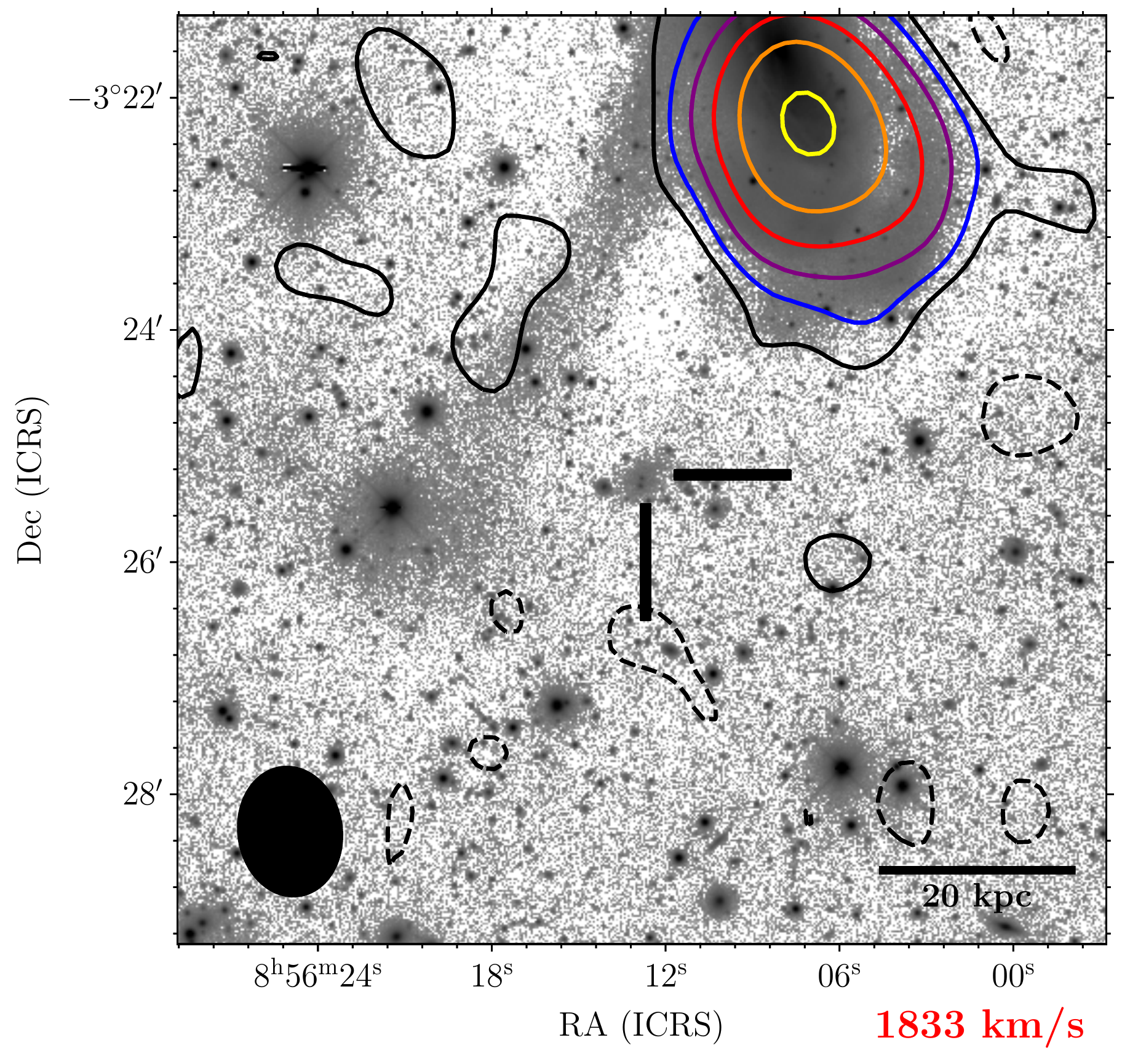}
    
    \includegraphics[width=0.32\textwidth]{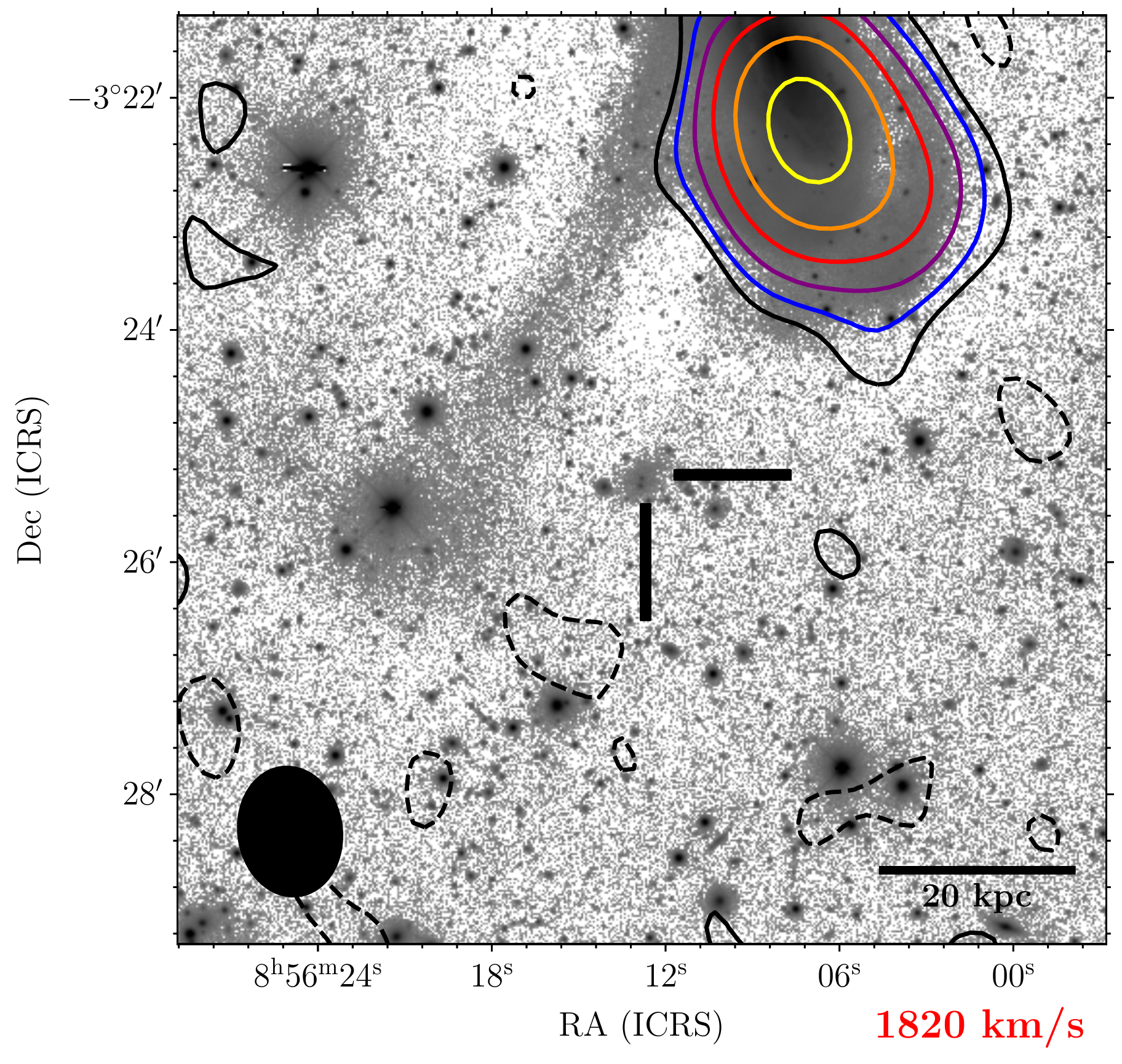}
    \includegraphics[width=0.32\textwidth]{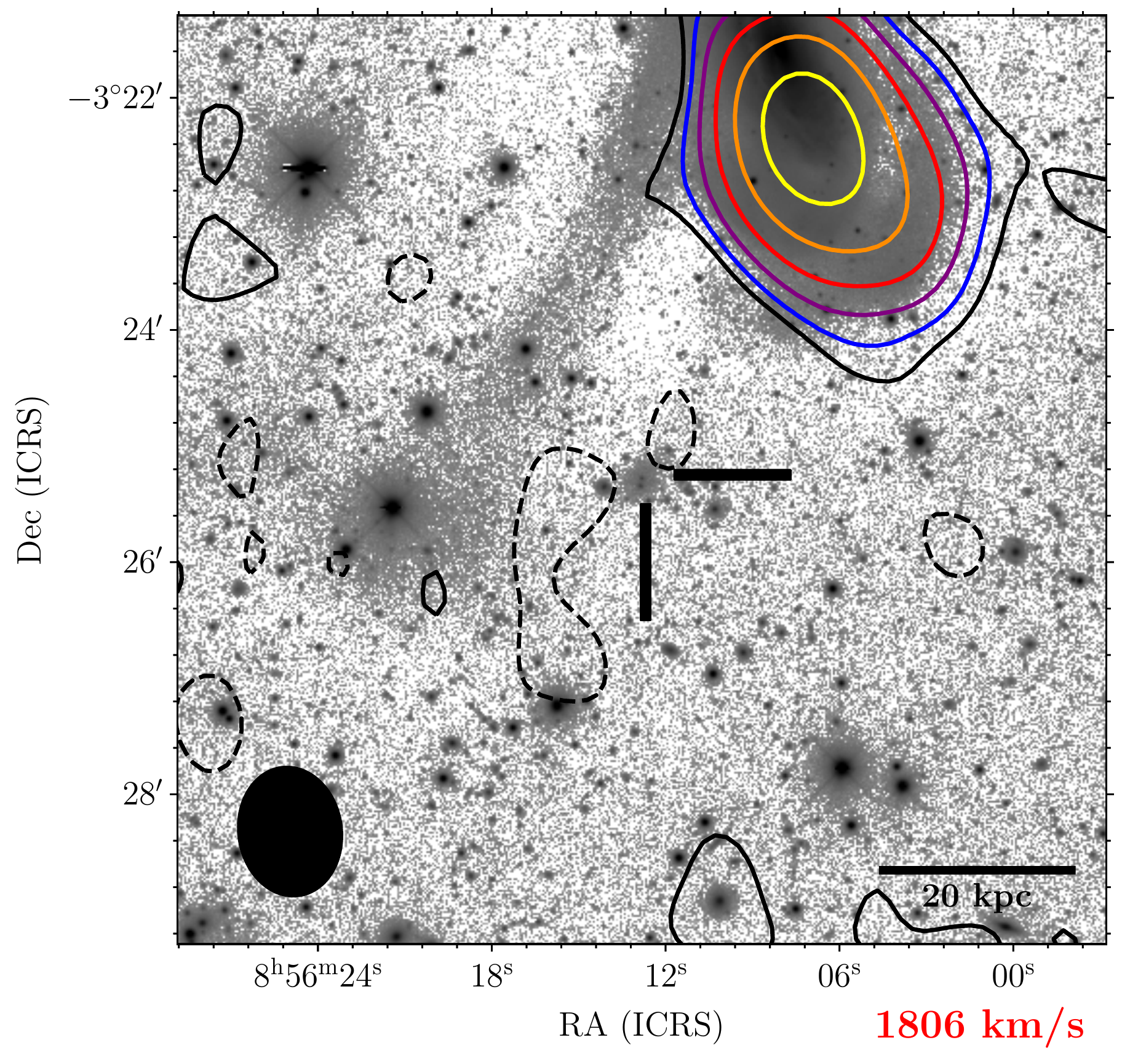}
    \includegraphics[width=0.32\textwidth]{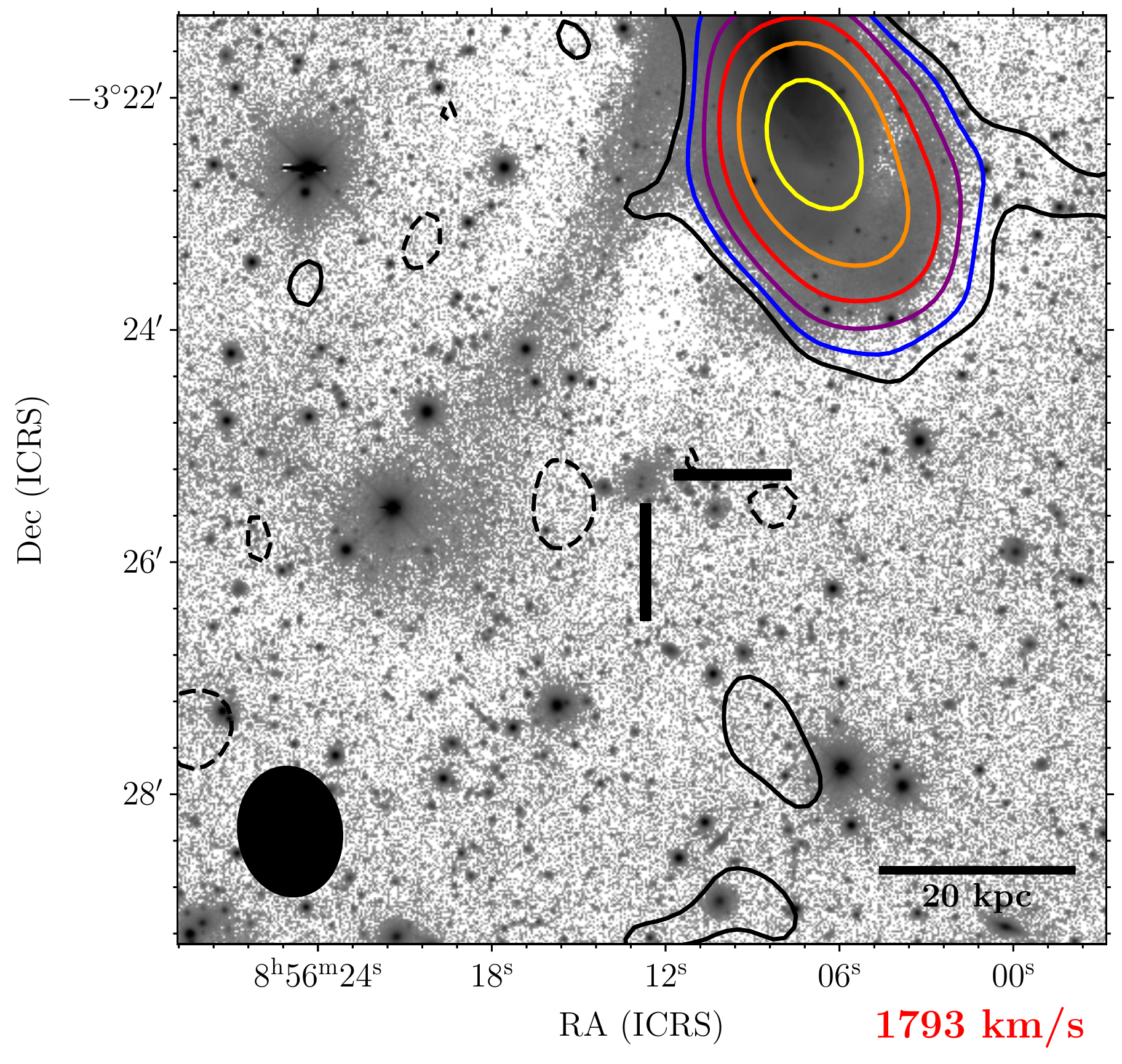}
\end{figure}

\begin{figure}
    \centering
    \includegraphics[width=0.32\textwidth]{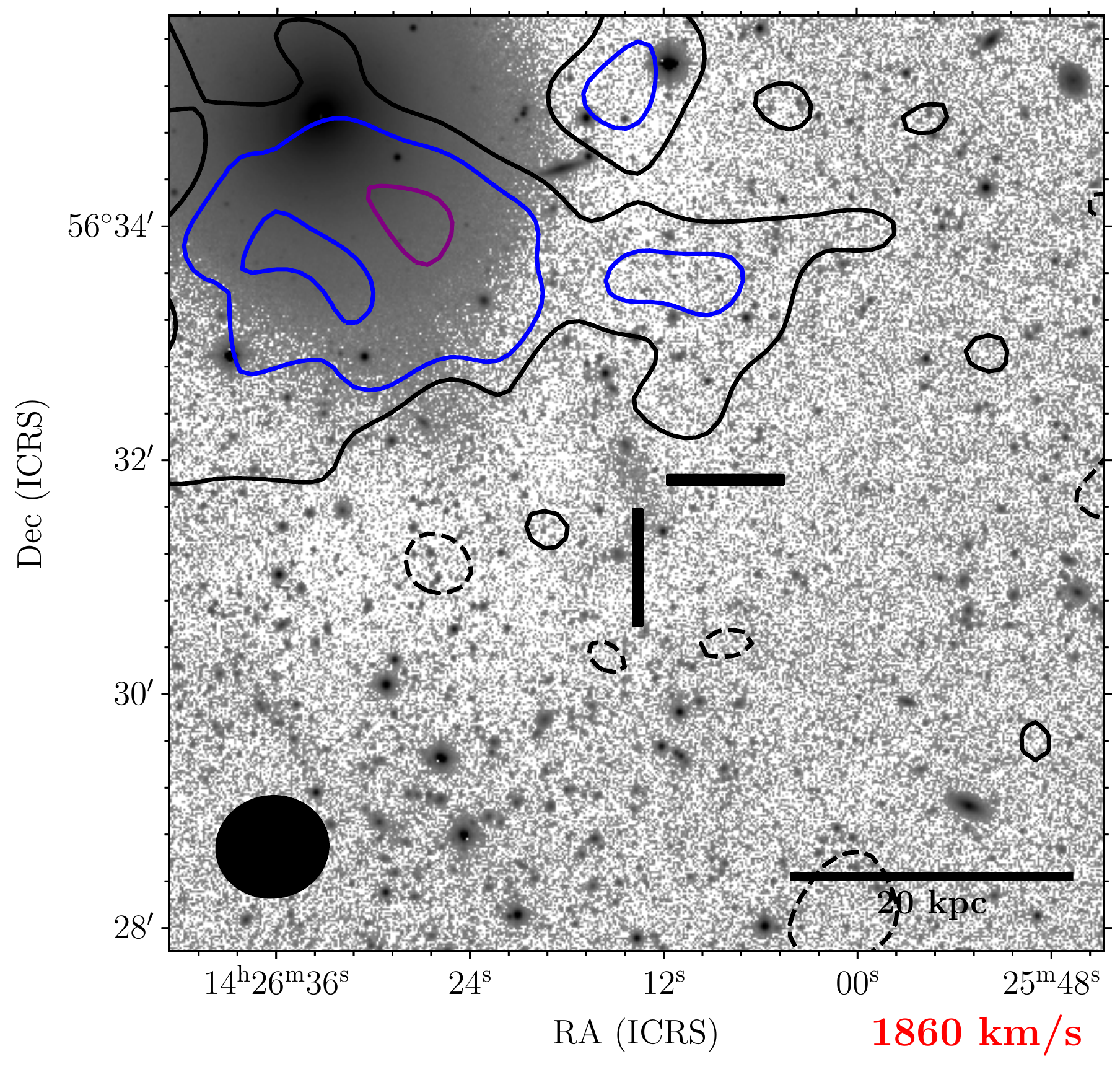}
    \includegraphics[width=0.32\textwidth]{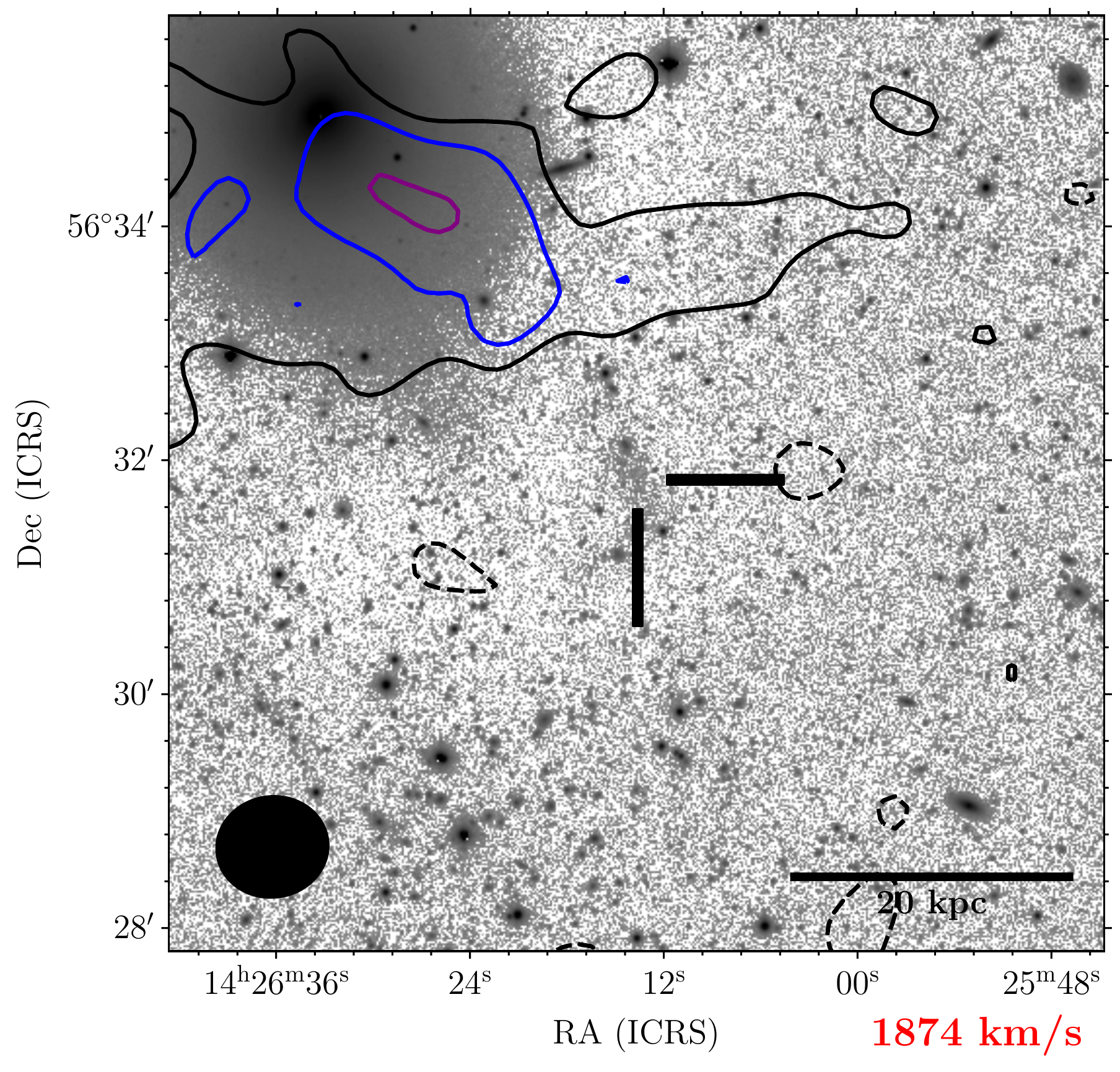}
    \includegraphics[width=0.32\textwidth]{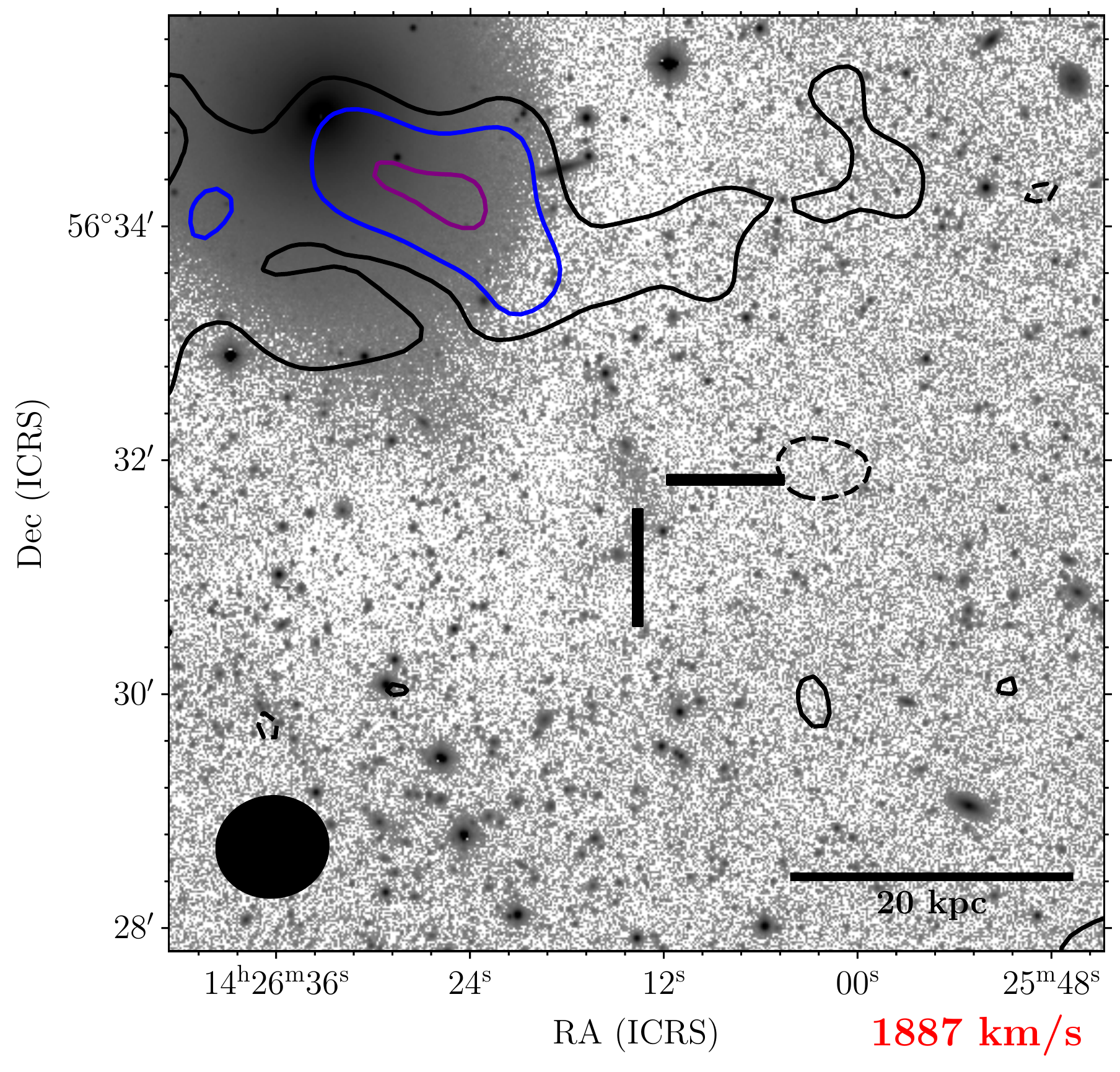}
    
    \includegraphics[width=0.32\textwidth]{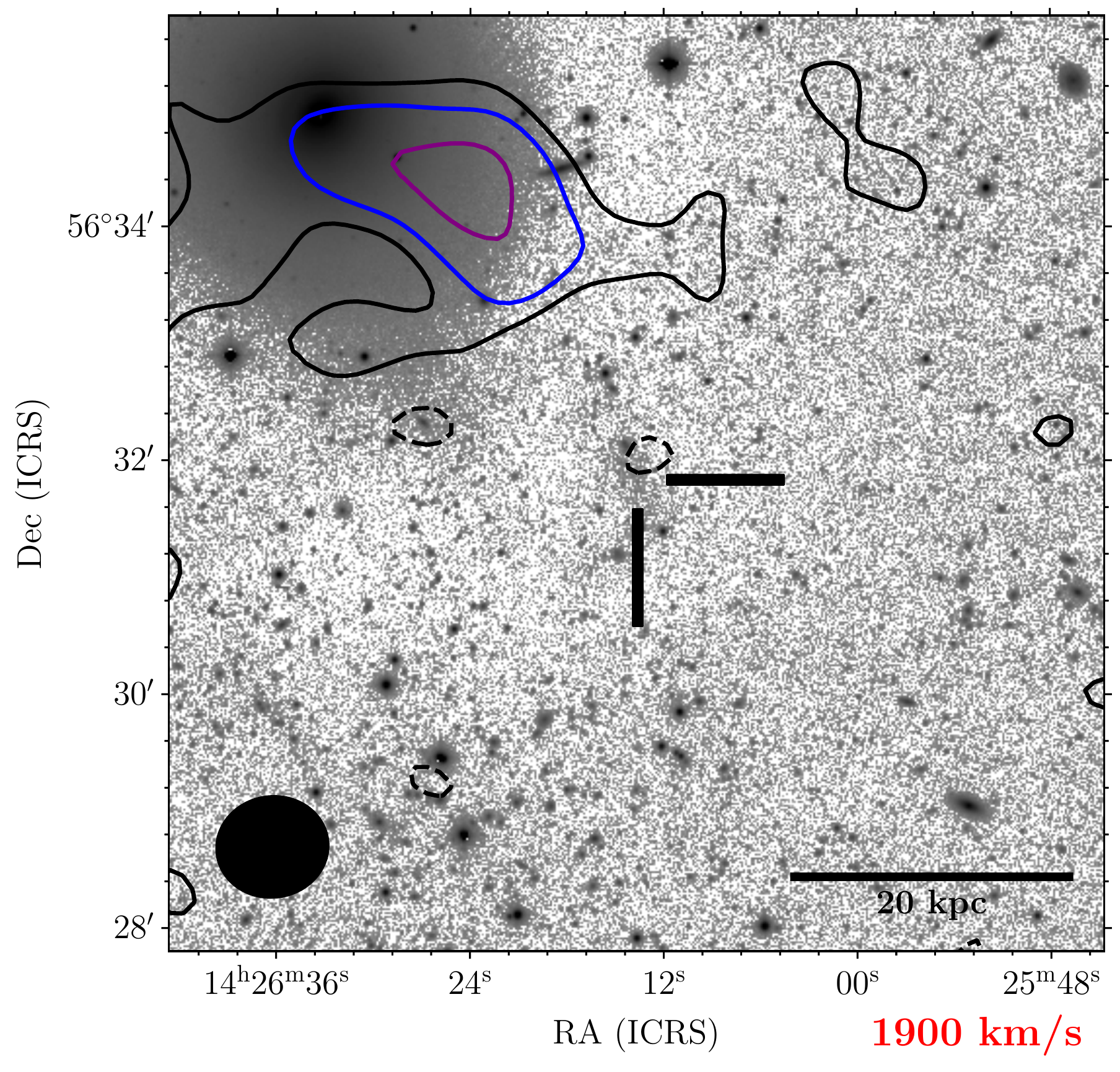}
    \includegraphics[width=0.32\textwidth]{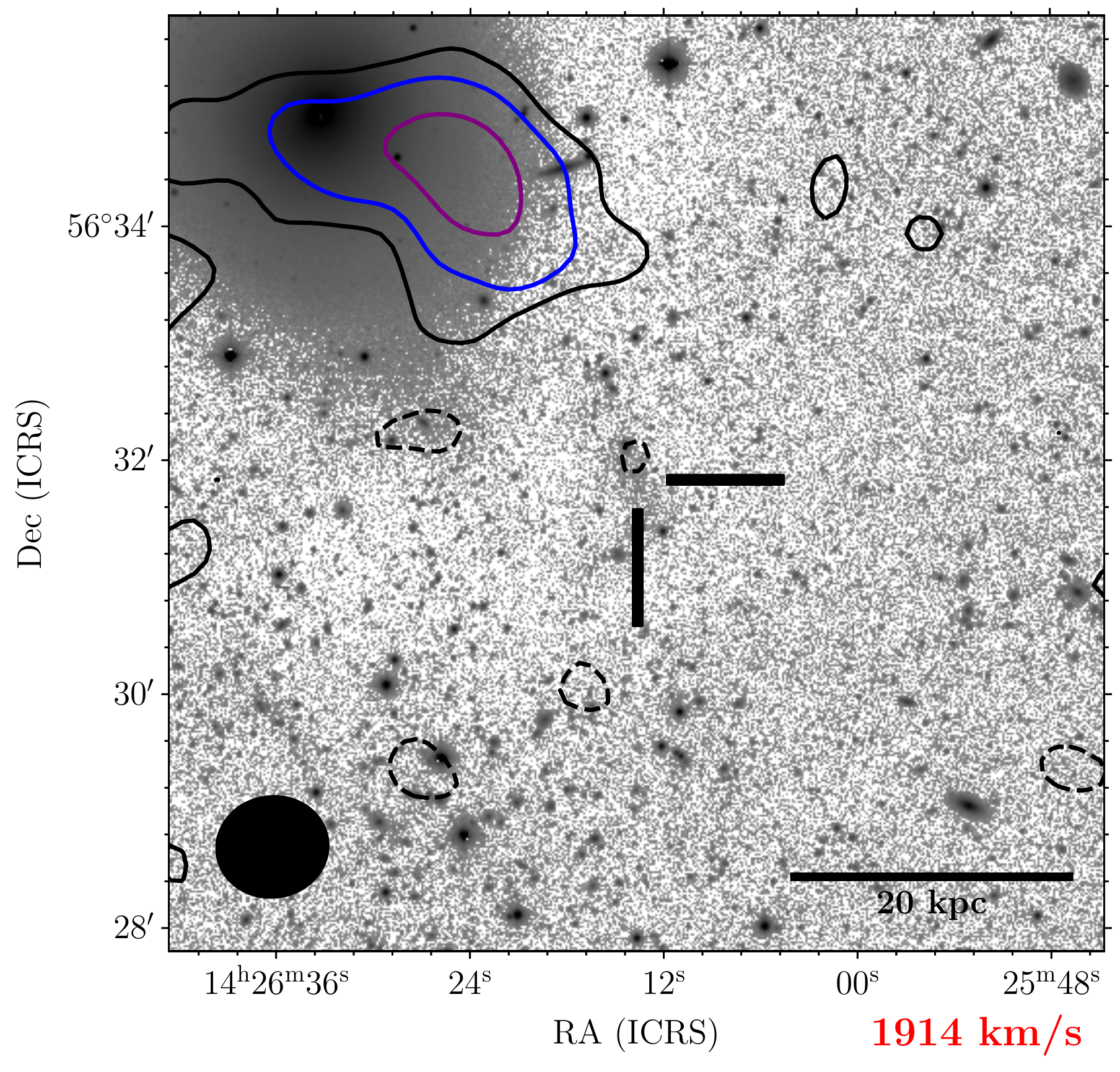}
    \includegraphics[width=0.32\textwidth]{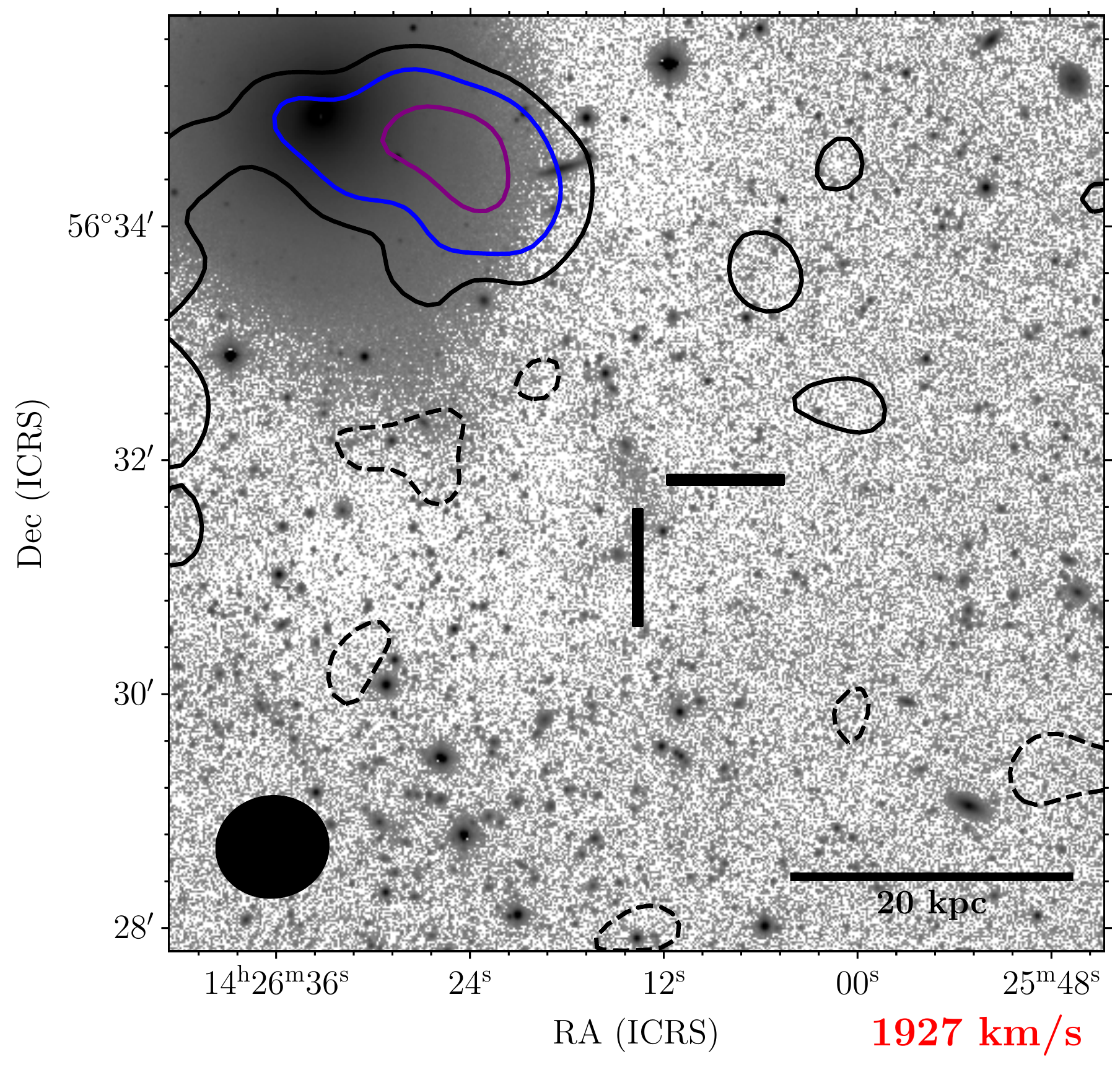}
    
    \includegraphics[width=0.32\textwidth]{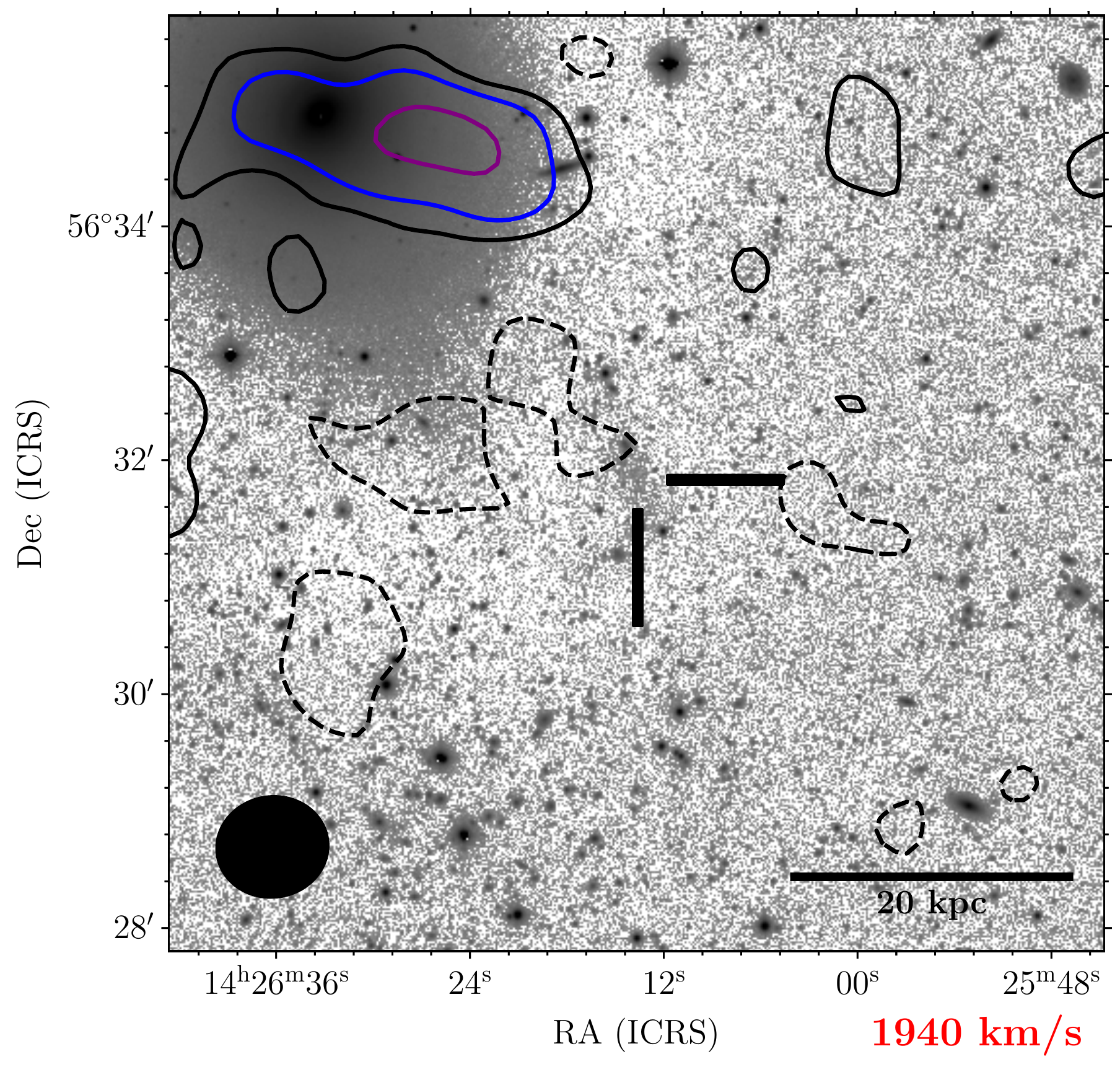}
    \includegraphics[width=0.32\textwidth]{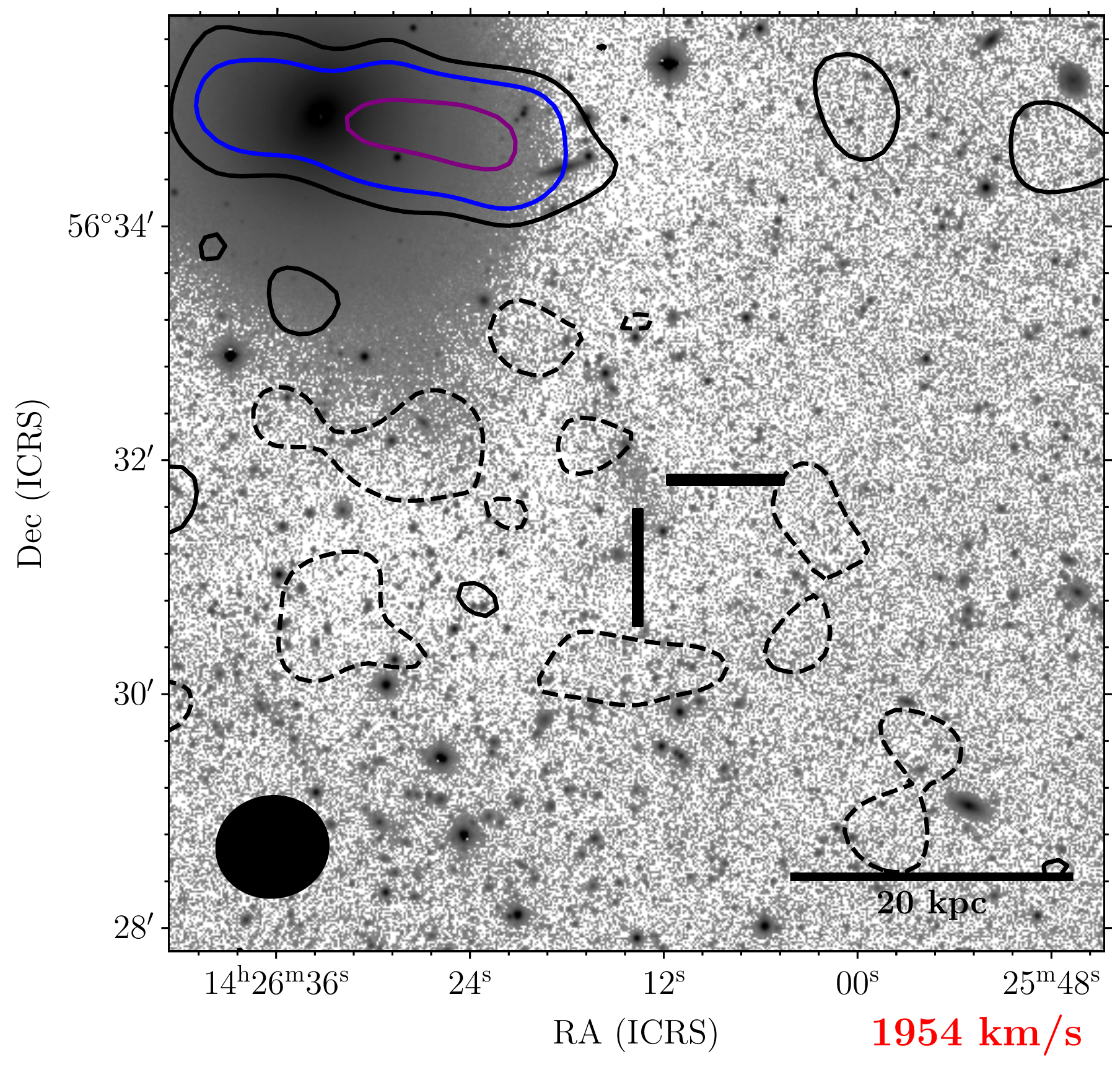}
    \includegraphics[width=0.32\textwidth]{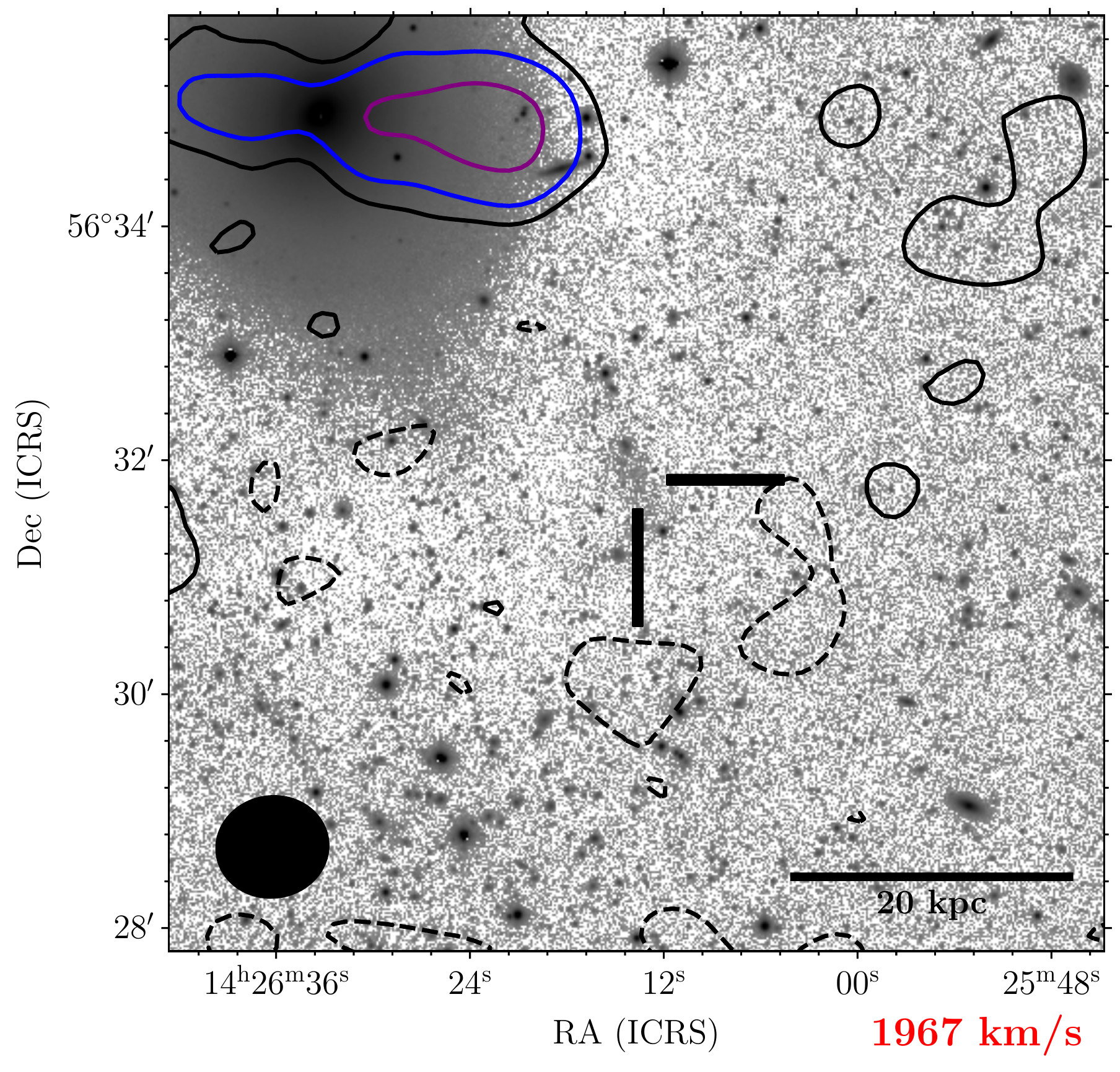}
\end{figure}

\section{VLA flux verification}
\label{sec:VLA_flux}

\begin{figure}
    \centering
    \includegraphics[width=0.5\columnwidth]{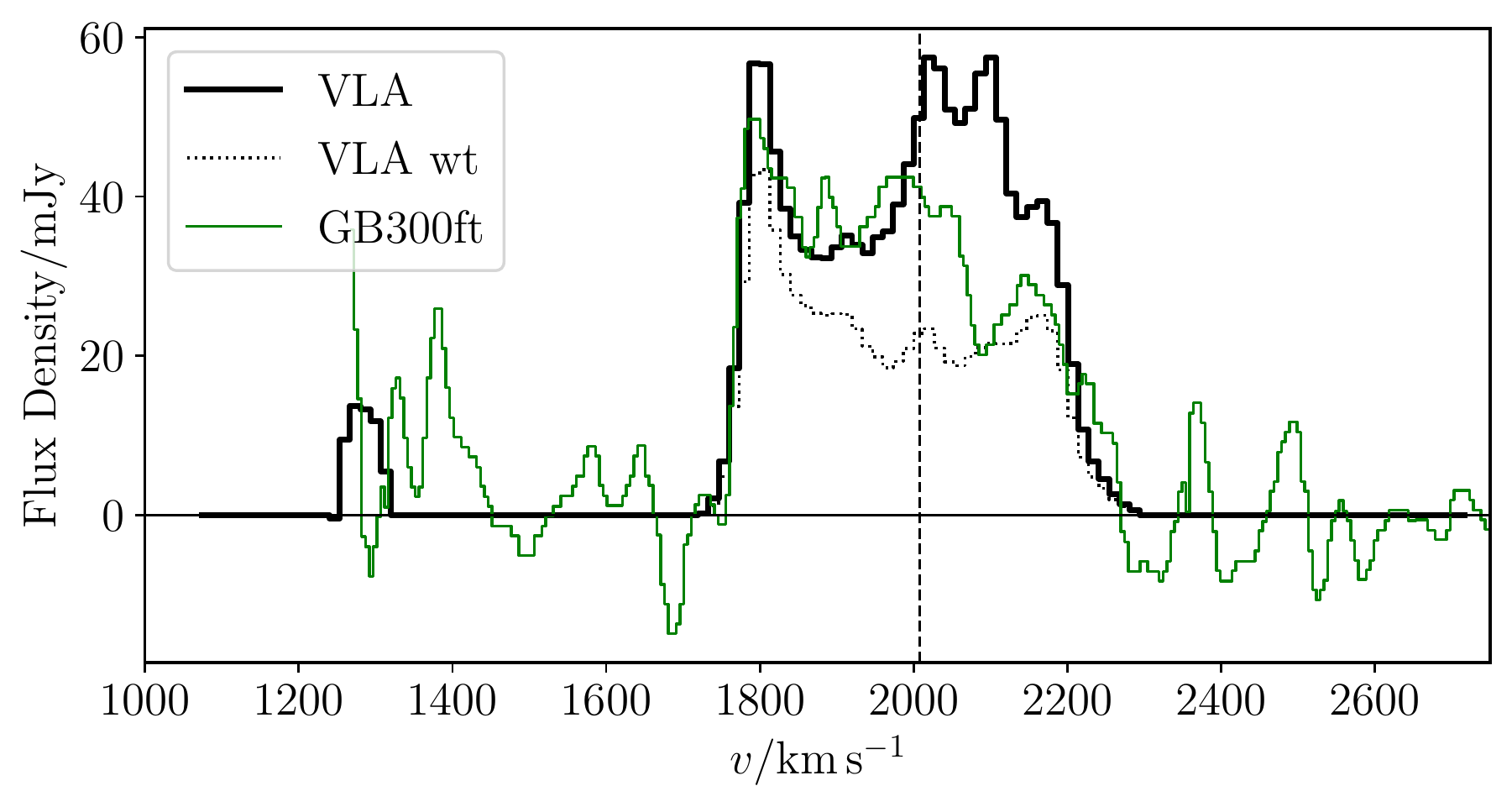}
    \caption{\hi \ spectrum of NGC 2708 from our observations with the VLA (thick black line) compared to the literature GB 300ft spectrum (green line), and the VLA observation accounting for the approximate beam response of the GB 300ft telescope (dotted black line). The vertical dashed line indicates the central velocity of NGC 2708.}
    \label{fig:NGC2708_spec}
\end{figure}

For verification purposes the integrated (with the 4$\sigma$ \texttt{SoFiA} mask) \hi \ spectrum of NGC 2708 was compared to an existing literature spectrum taken with the Green Bank (GB) 300 ft telescope \citep{Shostak1978}. As shown in Figure \ref{fig:NGC2708_spec} our VLA spectrum is above the single dish spectrum, which would usually point to a flux calibration issue. However, in this case it is more likely because this is a highly extended source that would have been partially resolved even with the GB 300 ft telescope. Approximating the GB 300 ft beam as a Gaussian of 10\arcmin \ HPBW and weighting the VLA cube by this beam model gives a spectrum that is in better agreement with the GB 300 ft spectrum. The central portion of the GB 300 ft profile sits between the weighted and unweighted VLA profiles, however, this is likely because a Gaussian is an oversimplification of the true beam response and a considerable fraction of the flux lies near the edge of the beam where this approximation is most unreliable.

\end{document}